\documentclass[12pt]{iopart}

\usepackage{amssymb}
\usepackage{graphicx}
\usepackage{stackengine}
\usepackage{geometry}
\usepackage{tikz}
\usepackage{subfig}

\begin{document}

\title[]{Large deviations of glassy effective potentials}

\author{Silvio Franz}%
\address{LPTMS, Universit\`{e} Paris-Sud 11, UMR 8626 CNRS, B\^{a}t. 100, 91405 Orsay Cedex, France}
\address{Dipartimento di Fisica Universit\`{a}, La Sapienza, Piazzale Aldo Moro 5, I-00185 Roma, Italy}

\author{Jacopo Rocchi}
\address{LPTMS, Universit\`{e} Paris-Sud 11, UMR 8626 CNRS, B\^{a}t. 100, 91405 Orsay Cedex, France}

\vspace{10pt}
\begin{indented}
\item[]
\end{indented}

\begin{abstract}
The theory of glassy fluctuations can be formulated in terms of
disordered effective potentials. While the properties of the average
potentials are well understood, the study of the fluctuations
has been so far quite limited.
Close to the MCT transition, fluctuations induced by the dynamical 
heterogeneities in supercooled liquids can be described by a cubic 
field theory in presence of a random field term.
In this paper we set up the general
problem of the large deviations going beyond the assumption of the vicinity to
$T_{MCT}$ and analyze it in the paradigmatic case of
spherical ($p$-spin) glass models. 
This tool can be applied to study the probability of 
the observation of a dynamics with memory of the initial condition in regimes where, typically, 
the correlation $C(t,0)$ decays to zero at long times, at finite $T$ and at $T=0$.


\end{abstract}

%
%
%
%
%

\section{Introduction}
The last two decades of research have underlined the deep role of
space-time fluctuations, a.k.a. dynamical heterogeneities, in the
relaxational dynamics of glassy systems \cite{berthier2011dynamical}.
Unfortunately, 
despite many efforts \cite{franz2000non,donati2002theory,bouchaud2004adam,biroli2004diverging,bouchaud2005nonlinear,berthier2007spontaneous}, an accomplished
theory of such fluctuations is still lacking. In \cite{franz2011field} it was proposed that relevant information could be gained
renouncing to describe the temporal dimension. Under a local
equilibrium hypothesis it was proposed to use an effective field
theory where the role of the order parameter is played by the space
dependent overlap of the actual system with a random thermalized
configuration. 
The construction provided a field theoretical extension
of the effective potential often used to describe the glass transition
in terms of a Landau-like theory. 
In that paper it was developed a theory of critical
fluctuations close to the putative Mode Coupling (MC) transition
temperature $T_{MCT}$, where MC theory predicts a growing dynamical length, see \cite{reichman2005mode} for a review. 
Under the strong hypothesis that any activated
processes could be neglected, a disordered effective field theory emerged, identical to that of
the spinodal point of a Random Field Ising Model.
The appearance of disorder in the description provided a foundation to the idea of “self-induced disorder”, a concept often advocated to rationalize the similarity between structural glasses and spin glasses \cite{binder1986spin,kirkpatrick1987p,mezard1987spin,kirkpatrick1988comparison,kirkpatrick1989scaling,fischer1993spin,parisi1994slow,franz1995glassy,bouchaud1998out}.
Similarly, an effective disorder was later found in the perturbative description of phase diagram of coupled glassy systems and systems where a fraction of the particles are frozen in random positions \cite{franz2013glassy,franz2013universality,biroli2014random}.
It is clear that to study the system away from -true or approximate- glassy critical points the analysis of fluctuations has to be extended to a non-perturbative level. In two recent papers, a brave attempt to describe the self-induced disorder at the non-perturbative level was undertaken. Unfortunately, the analysis is based on several approximations whose range of validity is difficult to assess \cite{biroli2018randomA,biroli2018randomB}.

We study the problem of non perturbative fluctuations in
the class of long-range spin glass models such as the spherical $p$-spin models in the pure
and the mixed versions.
These models are at the basis of the Random First Order Transition (ROFT), see \cite{biroli2012random} for a review.
Our basic object of study is the
effective potential function, defined as the large deviation
function of the overlap Probability Density Function (PDF) from a random reference configuration
chosen with the Boltzmann-Gibbs probability. In the thermodynamics limit,
this function is self-averaging with respect to the choice of the reference. Its shape reflects the properties of
the Gibbs measure, with a single minimum in the ergodic phase, and two
minima in glassy phases. For large but finite volumes, however, the
effective potential fluctuates from reference to reference and its
large fluctuations are themselves described by a
large deviation principle. 
Ideally we would like to study the fluctuation in shape of a
function. While theoretically conceivable, this is unfortunately a
technically formidable task. As a first step, we study
the fluctuations of neighboring pairs of points of the function. This
allows in particular to study the probability that the glassy
effective potential has a local minimum in regions where on average it
is either increasing or decreasing. In particular this gives the
probability of the existence of the secondary minimum in the high temperature
ergodic region. 
In terms of the dynamics, in the infinite size limit, the long time limit of the correlation with the initial condition $\lim_{t\rightarrow \infty}C(t,0)$ decays to zero at large temperatures.
On the other hand, in finite systems, as well as in numerical simulations, this may not be always true. 
The tools developed in the present paper allow to study the probability to observe a dynamics that does not lose memory of the initial configuration and to describe the features of such atypical dynamics.
After a short introduction of the glassy effective potential, we review the theory of fluctuations developed in \cite{franz2011field} and we describe how to extend it far from the MCT transition.
Finally we discuss the results at finite temperature and at $T=0$.
The more technical aspects of the computation are provided in the Appendix.

\section{Effective Potential: a Short Introduction}

A good starting point to study fluctuations in glasses is provided by the glassy effective potential, and its field theoretical generalizations, which have been described several times in the literature. Given a system specified by all its coordinates $X$ and a Hamiltonian
$H(X)$, and some notion of local similarity -or overlap- between configurations
$q_x(X,Y)$, one defines \cite{franz1995recipes}
\begin{eqnarray}
  \label{eq:1}
&  V_N(\{q(x)\}|X) & = -\frac T N (\log Z[\{q(x)\}|X] - \log Z_u )\:,\\
&  Z[\{q(x)\}|X] & = \sum_Y  e^{-\beta H[Y]}\mathbb{I}_{q_x(X,Y)=q(x)}\:,
\end{eqnarray}
where $Z_u$ is the unconstrained partition function $Z_u=\sum_Y  e^{-\beta H[Y]}$ and $\beta$ the inverse temperature. For reasonable choices of the configuration $X$, this is an effective field theoretical action for the overlap field $q(x)$. The
configuration $X$ is chosen randomly from an equilibrium measure at
the same or at a different temperature $T'$ from the one appearing in
(\ref{eq:1}),  $V(\{q(x)\}|X)$ is then a random object.
The $X$-average of $V$ has been well studied in mean-field models and numerical simulations \cite{coluzzi1998approach,parisi2009replica,cammarota2010phase,parisi2014liquid}. A few papers have studied fluctuations \cite{berthier2013overlap,berthier2015evidence}. In atomistic glass forming liquids, large fluctuations of the overlap emerge as temperature decreases, consistent with the existence of the random critical point that is predicted by effective field theories. From its
form one
can get important
insights of glassy behaviour; 
for example, point-to-set correlation functions of a set $B$ \cite{berthier2012static,biroli2008thermodynamic}, can be
obtained fixing $q(x)=1$ outside the set $B$, averaging over $X$ and
studying the overlap distribution inside $B$ \cite{franz2007analytic}. As we said, besides the mean,
the fluctuations provide crucial information on the nature of glassy
phases. 

A systematic approach starting from mean-field, and including the
space dimension in a controlled expansion seems desirable. 
As a preliminary step we study the finite volume fluctuations of 
the glassy effective potential in mean field in the large deviation regime.
To keep the level of difficulty to the minimum, we use the spherical $p$-spin model \cite{kirkpatrick1987p,crisanti1992sphericalp,crisanti1993sphericalp}, where metastable states can be studied with the TAP method \cite{thouless1977solution,rieger1992number} and replicas give rise to closed equations.
The fully connected spherical (mixed)
$p$-spin model, is defined by the Hamiltonian
\begin{equation}
\mathcal{H}[\sigma]=\sum_p  \frac{a_p}{p!}\sum_{i_{1}\ldots i_{p}}J_{i_{1}\ldots i_{p}}\sigma_{i_{1}}\ldots \sigma_{i_{p}}\:,
\end{equation}
where the $J_{i_{1}\ldots i_{p}}$  are independent Gaussian random variable with
mean $\overline{J_{i_{1}\ldots i_{p}}}=0$ and variance
$\overline{J_{i_{1}\ldots i_{p}}^{2}}=p!/(2N^{p-1})$ and $a_p\ge 0$
coefficient such that given two spin configurations $\sigma$ and $\tau$,
denoting $q_{\sigma,\tau}=\frac 1 N \sum_i \sigma_i \tau_i'$ their overlap, 
\begin{eqnarray}
  \label{eq:2}
&&  \overline{H[\sigma]H[\tau]}=N f(q_{\sigma,\tau})=\frac 1 2\sum_p a_p^2 q_{\sigma,\tau}^p.  
\end{eqnarray}
The $N$ spins $\sigma_i$ are continuous variables constrained to be on
the sphere $\sum_{i=1}^NS_{i}^{2}=N$. The pure $p$-spin model corresponds to a single non vanishing
coefficient $a_p$, while in mixed models, more than one coefficient is
non-zero. 
It can be studied within the framework of replicas and it may be described in terms of the so called one step Replica Symmetry Breaking (RSB) transition \cite{mezard1990spin}.
Moreover, its Langevin dynamics was shown to display the interesting behaviour known as aging and weak ergodicity breaking \cite{cugliandolo1993analytical,bouchaud1992weak}.

As described many times \cite{kirkpatrick1987dynamics,kurchan1993barriers,crisanti1995thouless,monasson1995structural,franz1995recipes}, see \cite{barrat1997p,castellani2005spin,zamponiRev} for reviews on the topic, while at large temperature the system is in the paramagnetic phase, an ergodicity breaking transition occurs at the dynamical temperature $T_d$, called dynamical temperature. The paramagnetic state disappears and the Boltzmann measure is replaced by an exponential number of
equilibrium states, whose number $e^{N\Sigma}$ is exponential in
the system size and $\Sigma$ is the complexity, or configurational entropy, and zero overlap among each other.
As mentioned above, this behavior can be interpreted in terms of the broader perspective of the RFOT \cite{biroli2012random}.
The dynamics of this model is described by a set of equations
formally identical to those obtained by MCT (mode coupling theory)
for liquids, reproducing the two steps relaxation behaviour of the
correlation function and the dynamical slowing down at $T_{MCT}$
\cite{crisanti1993sphericalp}. 

\begin{figure}
\centering{}\includegraphics[scale=0.5]{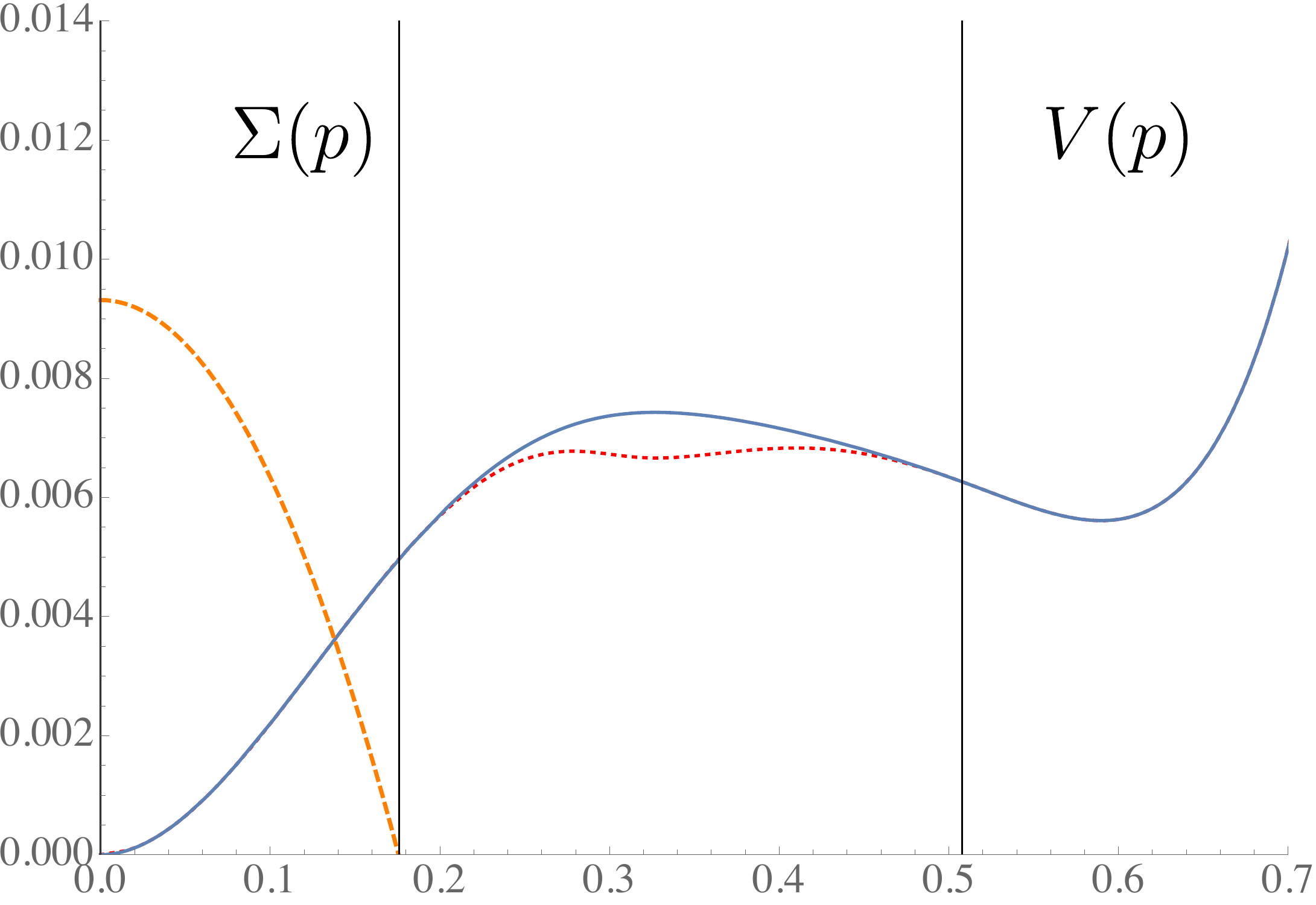}\caption{\label{Figpotdyn}$V(p)$ for the $3-$spin at $\beta=1.66$, larger than $\beta_{d}$, computed with the 1RSB anstaz (blu line) and the complexity $\Sigma(p)$ (orange dashed) as a function
of $p$. The RS potential is also plotted for comparison (red dotted). Vertical lines divide three regions: from left to right dynamic 1RSB, static 1RSB, RS.}
\end{figure}

Since there is no space in the model, the definition (\ref{eq:1})
reduces to the global effective potential 
\begin{eqnarray}
  \label{eq:3}
&& V_N(p|\tau) = -\frac T N (\log Z[p|\tau]-\log Z_u)\:,\\
&& Z[p|\tau]=\int D\sigma e^{-\beta H[\sigma]}\delta(q_{\sigma,\tau}-p)\:,
\end{eqnarray}
where $D\sigma$
denotes the uniform measure on the sphere. In the following we drop $N$ from $V_N(p|\tau)$ to ease the notation. We observe that the ($\tau$ dependent) potential can be written as $V(p|\tau) =F(p|\tau) - F$, where $F(p|\tau)$ is the ($\tau$ dependent) constrained free energy. We concentrate in this
paper to the case where the configuration $\tau$ is chosen from the
equilibrium distribution at temperature $T'$ larger than the Kauzmann
transition temperature $T_K$, where fluctuations with respect to the
couplings are very small and not only the free-energy, but the
partition function $Z_u$ has the self-averaging property and can be
computed in the annealed approximation. This property allows to introduce replicas only to
deal with the logarithm of the partition function. Thus, using the
relation $\overline{\log x}=\lim_{n\rightarrow0}\partial_{n}\overline{x^{n}}$, the constrained free energy reads
\begin{equation}
F(p)=\overline{\mathbf{\mathbb{E}_{\tau}}F(p|\tau)}=-\frac{1}{\beta N}\lim_{n\rightarrow0}\frac{\partial}{\partial n}\overline{\mathbb{E}_{\tau}\left(Z[p|\tau]\right)^{n}},\label{eq:deffcons1}
\end{equation}
where we get the limit from the analytic continuation from integer values of $n$. 
Details on the replica computation are provided in the Appendix.
The typical potential $V(p)=F(p)-F$ depends on the temperatures. For $T'=T$, at high
temperature, $T>T_d$ the potential has a single minimum
at $p=0$. Below $T_d$, one sees a characteristic 
two minima structure with a secondary minimum at a high value of
$p=q_{EA}$, where $q_{EA}$ is the Edward-Anderson parameters, that signals ergodicity breaking in an exponential number of
states. The difference between minima, the configurational
entropy times the temperature, tends to zero at the temperature
of the static phase transition $T_K$ and remains zero below.  
The profile of the entropy and of the potential in the dynamical phase is given in Fig. \ref{Figpotdyn}. 
It is possible to interpret the potential in terms of the PDF of the overlap.
In fact, the secondary minimum corresponds to the unlikely event that the second replica
is extracted from the same state of the first one. In this case, the
overlap between these two replicas would be that of the state where
the first replica is. Since the first replica is sampled at equilibrium,
the overlap of this state is $q_{EA}$. Moreover, since the typical
overlap between different equilibrium states is zero, one finds that
the global minimum continues to be at $p=0$.

As illustrated in Fig. \ref{Figpotdyn} and discussed in the Appendix, the interval $p\in [0,1]$ is divided into three regions \cite{barrat1997temperature,capone2006off}.
Their physical interpretation is the following: while at small $p$ the second replica feels a small constraint due to the first one, and it can explore an exponential number of states in the dynamical phase, this is not true anymore when $p$ is increased and $p>p_K$. For $p<p_K$ the complexity is larger than zero while for $p>p_K$ the system can explore only a sub-exponential number of states. Nevertheless, when the constraint is too strong, the second replica is forced to stay in the same state as the first one. This last regime must be Replica Symmetric (RS). From now on we denote by $p_{RS}(T)$ the point at the frontier between these two regions and we will discuss its properties afterwards.
For large enough temperatures, $\beta<\beta_{RS}$, the replica symmetric instability disappears and the potential is always replica symmetric. On the other hand, when $\beta>\beta_{RS}$ the replica symmetric ansatz does not estimate the potential correctly at intermediate values of $p$ and for $\beta>\beta_d$ spurious stationary points appear, see Fig. \ref{Figpotdyn}.
For $\beta<\beta_d$, before the $3$ regions observable in Fig. \ref{Figpotdyn}, there is a preliminary replica symmetric region as discussed in the Appendix.

In the thermodynamic limit, $V$ is self-averaging, both with respect
to the extraction of $\tau$ and the quenched couplings $J$ of the
model. However, for finite $N$, fluctuations are present. It was shown
in \cite{franz2011field} that thanks to the self-averaging of $Z_u$, the
fluctuations with respect to the $J$'s are much weaker than those with
respect to $\tau$ (this is one of the reasons why $p$-spin models are
good models of structural glasses). In the following we then
concentrate in the study of the large deviations with respect to the
reference configuration $\tau$.

\section{Fluctuations}


In this section we review the method of the effective potential to describe fluctuations in glasses. 
Leaving aside the difficulties arising with the mixed model in the $T'\neq T$ case \cite{franz1995glassy,barrat1997temperature,sun2012following,folena2019memories} that we will discuss afterwards, the shape of potential can be interpreted dynamically. The existence of a unique minimum is associated with ergodic behavior. If we consider relaxation dynamics with the reference replica 
as initial condition, the system will evolve till it will have zero correlations with the initial configuration. Conversely, if there are two minima, the system is not ergodic and it will remain confined in the vicinity of the initial state, at the value of the overlap of the secondary minimum. 
 The configuration space is split onto ergodic components, and the difference between the two minima just measures the configurational entropy (multiplied by temperature) of the ergodic components. 
For large $N$ the secondary minimum appears at a well defined temperature $T_d$, 
signaling a sharp breaking of ergodicity. In this paper we are concerned with the finite $N$ fluctuations, and we ask what is the probability of initial conditions capable to confine the system for an exponentially large time, even in the paramagnetic phase. Namely, we want to compute the probability that 
the potential has a minimum for $T>T_d$.  This is a large deviation regime where the probability is exponentially small in $N$ and we are interested in the rate function   
 $I(T)$.
We will use two strategies. The first
one, valid close to $T_d$, both from below and from above, is based on the perturbative theory of glassy fluctuations developed in \cite{franz2011field}. The second will use full fledged large deviation analysis and will be valid for all $T>T_d$. 

\subsection{Small fluctuations}

In this section we consider the temperature $T$ to be close to $T_d$. 
The central quantity we need is the covariance of the potential function for two different 
values of the overlap, with respect to the extraction of the
first replica 
\begin{equation}
W_{het}^{(2)}=\overline{\mathbf{\mathbb{E}_{\tau}}\left(V(p_{1}|\tau)V(p_{2}|\tau)\frac{}{}\right)-\mathbf{\mathbb{E}_{\tau}}\left(V(p_{1}|\tau)\right)\mathbf{\mathbb{E}_{\tau}}\left(V(p_{2}|\tau)\right)}.
\label{Whet0}
\end{equation}
Being small fluctuations Gaussian, this quantity specifies completely their statistics.
The covariance  $W_{het}^{(2)}$ can be computed within mean-field theory using replicas, starting from the identity
\begin{eqnarray}
W_{het}^{(2)} & =  \overline{\mathbf{\mathbb{E}_{\tau}}\left(\delta V(p_{1}|\tau)\delta V(p_{2}|\tau)\frac{}{}\right)}\nonumber  \\ & = \frac{T^{2}}{N^{2}}\lim_{n_{1}\rightarrow0}\lim_{n_{2}\rightarrow0}\frac{\partial^{2}}{\partial n_{1}\partial n_{2}}\log \overline{\mathbf{\mathbb{E}_{\tau}}\left(Z^{n_{1}}[p_{1}|\tau]Z^{n_{2}}[p_{2}|\tau]\frac{}{}\right)}\:
\label{Whet11}
\end{eqnarray}
and getting the limit from the analytic continuation from integer values of $n_1$ and $n_2$. This identity is discussed in the Appendix. 
We will not consider other sources of fluctuations. In fact, disorder or sample to sample fluctuations 
\begin{equation}
W_{dis}^{(2)}=\overline{\mathbf{\mathbb{E}_{\tau}}\left(V(p_{1}|\tau))\mathbb{E}_{\tau}(V(p_{2}|\tau)\right)}-\overline{\mathbf{\mathbb{E}_{\tau}}\left(V(p_{1}|\tau)\right)}\:\overline{\mathbf{\mathbb{E}_{\tau}}\left(V(p_{2}|\tau)\right)}\label{eq:disflucsamtosamdef}
\end{equation}
can be shown to be subdominant with respect to fluctuations in $\tau$ \cite{franz2011field}. From now on,
we drop the subscript ``het'' and we denote eq. (\ref{Whet11}) by $W^{(2)}_{p_1,p_2}$. Connected correlations are computed respect to the measure $\mathbb{E}_{\tau}$,
e.g. $\mathbb{E}_{\tau}[AB]_{c}=\mathbb{E}_{\tau}AB-\mathbb{E}_{\tau}A\:\mathbb{E}_{\tau}B$,
describing fluctuations with respect to the first replica. Eventually
we will consider quantities averaged over the disorder, $\overline{\mathbb{E}_{\tau}[AB]_{c}}=\overline{\mathbb{E}_{\tau}AB-\mathbb{E}_{\tau}A\:\mathbb{E}_{\tau}B}$.
We also denote by $\left<A\right>=\overline{\mathbb{E}_{\tau}(A)}$
and $\left<AB\right>_{c}=\overline{\mathbb{E}_{\tau}(AB)}-\overline{\mathbb{E}_{\tau}(A)\mathbb{E}_{\tau}(B)}$.

Given an integer number $n$, we introduce the fixed overlap
replica action
\begin{equation}
e^{\frac{N}{2}S[Q,n]}=\overline{\int\prod_{a=0}^{n}Ds^{a}e^{-\beta\sum_{a=0}^{n}H(s^{a})}\prod_{a,b}\delta(Q_{ab}-q(s^{a},s^{b}))}
\end{equation}
where we defined $\tau=s^{0}$ , $q(s,s')=N^{-1}\sum_{i}s_{i}s_{i}'$ and $Q_{ab}$
is a square symmetric matrix of size $(1+n)\times (1+n)$. Correlation functions 
can be expressed in terms of this action. Let us first observe that if we take all the overlaps $Q_{0a}=p$, ($a=1,...,n$), we can get the effective potential as 
\begin{equation}
e^{\frac{N}{2} S^{(1)}[p] }=\left< \left(Z[p|\tau]\right)^n \right> =\int\mathcal{D}Q_{ab}e^{\frac{N}{2}S[Q,n]}\prod_{a=1}^{n}\delta(p-Q_{0a})\:,
\label{eq:defS1firstappea}
\end{equation}
where $\mathcal{D}Q_{ab}$ denotes the integration over the parameters of
the overlap matrix, and 
\begin{equation}
V(p)=F(p)-F=-\frac{1}{N\beta}\lim_{n\rightarrow0}\frac{\partial}{\partial n}\frac{N}{2}S^{(1)}[p]-F\,.
\end{equation}
As usual, the integration over the non-constrained overlaps can be performed by taking the saddle point. As described in the Appendix, depending on the temperature and on the value of the overlap $p$, the saddle point can have a replica symmetric structure $Q_{ab}=q$ for $a\ne b$ $a,b=1,..., n$, or a 
1RSB one, described by three parameters $(q^1,q^0,x)$. 

Analogously, writing $n=n_1+n_2$ and imposing $Q_{0a}=p_1$ for $a=1,...,n_1$ and $Q_{0a}=p_2$ for $a=n_1+1,...,n_1+n_2$ we get the potential covariance from eq. (\ref{Whet11}),
\begin{equation}
W^{(2)}_{p_1,p_2} =\frac{T^{2}}{2N}\lim_{n_{1}\rightarrow0}\lim_{n_{2}\rightarrow0}\frac{\partial^{2}S^{(2)}[p_1,p_2]}{\partial n_{1}\partial n_{2}}
\end{equation}
where
\begin{eqnarray}
\label{eq:defS2actiondeff} e^{\frac{N}{2}S^{(2)}[p_1,p_2]} & = \left< \left(Z[p_1|\tau]\right)^{n_1} \left(Z[p_2|\tau]\right)^{n_2} \right> =    \\ & = \int\mathcal{D} Q_{ab}  e^{\frac{N}{2}S[Q,n_{1}+n_{2}]}  \prod_{a=1}^{n_{1}}\delta(p_1-Q_{0a})\prod_{a=n_{1}+1}^{n_{1+}n_{2}}\delta(p_2-Q_{0a}), \nonumber 
\end{eqnarray}
which again is computed by saddle point.
Let us notice that if $p_1=p_2=p$, 
\begin{equation}
e^{ \frac{N}{2} S^{(2)}[p,p]} =  \left<\left(Z[p|\tau]\right)^n \right> = e^{\frac{N}{2}S^{(1)}[p]} \:.
\label{eqduetouno}
\end{equation}
Thus, when $p_1$ and $p_2$ are close enough to $p$, we expect the saddle point values of $e^{ \frac{N}{2} S^{(2)}[p_1,p_2]} $ to be close to those of  $e^{\frac{N}{2}S^{(1)}[p]}$.
 In order to compute small fluctuations of the overlap close to the minimum that emerges at the dynamical transition, we can expand the action around the saddle point values found at $T_d$. This expansion leads to
\begin{equation}
NW^{(2)}_{\phi_1,\phi_2}=\frac{T_{d}^{2}}{2}\lim_{n_{1}\rightarrow0}\lim_{n_{2}\rightarrow0}\frac{\partial^{2}S^{(2)}[\phi_1,\phi_2]}{\partial n_{1}\partial n_{2}}=A+B(\phi_1+\phi_2)+C\phi_1\phi_2\:,
\label{eq:W2expphi1phi2}
\end{equation}
where $\phi_1=p_1-p_d$ and $\phi_2=p_2-p_d$.
For $T$ close enough to $T_d$, the potential has the linear behavior
$V(p=p_{d}+\phi)=V(p_{d})+\mu(T-T_{d})\phi + O(\phi^3)$, with $\mu>0$.
The condition $\mu>0$ insures that the secondary minimum disappears for $T>T_d$. 
On the other hand, eq. (\ref{eq:W2expphi1phi2}) suggests that the $\tau$-dependent potential can be rewritten
as the typical value $NV(p=p_{d}+\phi)$ plus a small correction
given by

\begin{equation}
\sqrt{N}\delta V(p=p_{d}+\phi|\tau)=\eta\phi+\alpha\:,
\end{equation}
where $\eta$ and $\alpha$ are correlated Gaussian random fields and, comparing with eq. (\ref{eq:W2expphi1phi2}), $\left<\eta^2\right>=C$.
While $\alpha$ represents a random correction to the value of the
free energy, $\eta$ is a random temperature term. 
This result leads to the observation that the secondary minimum can exist even for $T>T_{d}$ if $\sqrt{N}\mu(T-T_{d})+\eta=0$. In other words, the random fluctuation $\eta$ allows the re-appearance of the secondary minimum, even when typically it does not exists. The probability of this event can be written in terms of the rate function $I(T)$ computed in the small fluctuations regime,
\begin{equation}
e^{-NI_{SF}(T)}=\int_{-\infty}^{-\sqrt{N}\mu(T-T_{d})}d\eta P(\eta)
\end{equation}
where, denoting by $\sigma^{2}$ the variance of $\eta$,
\begin{equation}
I_{SF}(T)=\frac{1}{2}\frac{\mu^{2}(T-T_{d})^{2}}{\sigma^{2}}\:.\label{eq:smallflucIT}
\end{equation}
The parameters $\mu$ and $\sigma$ are given by 
\begin{equation}
\mu(T-T_{d})=\left. \frac{dV(p)}{dp} \right|_{p=p_{d}}
\end{equation}

\begin{equation}
\sigma^{2}=\left. \lim_{n_{1}\rightarrow0}\lim_{n_{2}\rightarrow0} \frac{T^{2}}{2}\frac{d}{d\phi_1}\frac{d}{d\phi_2}\frac{\partial^{2}S^{(2)}[\phi_1,\phi_2]}{\partial n_{1}\partial n_{2}}\right|_{\phi_1=0,\phi_2=0},
\label{eq:computemuIsd}
\end{equation}
and their computation is described in the Appendix. This result can be extended also for $T<T_d$, where typically the secondary minimum exists. In this case, random fluctuations play the opposite role, creating a saddle. The condition for this to happen is still $\sqrt{N}\mu(T-T_{d})+\eta=0$, where now $\mu<0$.
Finally we observe that generally this computation cannot be extended for a finite $|T-T_d|$ because there are no guarantees that the disorder induced by the initial condition for the dynamics can be described in terms of a Gaussian random term in a cubic field theory well above the dynamical temperature. In order to tackle this problem, we develop a novel technique, independent of this theory and based on a first principles computation. This technique is general and it may be applied in a general context, when one is interested in the computation of the rate function of the probability of the existence of stationary points in functional depending on some source of randomness.

\subsection{Large fluctuations}
In this section we extend the results discussed previously by looking at the probability of the existence of a secondary minimum in the effective potential for an arbitrary $T$.
This task requires controlling the values of the potential in multiple points. 
Our strategy is based on taking two points $p_1$ and $p_2$, distant $\delta p$, and considering the difference $V_{2}-V_{1} = V(p_2|\tau)-V(p_1|\tau)$. In the absence of stationary points, $V$ is locally linear and $V_{2}-V_{1}$ of order $O(\delta p)$. This is the typical situation above $T_d$ and $p_1\neq 0$. Conversely, close to a stationary point $V$ is quadratic. In order to describe the rare appearance of a stationary point in this regime we then look for the probability that $V_{2}-V_{1} = V(p_2|\tau)-V(p_1|\tau) \sim O(\delta p^2)$, which has a large deviation form.

We first introduce our notation and then we discuss the computation of the large deviation functional. Even if fluctuations induced by the quenched disorder are sub-dominant with respect to sample-to-sample fluctuations, to be as clear as possible we maintain the subscript $J$ in quantities that are formally dependent of the quenched disorder.
We consider  the probability  $P(V_{2}-V_{1},p_{1},p_{2})$ that, given a configuration $\tau$,
the difference between $V(p_{2}|\tau)$ and $V(p_{1}|\tau)$ is $V_{2}-V_{1}$,
\begin{equation}
P(V_{2}-V_{1},p_{1},p_{2})=e^{-NG_J(V_{2}-V_{1},p_{1},p_{2})}\:,
\label{eq:PV2V1p1p2definit}
\end{equation}
where $G_J$ is the rate function of this probability at given quenched disorder. By definition,
\begin{equation}
P(V_{2}-V_{1},p_{1},p_{2})=\mathbb{E}_{\tau} \delta\left[ \frac{}{} V_2 - V_1 - (V(p_2|\tau) - V(p_1|\tau))\right]\:,
\end{equation}
and using the exponential representation of the delta function we can write it in the form
\begin{equation}
P(V_{2}-V_{1},p_{1},p_{2})=e^{-N m (V_2-V_1)} Z_{J}(m,p_{1},p_{2})\:,
\label{P1P2deffirstquadsssds}
\end{equation}
where the last term is 
\begin{equation}
Z_{J}(m,p_{1},p_{2})=\mathbb{E}_{\tau}e^{Nm(V(p_{2}|\tau)-V(p_{1}|\tau)}=e^{N\Gamma_{J}(m,p_{1},p_{2})}\:.
\label{eq:ZmdeltaVLD}
\end{equation}
and it can be written in terms of the generating function of the connected correlation functions $\Gamma_{J}(m,p_{1},p_{2})$.
Taking the average over $J$, we finally obtain the average rate function
\begin{equation}
G(V_{2}-V_{1},p_{1},p_{2})=\left[(V_{2}-V_{1})m^{*}\frac{}{}-\Gamma(m^{*},p_{1},p_{2})\right]\:,
\label{eq:lagrangeTV2V1p2p1}
\end{equation}
where $\Gamma(m^{*},p_{1},p_{2})=\overline{\Gamma_{J}(m^{*},p_{1},p_{2})}$
and the parameter $m^{*}$ is the solution of the equation 
\begin{equation}
V_{2}-V_{1}=\frac{\partial\Gamma(m^{*},p_{1},p_{2})}{\partial m}\:.
\label{eq:legendretranscond1}
\end{equation}
$G$ can be related to the large deviation function of the existence of the secondary minimum of the potential $I$, as illustrated below. For the reasons discussed above, we set $p_2$ close to $p_1$, $p_{1}=p-\delta p/2$, $p_2=p+\delta p/2$.
The absolute value of $m^*$ grows as we require $V_2-V_1$ to be far from its equilibrium value, being exactly $0$ when this difference is chosen to be $\left<V(p_2|\tau) - V(p_1|\tau)\right>$.  This could lead to the conclusion that it is sufficient to look at a low order truncation of the series of $\Gamma$ in powers of $m$, where successive terms appears to be of higher and higher order in $\delta p$:
\begin{eqnarray}
\Gamma_{J}(m,p_{1},p_{2}) & = \sum_{k=1}^{\infty}\frac{m^{k}}{Nk!}\frac{\partial\log Z_{J}(0,p_{1},p_{2})}{\partial m^{k}} \nonumber \\ & = \sum_{k=1}^{\infty}\frac{(mN)^{k}}{Nk!}\mathbb{E}_{\tau}\left[\left(\frac{}{}V(p_{2}|\tau)-V(p_{1}|\tau)\right)^{k}\right]_{c},\label{eq:gammaJexpansionmpp}
\end{eqnarray}
where the subscript $c$ indicates the connected component of the correlation function. Unfortunately, since we have to take a saddle point in $m$, which therefore  depends on $\delta p$, the nominal order $\delta p^k$ of the $k$-th term in the expansion, does not coincide with the effective order. 
To see this let's first truncate the series to the second order in $\delta p$,
\begin{eqnarray}
\Gamma_{2}\left(m,p-\frac{\delta p}{2},p+\frac{\delta p}{2}\right)=  m\left< \frac{}{} V'(p|\tau)\delta p \right> + \frac{N m^{2}}{2}\left<\left(\frac{}{}V'(p|\tau)\delta p\right)^{2}\right>_{c} \nonumber
\label{eq:gamma2expsecord}
\end{eqnarray}
and optimizing eq. (\ref{eq:lagrangeTV2V1p2p1}) over $m$ leads to
\begin{equation}
m^{*}=\frac{(V_2-V_1)-\left<\frac{}{}V'(p_{1}|\tau)\right>\delta p}{N\left<\left(\frac{}{}V'(p_{1}|\tau)\right)^{2}\right>_{c} \delta p^2}\:.\label{eq:gamma2solmstar}
\end{equation}
In the case of our interest,  $V_2-V_1=O(\delta p^2)$, leading to $m^*=O(\delta p^{-1})$,  and the expansion cannot be truncated as all the terms are of
the same order. 
If we blindly ignore this issue, the expression for $G$ to the second order is
\begin{equation}
G_{2}\left(V_2-V_1,p-\frac{\delta p}{2},p + \frac{\delta p}{2}\right)=\frac{1}{2 N}\frac{\left(V_2-V_1-\left<\frac{}{}V'(p|\tau)\right> \delta p \right)^{2}}{\left<\left(\frac{}{}V'(p|\tau) \delta p \right )^{2}\right>_{c}\:}\:.
\end{equation}
If on the other hand we set $V_2-V_1$ exactly equal to zero, we get an expression similar to eq. (\ref{eq:smallflucIT}).
Let us denote by $G_2(p)$ the zero order term of the Taylor expansion of $G_2(0,p-\delta p /2, p+\delta p /2)$ in $\delta p$,
\begin{equation}
G_2(p) = \lim_{\delta p \rightarrow 0} G_2\left(0,p-\frac{\delta p}{2}, p+\frac{\delta p}{ 2}\right)\:.
\label{eq:deffirstg2tobegeneral}
\end{equation}
The associated large deviation function $I(T)$ can be obtained minimizing $G_2(p)$ over $p$, 
\begin{eqnarray}
G_2(p) & = \frac{\left<\frac{}{}V'(p|\tau)\right>^{2}}{2\left<\left(\frac{}{}V'(p|\tau) \right)^{2}\right>_{c}  }\label{eq:computeILDSDl1}\\
 I_{LF,2}(T) & =\min_{p} G_2(p) \:.
 \label{eq:computeILDSDl2}
\end{eqnarray}
We discuss in the Appendix how to extend this computation to the third order in $m$. 


Having learned that for $V_2-V_1=O(\delta p^2)$ $m=O(1/\delta p)$,  we need to perform the large deviation computation without resorting to a power expansion in $m$. 
Our main focus is the computation of $\Gamma(m,p_1,p_2$). In fact, using eq. (\ref{eq:ZmdeltaVLD}), the large deviation function $G(V_2-V_1,p_1, p_2)$ can be computed from
\begin{equation}
e^{N \Gamma (m,p_1,p_2)} =  \left<\left(Z[p_1|\tau]\right)^{m/\beta} \left(Z[p_2|\tau]\right)^{-m/\beta}\right> 
\end{equation} 
by taking $n_1 \rightarrow m/\beta$ and $n_2 \rightarrow - m/\beta$ in the definition of $S^{(2)}[p_1, p_2]$, given in eq. (\ref{eq:defS2actiondeff}). 
With this replacement, 
\begin{equation}
\Gamma (m,p_1,p_2) = \left.  \frac{ S^{(2)}[p_1, p_2]}{2} \right|_{n_1=+\frac{m}{\beta } \atop n_2=-\frac{m}{\beta}} \:.
\label{eq:computeLDFfull}
\end{equation}
All the technical details of the computation of $S^{(2)}[p_1,p_2]$, which we perform with the replica method are provided in the Appendix.
For simplicity, we denote by $S^{(2)}(u,p)$ the zero order term of the Taylor expansion of $S^{(2)}[p_1=p-\delta p /2,p_2=p+\delta p /2]$ in $\delta p$, computed at $m=u/\delta p$,
\begin{equation}
S^{(2)}(u,p)=\lim_{\delta p \rightarrow 0} \left. S^{(2)}\left[p-\frac{\delta p}{2},p+\frac{\delta p}{2}\right]\right|_{n_1=+\frac{u}{\beta \delta p} \atop n_2=-\frac{u}{\beta \delta p}}\:.
\label{eq:computeLDFfull2}
\end{equation}
Its expression within a Replica Symmetric ansatz reads
\begin{eqnarray}
S^{(2)} (u,p)& = u^2  \frac{  f(q_1) +  f(q_2) -2  f(q_{12})}{ \delta p^2 } +  \beta u \frac{ f(q_2) - f(q_1) }{\delta p} + \nonumber 
\\ & \quad  + u \frac{ \log(1 - q_1) -  \log(1 - q_2)}{\beta \delta p} -2 u \beta' f'(p) +\nonumber 
\\ & \quad  - \log(1 - q_1) - \log(1 - q_2) + \nonumber
\\ & \quad + \log\left(u^2  \frac{ q_{12}^2 -2 p^2 q_{12} + p^2 q_1 +  p^2 q_2 - q_1 q_2 }{\beta^2 \delta p^2}\right) \:,
\end{eqnarray}
where $q_1$, $q_2$ and $q_{12}$ parametrize the overlap matrix in eq. (\ref{eq:defS2actiondeff}). 
We observe the presence of apparent divergences in this expansion. They come from the rescaling of the parameter $m$, that is necessary as long as we force $V_2-V_1\sim O(\delta p^2)$ for $T>T_d$, as observed in eq. (\ref{eq:gamma2solmstar}). However, these divergences are unphysical.
When $p_1$ and $p_2$ are close enough to $p$, saddle point values of the two-points function $S^{(2)}[p_1,p_2]$ can be written as perturbations of those of $S^{(1)}[p]$ (see the discussion below eq. (\ref{eqduetouno})).
The divergences disappear if we take into account the dependence of the parameters $q_1$, $q_2$ and $q_{12}$ on $\delta p$ at the saddle point
\begin{eqnarray}
\left\{ \begin{array}{cl}
q_{1}  =& q-(\delta p/2) \: \delta q \\
q_{2}  =& q+(\delta p/2)\: \delta q \\
q_{12} =& q+\delta p^2 \: \delta q_{12}
\end{array}\right.\:,
\label{eq.rescspparS2} 
\end{eqnarray}
where we indicate with $q$ the saddle point value of the RS potential for a value of the mutual overlap $p$. These two quantities are related by the saddle point equation 
\begin{equation}
p = \sqrt {q - \beta^2 f'(q) (1-q)^2 }\:,
\label{sppqRepsymm}
\end{equation}
see eq. (\ref{eq.qpRSVpot}). The perturbation in $ q_{12}$ is $O(\delta p^2)$ because the two $O(\delta p)$ contributions induced by $\delta p_1 = -\delta p/2$ and $\delta p_2 = \delta p/2$ cancel out. More details are provided in the Appendix.  
The expression of $S^{(2)}$ when evaluated on eq. (\ref{eq.rescspparS2}) reads
\begin{eqnarray}
S^{(2)} (u,p)   = &
\frac{ u \delta q}{\beta(1 - q )}  + 
   u (  \beta \delta q - 2 u \delta q_{12}) f'(q) -  2 \log\beta - 2 \log(1 - q) \nonumber 
   \\ &   + \log \Bigg[   q \Big(u^2 - 2 p u \beta - 2 \beta^2 + 2 u^2 \delta q_{12}\Big) + 
        p u \Big(2 \beta - u \delta q\Big) +  \nonumber
   \\ & \quad \qquad + (q \beta)^2 + \frac{1}{4}\Big(2 \beta - u \delta q\Big)^2 + \nonumber
   \\ & \quad \qquad + p^2 u \Big(-2 u \delta q_{12}  + \beta \delta q\Big) \frac{}{} \Bigg]     \nonumber
   \\ &    -  2 u \beta' f'(p) + \frac{1}{4} u^2 \delta q^2 f''(q) 
   \label{eq:exprescompletefullladevfuff}   
\end{eqnarray}
Finally we need to optimize over the parameters $\delta q$ and $\delta q_{12}$ and $u$. Thus we solve the system of equations
\begin{eqnarray}
\left\{ \begin{array}{ll}
\partial_{\delta q} S^{(2)}(u,p) & = 0 \\ 
\partial_{\delta q_{12}} S^{(2)}(u,p) & = 0 \\ 
\partial_{ u} S^{(2)}(u,p) & = 0 
\end{array}\right.\:,
\label{eq:saddlepGLFD11}
\end{eqnarray}
The solution of this system of equations leads to an expression that depends only on $q$ and $p$. Using eq. (\ref{eq:computeLDFfull}), we define $\Gamma(u^*,p)=S^{(2)}(u^*,p)/2$, where $*$ denotes the solution of eq. (\ref{eq:saddlepGLFD11}),
\begin{eqnarray}
\Gamma(u^*,p)= & \frac{1}{2} \left[(q-p^2 )[ f'(q) \frac{}{} + q f''(q)]\right]^{-1} \Bigg \{ \frac{}{} (q-p^2) f'(q) \times \nonumber  \\
 & \times \left(1 + \log\left[\frac{\frac{}{}(q-p^2) }{\beta^2 f'(q)(1 - q)^2}\right] + ( q-1)^2 \beta^2 f''(q)\right) \nonumber \\
 & -\Bigg [ \beta ' f'(p) \left(\frac{}{} \beta \beta ' (q-1)  f'(p)+2 p\right)+\beta  (q-1) f''(q) \times \nonumber  \\
 & \times  \left(  q -q \log \left[\frac{ \left(q-p^2\right)}{\beta ^2
   f'(q)(1-q)^2}\right]+2 \beta  \beta ' p (q-1) f'(p)\right) \Bigg] \times \nonumber \\
 & \times ( q-1) \beta f'(q) - q f'(q) + \nonumber \\
 & + (p^2 - q) ( q-1)^2 \beta^2 \beta'^2 f'(p)^2  f''(q)  \Bigg \}
 \label{rsexpgrandideviazionifinzp}
\end{eqnarray}
and using eq. (\ref{eq:lagrangeTV2V1p2p1}) we also define $G(p) = -\Gamma(p)$. The large deviation function $I(T)$ is obtained by taking the minimum over $p$ of $G(p)$, as in eq. (\ref{eq:computeILDSDl2}). 
\begin{equation}
I_{LF}(T) =\min_{p} G(p) 
\label{fulldevilargminGp1rsb}
\end{equation}
A comparison between $G(p)$ and $G_2(p)$ can be found in the Appendix. In Fig. \ref{GV3} we plot the potential and the large deviation function for the $3$-spin in the dynamical phase, where the potential has a local minimum. 
The first minimum of $G(p)$ on the right of the figure corresponds to the local minimum of the potential and, since $\beta>\beta_d$, $G(p)=0$ in this point. In fact, the probability of having a stationary point on this point is $1$.
We observe in the figure that for each stationary point of the potential, $G(p)$ has a zero. We observe that $G(p)$ is zero also on the spurious stationary points of the RS potential (the first two from the left).

\begin{figure}
\centering{}\includegraphics[scale=0.5]{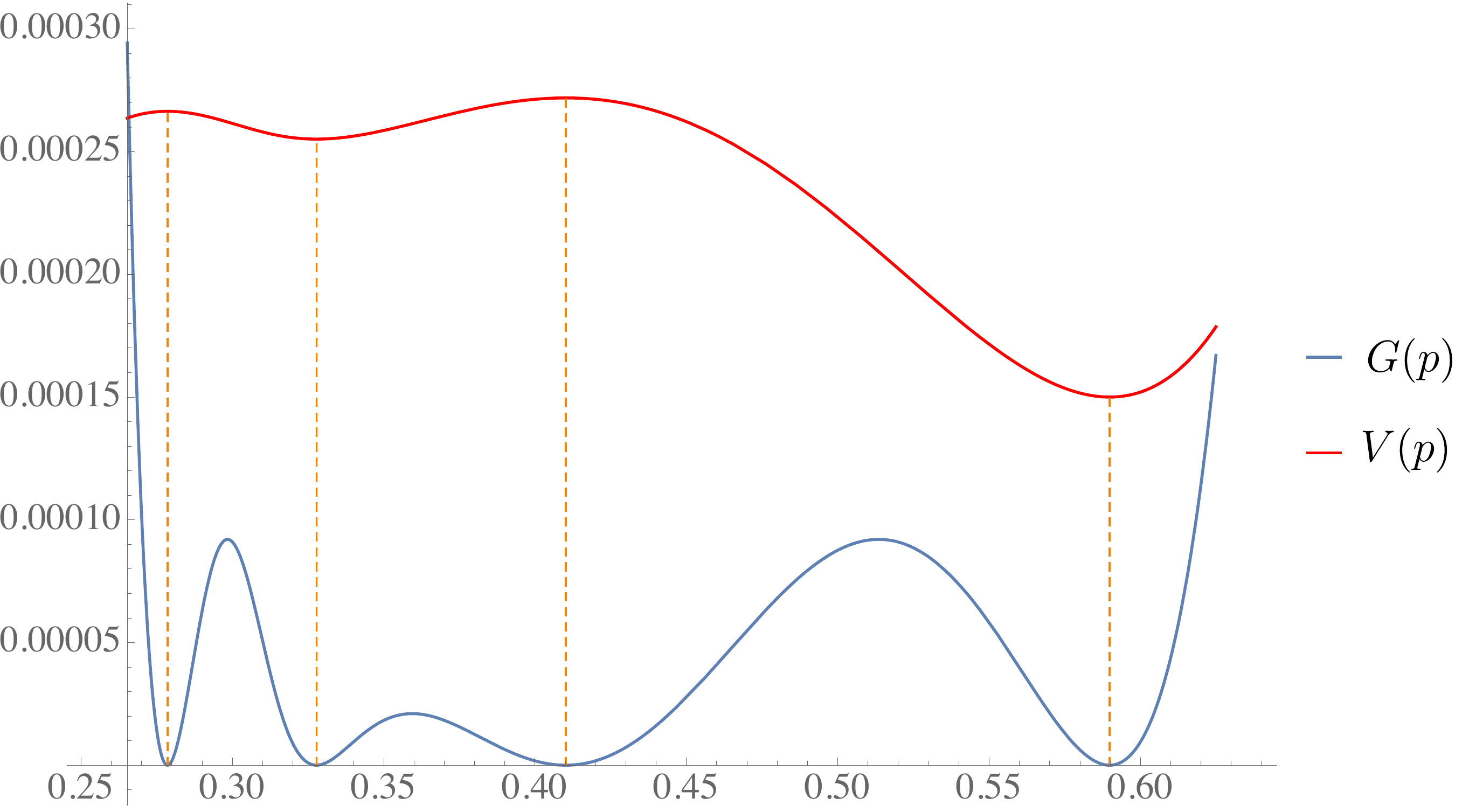}\caption{\label{GV3}Potential $V(p)$ and the large deviation function $G(p)$ computed at the RS level for the $3$-spin at $\beta=1.66$, larger that $\beta_d$. The potential has been rescaled in order to fit in the figure. The first two minima of $G(p)$ from the left are spurious, as they correspond to the first two stationary points of the potential due to the RS ansatz.}
\end{figure}

\begin{figure}
\centering{}\includegraphics[scale=0.5]{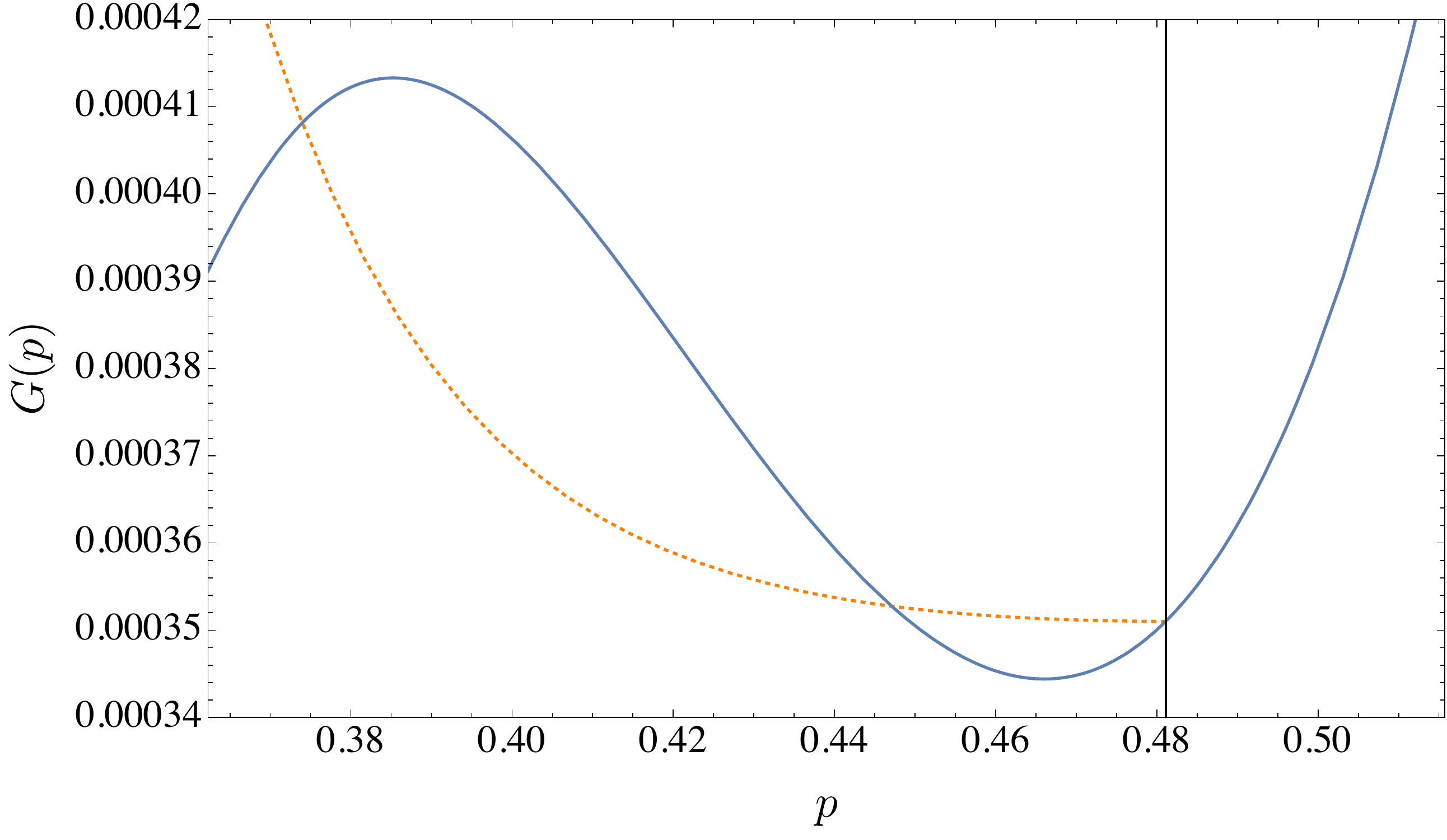}\caption{\label{G3}$G(p)$ for the $3$-spin at $\beta=\beta'=1.58$, larger than $\beta_{d}$, computed with the 1RSB anstaz (dots) and the RS ansatz (line). The vertical line indicates $p_{RS}$.}
\end{figure}

\begin{figure}
\centering{}\includegraphics[scale=0.5]{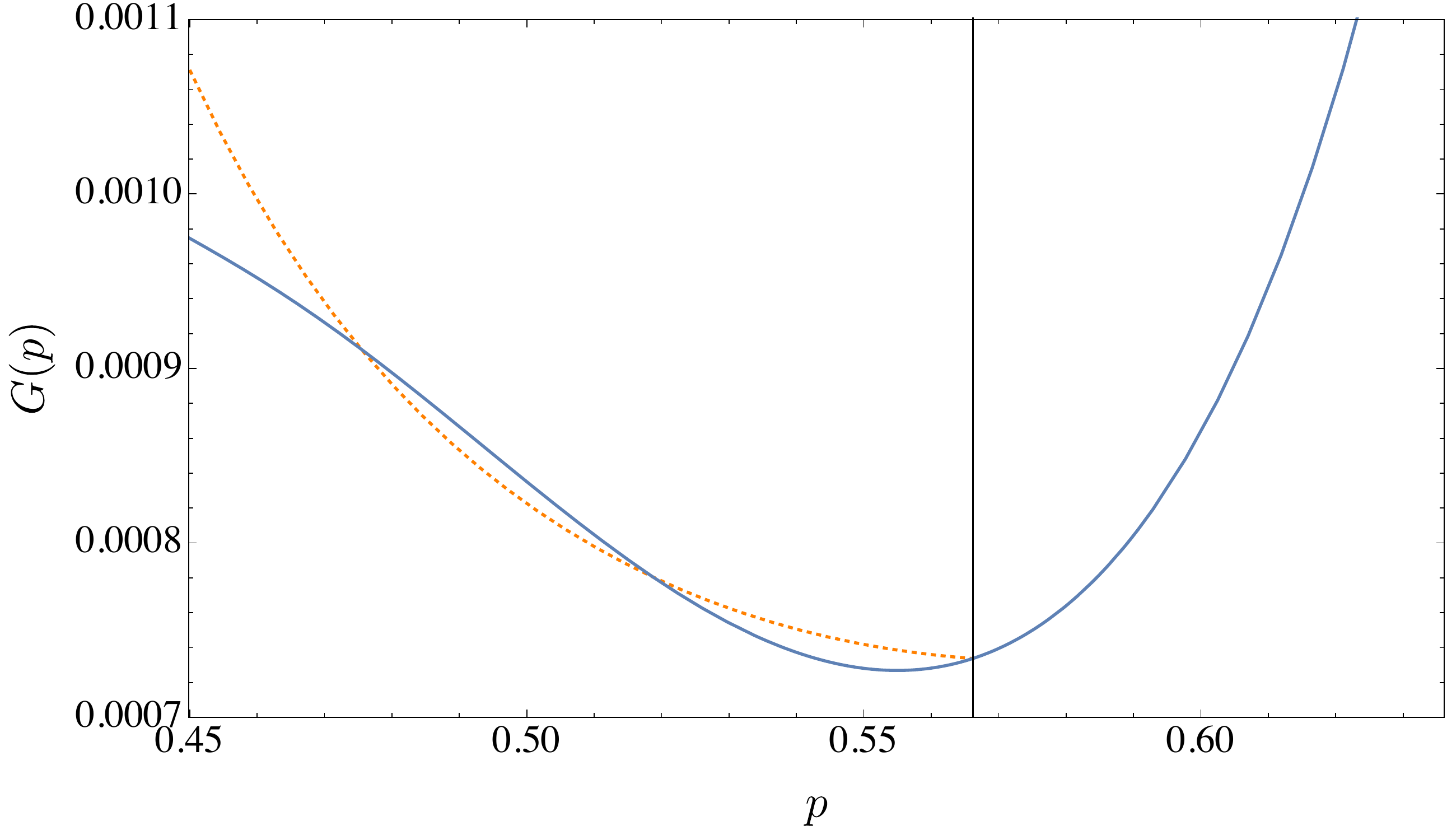}\caption{\label{G34}$G(p)$ for the $3+4$-spin model at  $\beta=\beta'=1.19$, larger than $\beta_{d}$, computed with the 1RSB anstaz (dots) and the RS ansatz (line). The vertical line indicates $p_{RS}$.}
\end{figure}

\begin{figure}
\centering{}\includegraphics[scale=0.5]{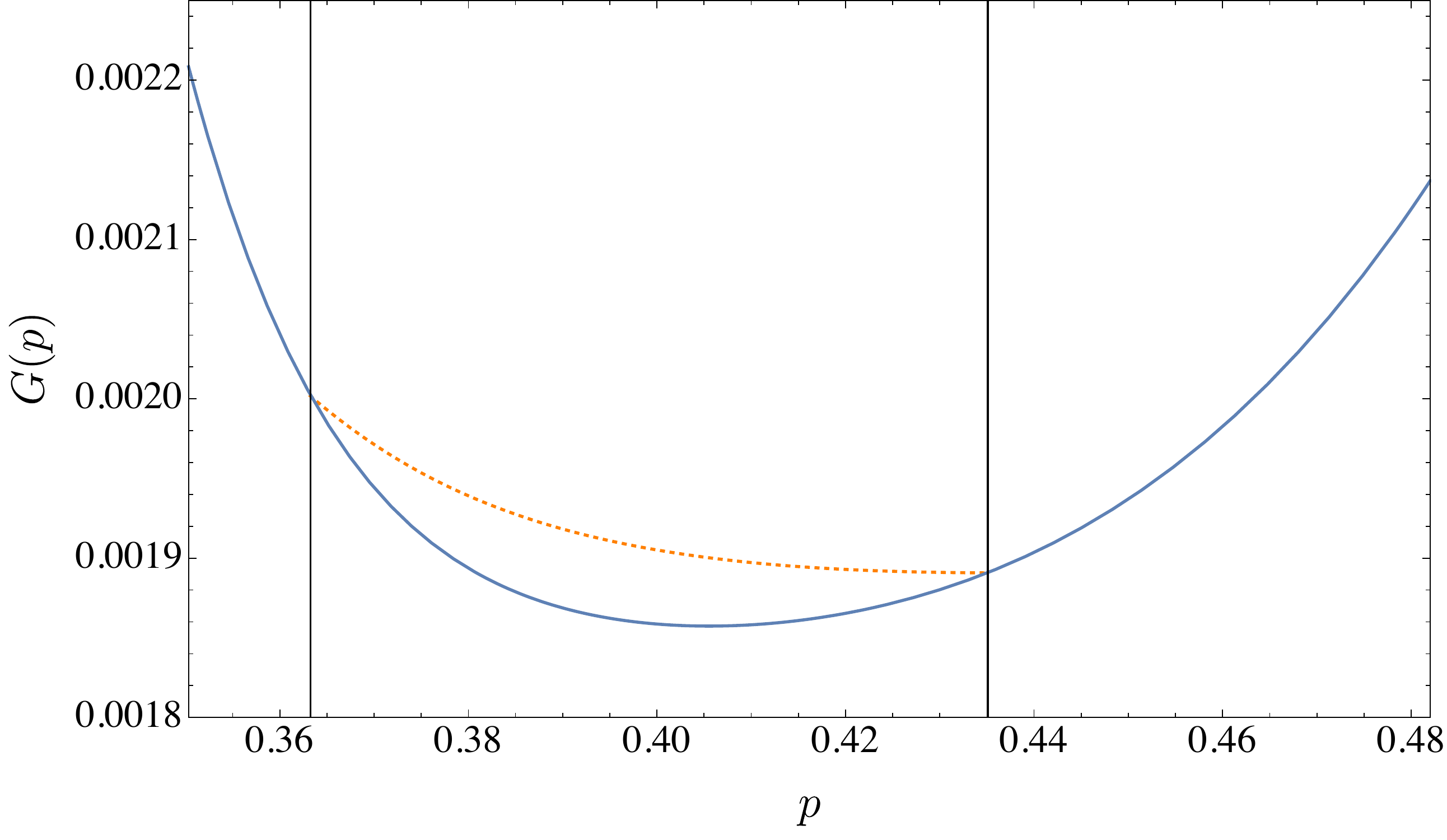}\caption{\label{G3l}$G(p)$ for the $3$-spin at $\beta=\beta'=1.51$, larger than $\beta_{d}$, computed with the 1RSB anstaz (dots) and the RS ansatz (line). The vertical lines indicate, from left to right, $p_K$ and $p_{RS}$.}
\end{figure}

\begin{figure}
\centering{}\includegraphics[scale=0.5]{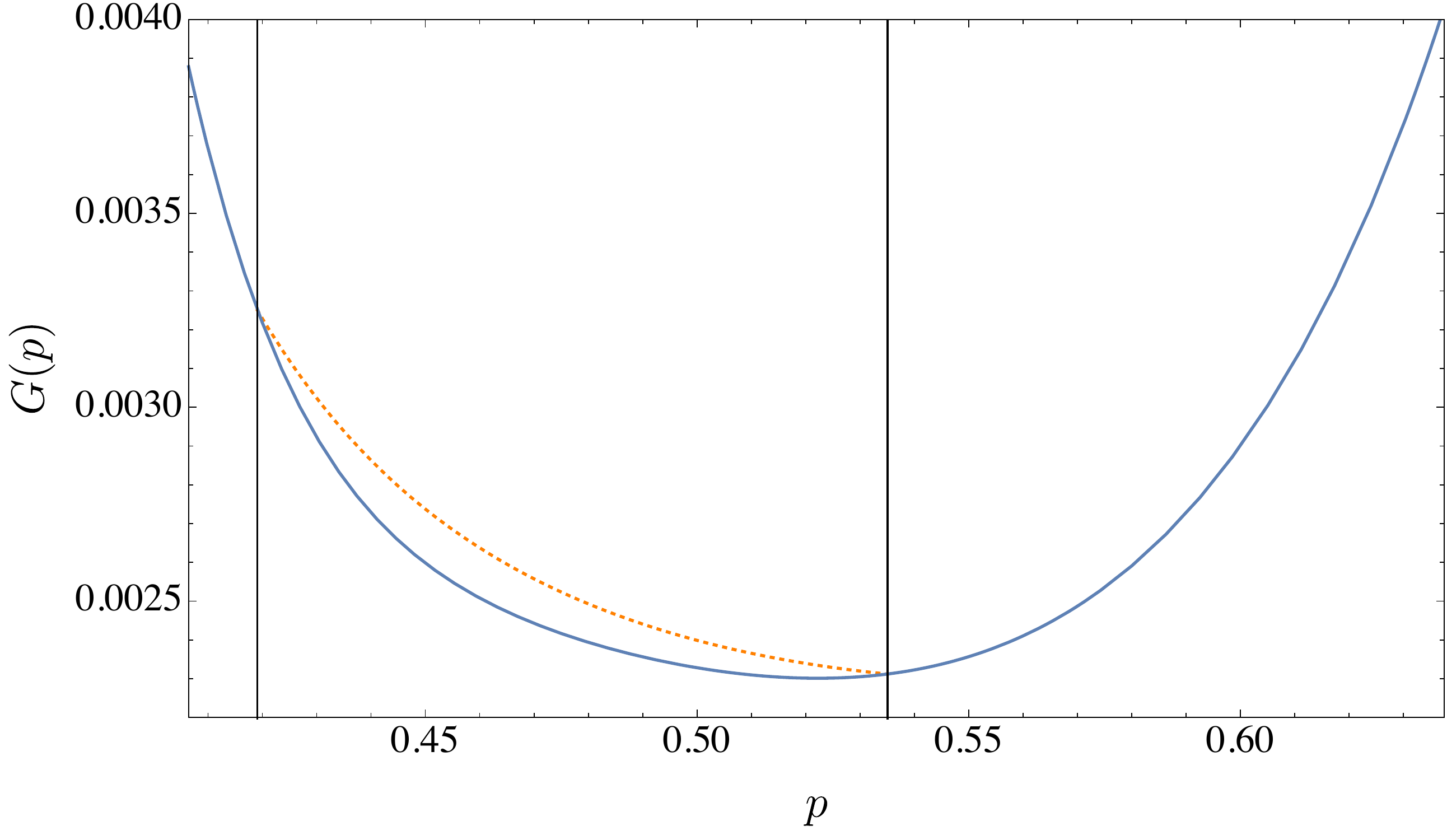}\caption{\label{G34l}$G(p)$ for the $3+4$-spin model at  $\beta=\beta'=1.15$, larger than $\beta_{d}$, computed with the 1RSB anstaz (dots) and the RS ansatz (line). The vertical lines indicate, from left to right, $p_K$ and $p_{RS}$.}
\end{figure}

\begin{figure}
\centering{}\includegraphics[scale=0.5]{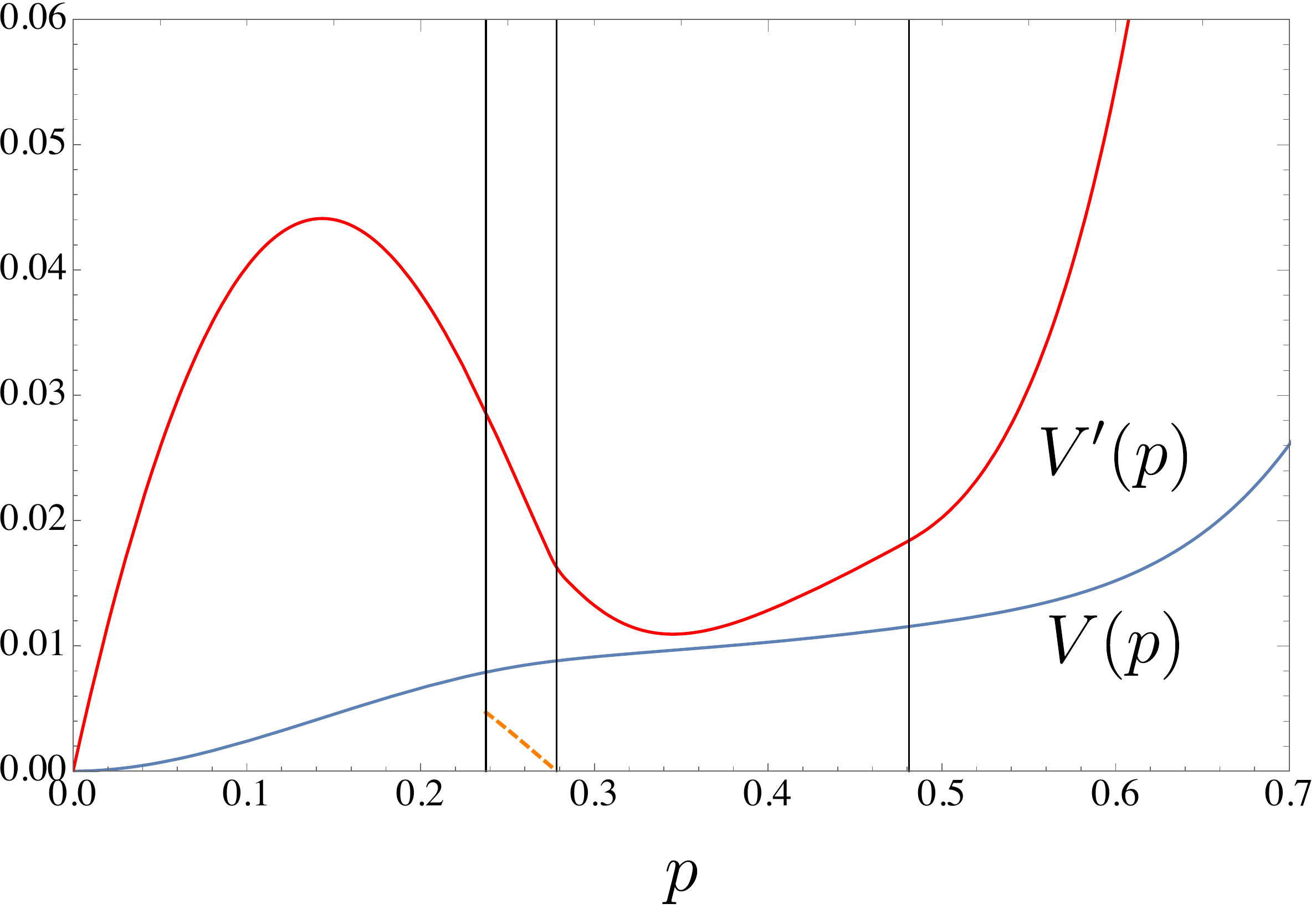}\caption{\label{VVp3spin}$V(p)$ and $V'(p)$ computed at 1RSB level for the $3$-spin at $\beta=1.58$, in the paramagnetic phase. The three vertical lines separate, from the left, a first RS region followed by the dynamic 1RSB, the static 1RSB and the last RS region.}
\end{figure}

The broken replica 1RSB computation is similar. As with the RS computation, the rescaling of $m$ leads to apparent divergences terms:
\begin{eqnarray}
S^{(2)}(u,p) =& u^2  \frac{  f(q^0_1) +  f(q^0_2) -2  f(q_{12})}{ \delta p^2 } + \beta u \frac{f(q^1_2)-f(q^1_1)}{\delta p}+\nonumber \\ &  + \beta u \frac{x_2 f(q^0_2)-x_1 f(q^0_1)}{\delta p}  - \beta u \frac{x_2 f(q^1_2)-x_1 f(q^1_1)}{\delta p}+ \nonumber
\\ & + u \frac{ \log(1 - q^1_2)/x_2 -  \log(1 - q^1_1)/x_1}{\beta \delta p} -2 u \beta' f'(p) + \nonumber
\\   &  + u \frac{\log(1 + q^1_1(x_1-1) -q^0_1 x_1)}{x_1 \beta \delta p} +  \nonumber 
\\ & -  u \frac{\log(1 + q^1_2(x_2-1) -q^0_2 x_2)}{x_2 \beta \delta p} + \nonumber 
\\ & + u \frac{ \log(1 - q^1_1) -  \log(1 - q^1_2)}{\beta \delta p} + \nonumber 
\\ & - \log(1 + q^1_1(x_1-1) -q^0_1 x_1) - \log(1 + q^1_2(x_2-1) -q^0_2 x_2)  \nonumber 
\\ & + \log\left(u^2  \frac{ q_{12}^2 -2 p^2 q_{12} + p^2 q^0_1 +  p^2 q^0_2 - q^0_1 q^0_2 }{\beta^2 \delta p^2}\right) \:. 
\label{eq:s21rsbfirstexpan}
\end{eqnarray}
In this case, for each $k=1,2$, $q^0_k$, $q^1_k$ and $x_k$ need to be shifted when $p_{1}$ and $p_{2}$ are shifted from $p$. These order parameters parametrize the 1RSB overlap matrix in eq. (\ref{eq:defS2actiondeff}). The apparent divergences are unphysical and in order to re-absorb them we set:
\begin{eqnarray}
\left\{ \begin{array}{cl}
q^1_{1}  =& q^1-(\delta p/2) \: \delta q_1 \\
q^1_{2}  =& q^1+(\delta p/2)\: \delta q_1 \\
q^0_{1}  =& q^0-(\delta p/2) \: \delta q_0 \\
q^0_{2}  =& q^0+(\delta p/2)\: \delta q_0 \\
x_{1}  =& x-(\delta p/2) \: \delta x \\
x_{2}  =& x+(\delta p/2)\: \delta x \\
q_{12} =& q_{12} +\delta p^2 \: \delta q_{12}
\end{array}\right.\:,
\label{eq.rescspparS21rsb} 
\end{eqnarray}
where we indicate with $q^1$, $q^0$ and $x$ the saddle point values of the 1RSB potential for a value of the mutual overlap $p$.
These quantities are determined by eqs. (\ref{eq.qpRSVpot1})-(\ref{eq.qpRSVpot3}) in the Appendix. When evaluating eq. (\ref{eq:s21rsbfirstexpan}) on eq. (\ref{eq.rescspparS21rsb}), the expression for $S^{(2)}$ has to be optimized over $\delta q^1$, $\delta q^0$, $\delta x$ and $\delta q_{12}$. However, these variables are not independent. This can be observed by setting 
\begin{equation}
t = \delta q_{12} + \frac{ \beta (x-1) \delta q_1 + (q_1-q_0) \delta x }{2 u}
\end{equation}
With this replacement, $S^{(2)}$ depends only on $t$ and $\delta q_0$,
\begin{eqnarray}
S^{(2)}(u,p)=& \frac{u \delta q_0}{\beta (1-q^1) + x \beta ( q^1-  q^0)} + u ( x \beta \delta q_0-2 u t ) f'(q^0)  -2 \log \beta \nonumber 
\\ & - 2 \log(1 + q^1 ( x-1) - q^0 x)  \nonumber
\\ & + \log \Bigg[ q^0 \Big( u^2 - 2 p u x \beta + 2 x (-1 + q^1 - q^1 x) \beta^2 + 2 u^2 t \Big) + \nonumber
\\ & \qquad \quad +  p u \Big( 2 (1 + q^1 ( x-1)) \beta - u \delta q^0\Big) + (q^0 x \beta)^2 + \nonumber
\\ & \qquad \quad + \frac{1}{4} \Big(2 (1 + q^1 ( x-1)) \beta - u \delta q_0 \Big)^2 + \nonumber
\\ & \qquad \quad + p^2 u \Big(-2  u t + x \beta \delta q_0\Big) \Bigg] \nonumber
\\ & -  2 u \beta' f'(p)  + \frac{1}{4} u^2 \delta q_0^2 f^{''}(q^0)  \:,
\end{eqnarray}
and it can be optimized over these variables together with the optimization over $u$ as done in the RS computation: 
\begin{eqnarray}
\left\{ \begin{array}{ll}
\partial_{\delta q_0} S^{(2)}(u,p) & = 0 \\ 
\partial_{t} S^{(2)}(u,p) & = 0 \\ 
\partial_{ u} S^{(2)}(u,p) & = 0 
\end{array}\right.\:.
\label{eq:saddlepGLFD11rsb}
\end{eqnarray}
The solution to this system of equations leads to an expression that depends only on $q^1$, $q^0$, $x$ and $p$. As in the RS computation, we can compute $G(p)= - \Gamma(u^*,p)$ and the large deviation function $I(T)$.
For simplicity we do not report its expression and we study its behavior in Fig. \ref{G3}-\ref{G34l} for the $3$-spin and the $3+4$-spin, at different values of $\beta<\beta_d$ for $\beta'=\beta$. We denote by $p^*(T) = \arg \min G(p)$.
When $\beta'=\beta$ and $\beta<\beta_d$, the minimum over $p$ of $G(p)$ appears at $p_{RS}$, the frontier between the static 1RSB region and the RS region, as can be observed in Fig. \ref{G3}-\ref{G34l}.
When $p>p_{RS}$, the associated $q$, see eq. (\ref{sppqRepsymm}), is such that the replicon mode is positive
\begin{equation}
1-\beta^2 f''(q)(1-q)^2 > 0
\end{equation}
and the RS anzatz is correct. The 1RSB computation differs from the RS one only in the static 1RSB regime. As long as $\beta>\beta_{RS}$, the replica symmetric $G(p)$ does not estimate correctly $p^{*}$.  
For larger temperature, no RS instability exists. Nevertheless, as will be discussed afterwards, $p^*$ coincides with the point associated to $q$ for which the replicon mode $1-\beta^2 f''(q)(1-q)^2$ is minimum.
For any $\beta<\beta_d$, it is worthwhile noticing that  $p^{*}$ does not correspond to the point where the first derivative of the potential takes the smallest value, as observed in Fig. \ref{VVp3spin}.
Finally, we notice that for $\beta<\beta_d$, the value of $G$ in $p^*$ is not zero, differently from Fig. \ref{GV3}, because at $\beta<\beta_d$ the secondary minimum does not exist in the typical potential $V(p)$.

\section{Results}

\subsection{Finite temperature}

\begin{figure}%
    \centering
    \subfloat[\label{TsT3} $3$-spin.]{{\includegraphics[scale=0.24]{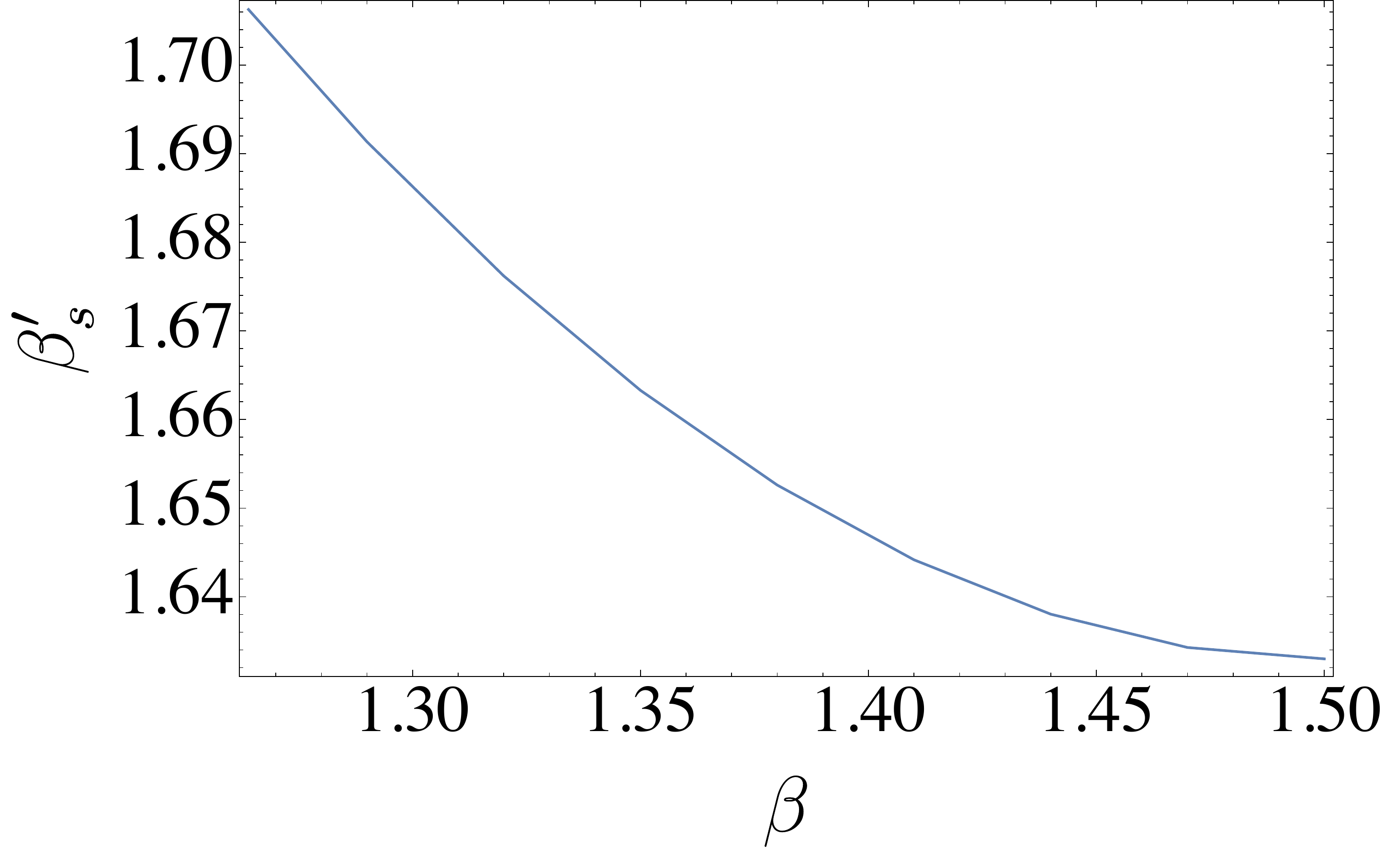} }}%
    \qquad
    \subfloat[\label{TsT34} $3+4$-spin.]{{\includegraphics[scale=0.24]{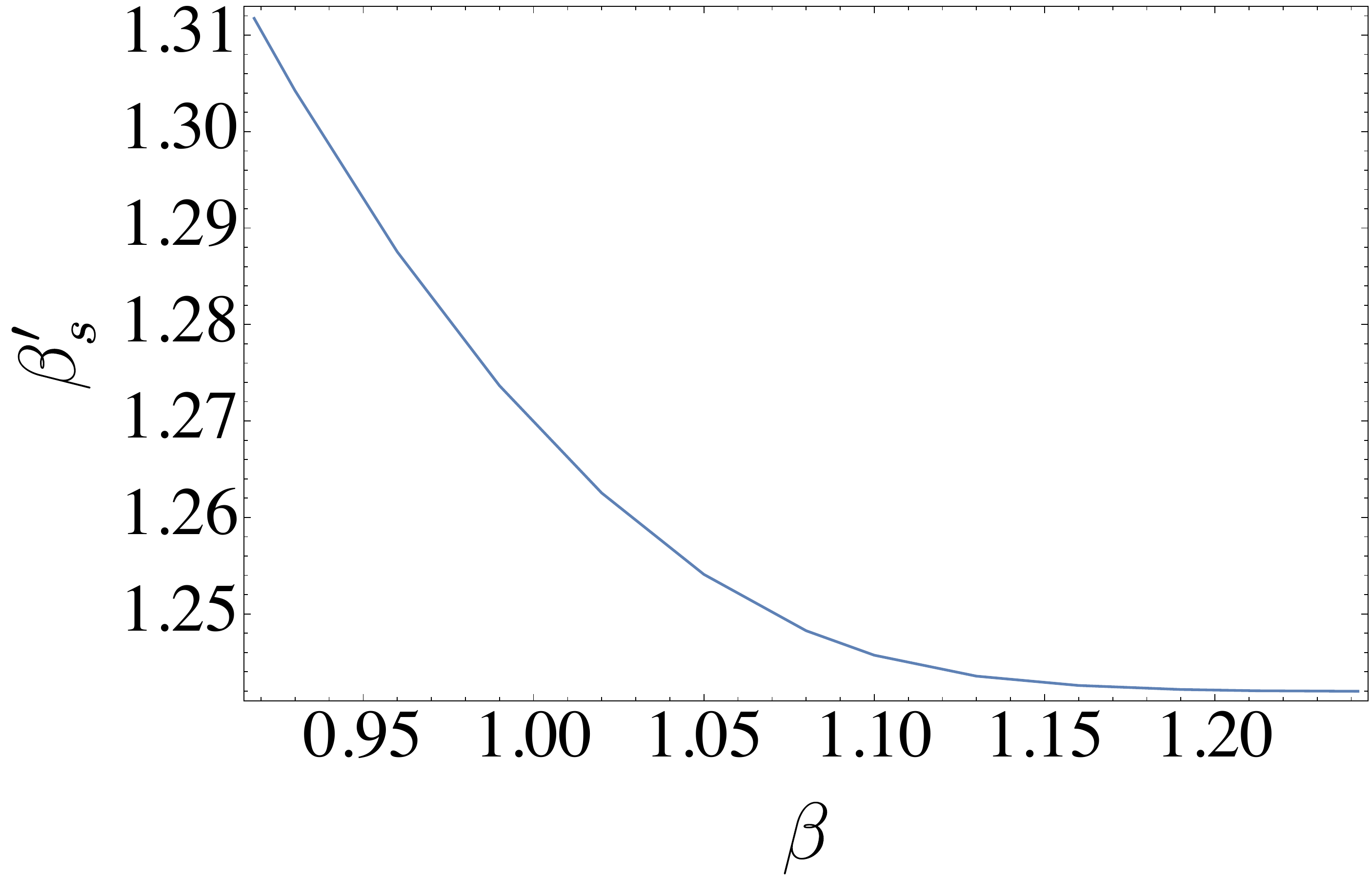} }}%
    \caption{\label{fig:bf3-34} $\beta'_s$ as a function of $\beta$. $\beta'_s$ reaches $\beta_K$ on the left side of the plots. In the $3$-spin, when $\beta=1.5$, $\beta'_s=\beta_d$, i.e. there is not need to increase $\beta$ from $\beta_d$ to find the secondary minimum as it still exists. This is true $\forall \beta : 1.5<\beta<\beta_d$. In the mixed model, as soon as $\beta<\beta_d$, $\beta'_s$ must be increased in order to find the secondary minimum. This is the reason why in the right border of the $x$-axis is $\beta=1.5$ in (a) and $\beta=\beta_d$ in (b). }%
\end{figure}

The results presented above suggest that the most likely event for the appearance of a stationary point in the potential $V(p)$ for $T>T_d$ is that the first replica is picked in one of the marginal states still surviving at $T$, i.e. in one of the out-of-equilibrium states for which the replicon mode is zero:  $\beta^2 f''(q)(1-q)^2=1 $. In the following we denote by $q_{RS}$ the largest solution to this equation.
$q_{RS}$ exists as long as $\beta>\beta_{RS}$. For smaller $\beta$, the RS ansatz is always stable and
the most likely event for the appearance of a stationary point in the potential $V(p)$
is that the first replica is chosen in one of the states for which the replicon mode is minimum.
 
In the pure model it is possible to interpret these results in terms of the potential where two temperatures are considered. Using the notation introduced previously, we denote by $T'$ the temperature of the first replica and by $T$ the temperature of the second replica. 
When $T'>T_d$, no local minimum exists.  
If $T'\leq T_d$, increasing $T$ the secondary minimum disappears at a certain point and, as long as it exists, it describes situations where the second replica is in a TAP state of equilibrium at temperature $T'$, followed at $T$ \cite{franz1995recipes,barrat1997temperature,krzakala2010following,sun2012following}. 
Following states at different temperatures can be given a dynamical meaning considering a system thermalized at $T'$ and whose Langevin dynamics is done at temperature $T$: in the long time limit the system relaxes inside one of the TAP states that was at equilibrium at $T'$, whose properties have changed since temperature has been shifted to $T$ \cite{franz1995recipes,barrat1997p,barrat1996dynamics}. 
The minimum of the potential represents the correlation between the state where the first replica is extracted from and the tame state at a different temperature $T \neq T'$.
When $T>T'$, depending on the model, different things happen. For the pure model, if $T'=T_d$, increasing $T$ the secondary stationary point continues to exist (as a saddle) until $T_{RS}=1/\beta_{RS}$ ($\beta_{RS}=1.5$ in the $3$-spin model). For larger temperatures, it disappears and in order to make it re-appear, the first temperature $T'$ must be decreased.
For $T>T_{RS}$ no $p_{RS}$ exists, i.e. no RS instability appears. At a given $T$, we denote by $T'_s$ the largest value of $T'$ for which the secondary minimum exists. For consistency, we set $T'_s=T_d$ for $T<T_{RS}$.
In the mixed model, if $T'=T_d$, as soon as $T>T_d$, the secondary minimum disappears. In order to make it re-appear, the first temperature $T'$ must be decreased from $T_d$. This behavior is described in Fig. \ref{fig:bf3-34}. 
In both the pure and the mixed spin models, when $T'=T=T_d$, the second stationary point at $p=q_{EA}$ describes marginal states, i.e. states with an overlap value $q$ for which the replicon mode is zero.
The description provided above can be rephrased by saying that using the potential $V(p)$ with two temperatures $T'$ and $T$, marginal states can be followed up in temperature in the pure model, but not in the mixed one. 
Let us denote by $p_{min}(T',T)$ the point where the potential has the secondary stationary point when the first temperature is $T'$ and the second $T$, by $V_{min}(T',T)$ the value of the potential in the local minimum and by $q_{min}(T',T)$ the self-overlap of the state described by the secondary stationary point. In the pure model, it turns out that 
\begin{equation}
p_{min}(T'_s,T)=p^*(T),
\label{eqpminpTpTpotFP}
\end{equation}
i.e. $p^*(T)$ corresponds to the secondary stationary point of the potential when $T'=T'_s$ and the second temperature is $T$. 
Moreover, for $\beta>\beta_{RS}$,
\begin{equation}
p_{RS}(T)=p_{min}(T_d,T)=p^*(T),
\end{equation}
i.e. these states are marginal. For $\beta<\beta_{RS}$, no $p_{RS}$ exists but eq. (\ref{eqpminpTpTpotFP}) is still valid. In this case, as mentioned above, the overlap $q$ associated to $p^*$ trough eq. (\ref{sppqRepsymm}) minimizes the replicon mode.

Before commenting these results, let us also define the tilted potential
\begin{equation}
\widetilde{V}(p_1|V_2-V_1)=\int dV_1 V_1 P(V_1|V_2-V_1)=  \int dV_1 V_1 \frac{P(V_1,V_2-V_1)}{P(V_2-V_1)}
\label{tiltpotdeffracVV2}
\end{equation}
where we use $P(V_1|V_2-V_1)$ to denote the probability that the  random potential $V(p_1|\tau)$ is equal to $V_1$ given that $V(p_2|\tau)-V(p_1|\tau)=V_2-V_1$, and $P(V_2-V_1)$ is the quantity defined in eq. (\ref{eq:PV2V1p1p2definit}) where, to simplify the notation, we omit $p_1$ and $p_2$ in the arguments.
$\widetilde{V}(p_1|V_2-V_1)$ is the value of the potential in $p_1$ conditioned on the value of the difference $V(p_2|\tau)-V(p_2|\tau)$. Without this condition, $\widetilde{V}(p_1)=V(p_1)$, the typical potential.
By definition
\begin{eqnarray}
& P(V_1,V_2-V_1) = \left< \delta(V_1-V(p_1|\tau)) \delta(V_2-V_1-V(p_1|\tau)+V(p_2|\tau)) \right> \nonumber \\
& \int dV_1 V_1 P(V_1,V_2-V_1) =  \left< V(p_1|\tau) \delta(V_2-V_1-V(p_1|\tau)+V(p_2|\tau))  \right> \nonumber = \\
& \qquad \qquad \quad \; \; \qquad \qquad = \lim_{z\rightarrow 0} N^{-1} \partial_z  e^{-N m^*  (V_2-V_1) }  Z(m^*,z,p_1,p_2)
\end{eqnarray}
where $m^*$ denotes the saddle point value of $m$ and where we defined
\begin{equation}
 Z(m,z,p_1,p_2) = \left< e^{N m(V(p_2|\tau)-V(p_1|\tau))} e^{N z V(p_1|\tau)} \right>
\end{equation}
generalizing eq. (\ref{eq:ZmdeltaVLD}). 
Taking the logarithm of this function
\begin{equation}
N \Gamma(m,z,p_1,p_2) = \log Z(m,z,p_1,p_2)
\end{equation}
we may write the integral as
\begin{eqnarray}
\int dV_1 V_1 P(V_1,V_2-V_1)  = \lim_{z \rightarrow 0} & e^{-N m^* (V_2-V_1)+N\Gamma(m^*,z,p_1,p_2)} \times \nonumber \\ 
&\times \partial_z \Gamma(m^*,z,p_1,p_2)\:.
\label{intdefsuV1}
\end{eqnarray}
Using eq. (\ref{eq:defS2actiondeff}) we easily obtain
\begin{equation}
\Gamma(m,z,p_1,p_2)= \left. \frac{1}{2}S^{(2)}[p_1,p_2] \right|_{\:\:\:\:\: n_1=+\frac{m-z}{\beta} \atop n_2=-\frac{m}{\beta}} - z F
\end{equation}
and by taking $p_1=p-\delta p /2$ and $p_2=p+\delta p /2$, $m^*=u/\delta p$, denoting by
\begin{eqnarray}
\Gamma(u,z,p) & = \lim_{\delta p \rightarrow 0} \Gamma \left(u/\delta p,z,p-\frac{\delta p}{2},p+\frac{\delta p}{2}\right) \\
S^{(2)}(u,z,p) & =\left. \lim_{\delta p \rightarrow 0} S^{(2)}\left[p-\frac{\delta p}{2},p+\frac{\delta p}{2}\right] \right|_{\:\:\:\:\: n_1=+\frac{u/\delta p-z}{\beta} \atop n_2=-\frac{u/\delta p}{\beta}}
\end{eqnarray}
we finally obtain the expression for the tilted potential
\begin{equation}
\widetilde{V}(p|V_2-V_1) = \lim_{z\rightarrow 0} \partial_{z} \left[ \frac{1}{2} S^{(2)}(u,z,p) - z F \right] 
\end{equation}
where the prefactor in eq. (\ref{intdefsuV1}) cancels out with the denominator $P(V_2-V_1)$ in eq. (\ref{tiltpotdeffracVV2}), see also eq. (\ref{P1P2deffirstquadsssds}). $\widetilde{V}(p|V_2-V_1)$ formally depends on the difference $V_2-V_1=V(p+\delta p/2|\tau)-V(p-\delta p/2|\tau)$. From now on we will set $V_2-V_1=0$ and denote by $\widetilde{V}(p)$ the value of the tilted potential for $V_2-V_1=0$, i.e. $\widetilde{V}(p)=\widetilde{V}(p|V_2-V_1=0)$.
An important observation to make is that $\widetilde{V}(p)$ is the value of the potential in $p$ conditioned that there is a stationary point in $p'$, with $p'=p$. Generalizing this computation to a generic $p'$ would give access to the shape of atypical potentials but it is out of the scope of the present work. 

\begin{figure}%
    \centering
    \subfloat[\label{pspm3} $3$-spin: for $\beta>\beta_{RS}$ the two quantities are equal to $p_{RS}(T)$. For smaller $\beta$, $p^*$ corresponds to the state with the smallest value of the replicon mode.]{{\includegraphics[width=10cm]{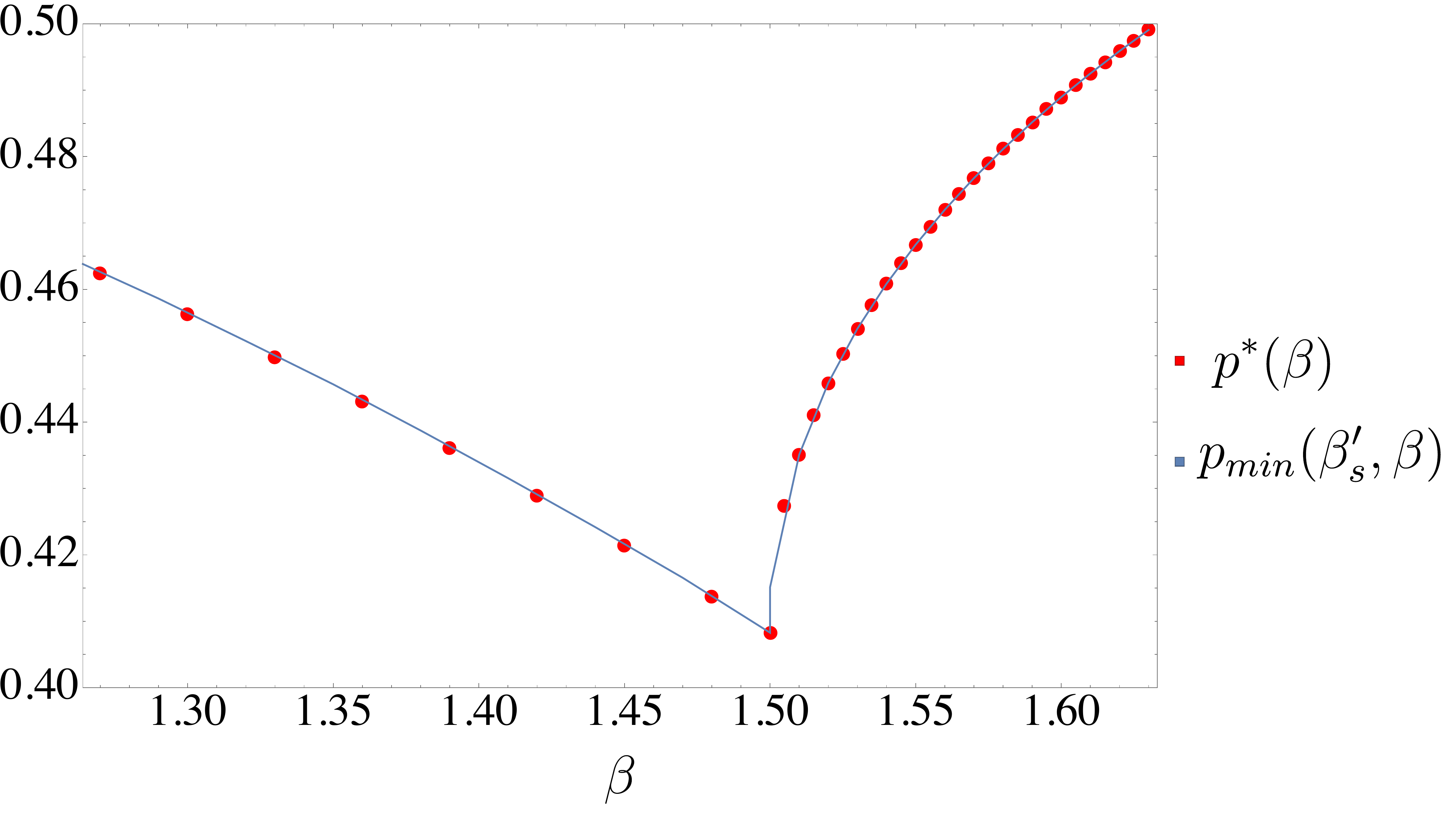} }}%
    \\
    \subfloat[\label{pspm34} $3+4$-spin: for $\beta>\beta_{RS}$, $p_{min}(T'_s,T)$ is equal to $p_{RS}(T)$. For smaller $\beta$, $p^*$ corresponds to the state with the smallest value of the replicon mode.]{{\includegraphics[width=10.5cm]{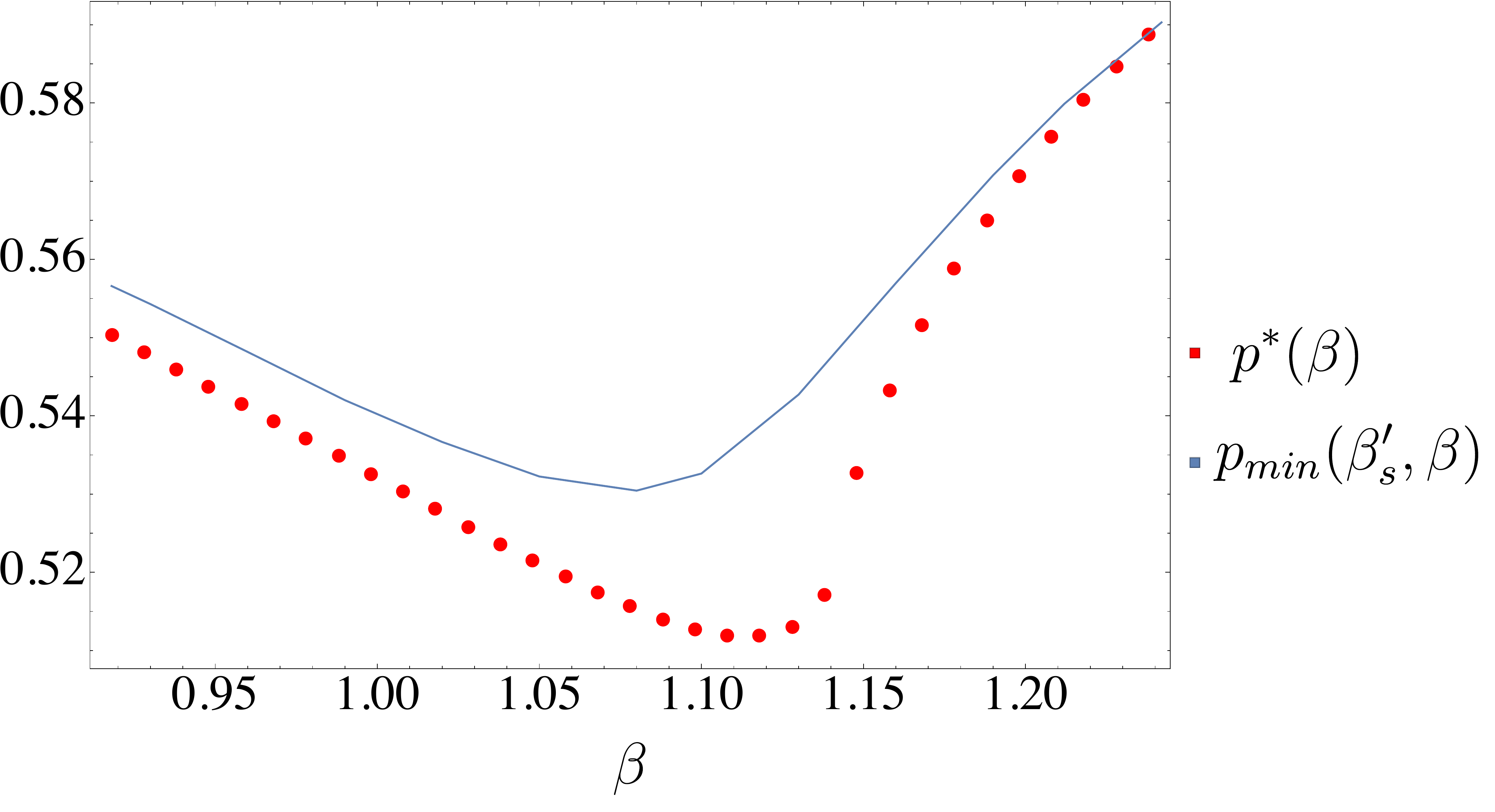} }}%
    \caption{\label{fig:pspm3-34} Comparison between $p_{min}(T'_s,T)$ and $p^{*}(T)$ for $\beta<\beta_d$. Contrarily to the pure spin, in the mixed model the two quantities do not match.}%
\end{figure}

\begin{figure}%
    \centering
    \subfloat[\label{Vtilt3} $3$-spin: $V_{min}(\beta'_s,\beta)$ and $V^{*}(\beta'_s,\beta)$ coincide, as discussed in the text.]{{\includegraphics[width=10cm]{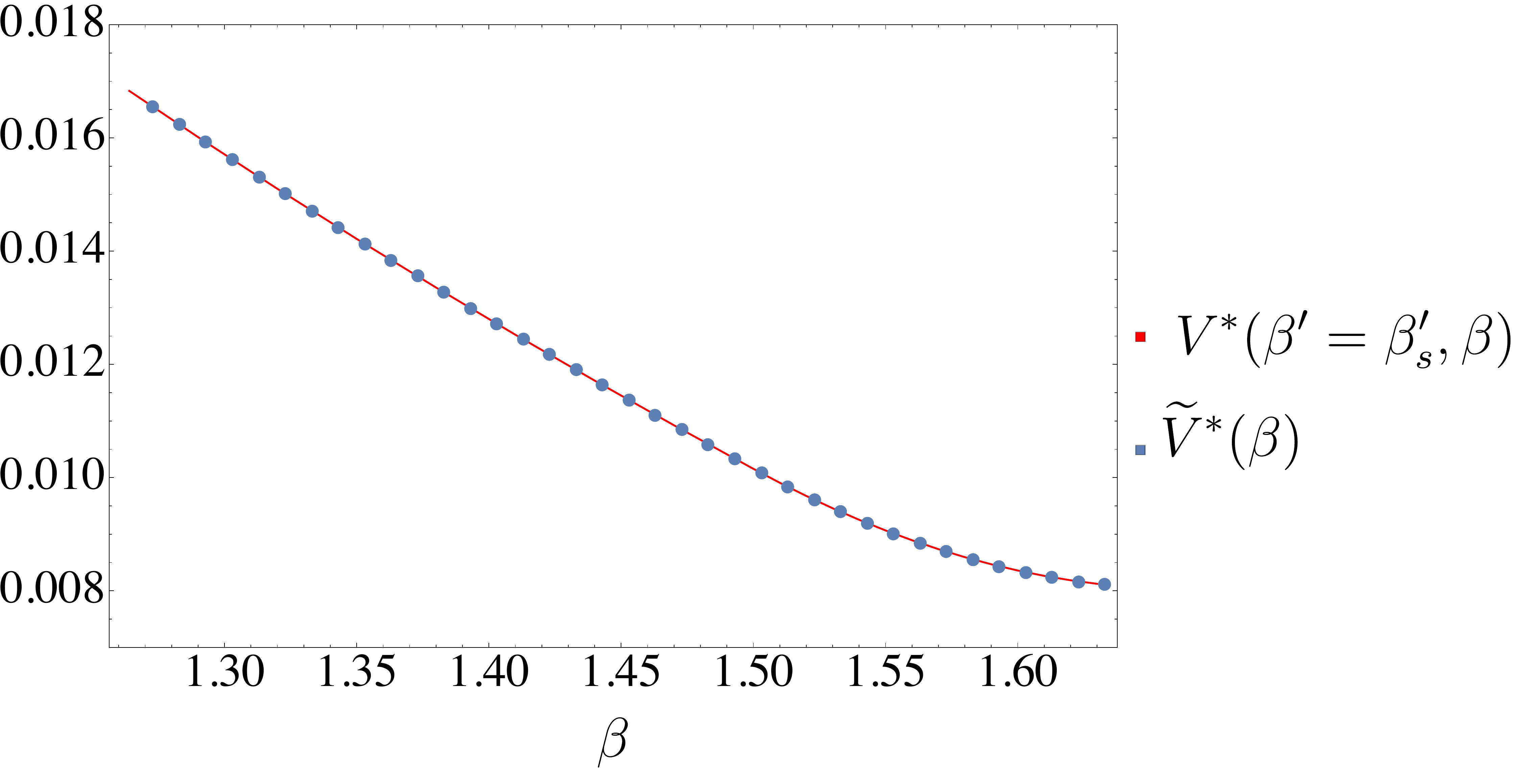} }}%
    \\
    \subfloat[\label{Vtilt34} $3+4$-spin: $V_{min}(\beta'_s,\beta)$ and $V^{*}(\beta'_s,\beta)$ do not coincide, as discussed in the text.]{{\includegraphics[width=10cm]{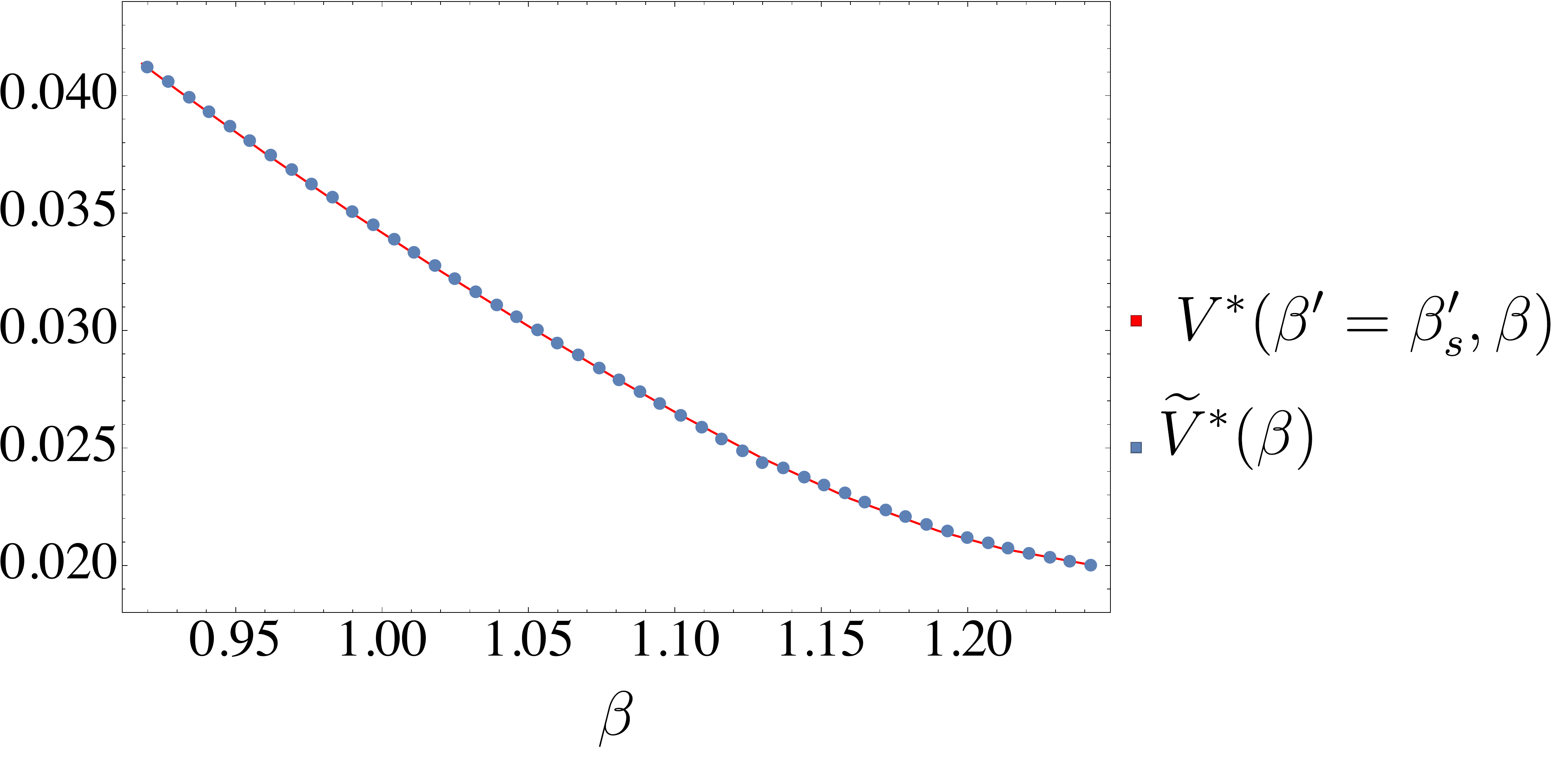} }}%
    \caption{\label{fig:Vtilt} Comparison between the value of the tilted potential in $p^*$ at inverse temperature $\beta$, $\widetilde{V}^*(\beta)$, and the value of the potential with first temperature equal to $T'_s$ and second temperature equal to $T$, in $p^*$, $V^*(\beta'_s,\beta)$. These two quantities match. As commented in the text, and shown in Fig. \ref{fig:pspm3-34}, while in the pure model the secondary minimum of the potential with these two temperatures occurs in $p^*(T)$, this is not true in the mixed model.}%
\end{figure}

Eq. (\ref{eqpminpTpTpotFP}) suggests that for $\beta<\beta_d$, lowering the first temperature of the potential to the first value where the stationary point appears, $T'_s=1/\beta'_s$, it is possible to describe the point where a stationary point is more likely to appear in the random potential $V(p|\tau)$ at temperature $T=1/\beta>T_d$. 
Thus we compare the value of the tilted potential $\widetilde{V}(p)$ in $p^*$ at temperature $T$, $\widetilde{V}^{*}(T)$, with $V_{min}(T'_s,T)$.
In the pure model, $V_{min}(T'_s,T) = V^{*}(T'_s,T)$, where the r.h.s. of the equation is the value of the potential with first temperature equal to $T'_s$ and second temperature equal to $T$ in $p^*$.
It turns out that 
\begin{equation}
V_{min}(T'_s,T)=\widetilde{V}^*(T)\:.
\end{equation}
This behavior is analyzed in Fig. \ref{pspm3}-\ref{Vtilt3} in the range $1.264<\beta<\beta_d$ for the $3$-spin. At $\beta=1.264$ the secondary minimum exists only if the first replica is taken at $\beta'_s=\beta_K$. Decreasing further $\beta$, we should increase the value of $\beta'_s$ of the first replica and our approximation would not be valid anymore.
The interpretation of these results is that when $T>T_d$ and no local minimum is present in $V(p)$, the most likely way to make it appear in an atypical realization $V(p|\tau)$ is that the first replica $\tau$ is taken inside one of the equilibrium states at $T'_s$, still existing at $T$ as an out-of-equilibrium state.
This state can be studied with the potential, lowering the first temperature from $T$ to $T'_s$ in order to make  the secondary minimum re-appear.
In other words, the rare appearance of the secondary minimum in the random potential $V(p|\tau)$ at temperature $T$ is due to the sampling of $\tau$, at temperature $T'=T$, among the configurations that are typical at $T'_s$. 

\begin{figure}%
    \centering
    \subfloat[\label{Sq3-1} $\beta=1.62$]{{\includegraphics[width=10cm]{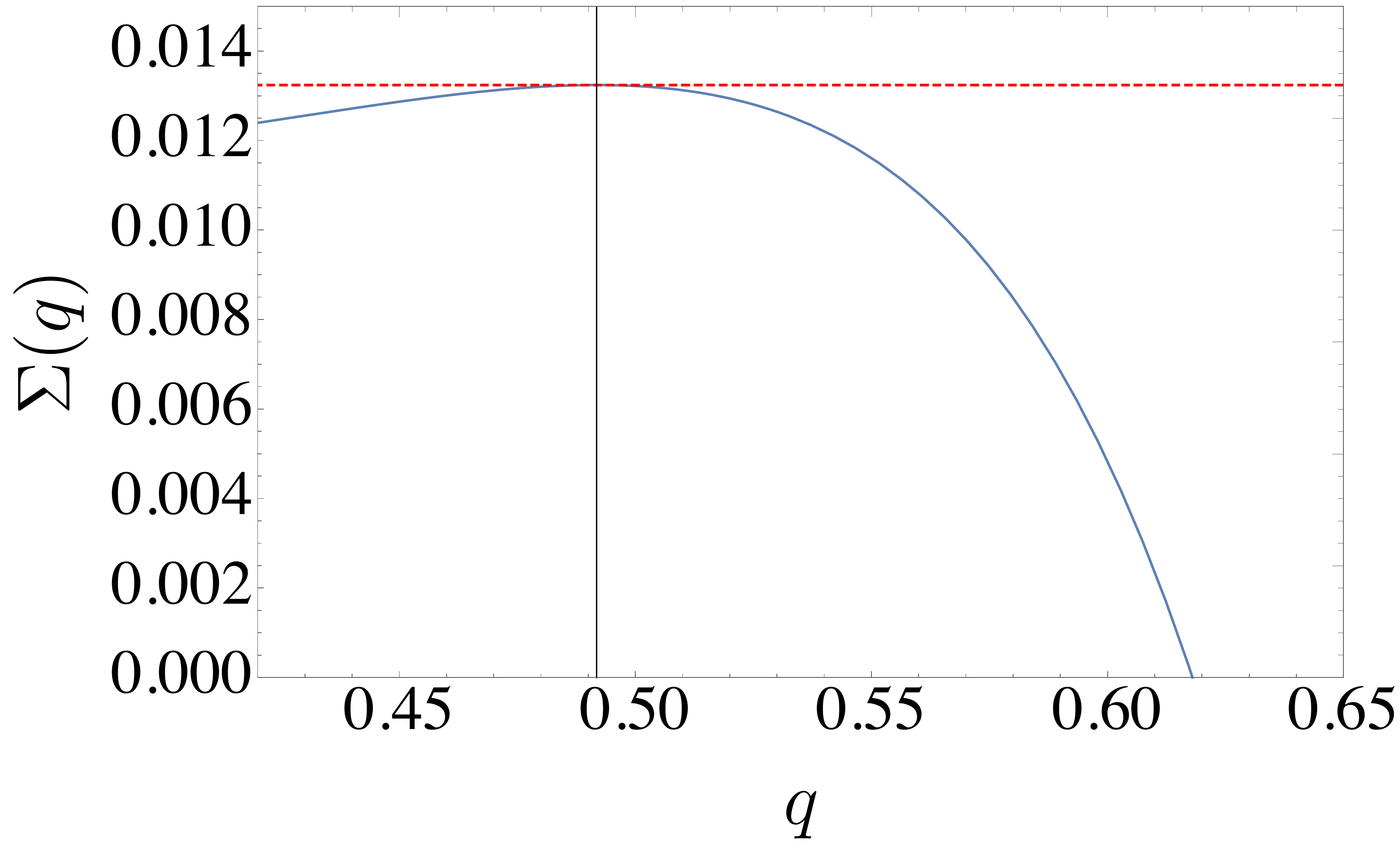} }}%
    \\
    \subfloat[\label{Sq3-2} $\beta=1.55$]{{\includegraphics[width=10cm]{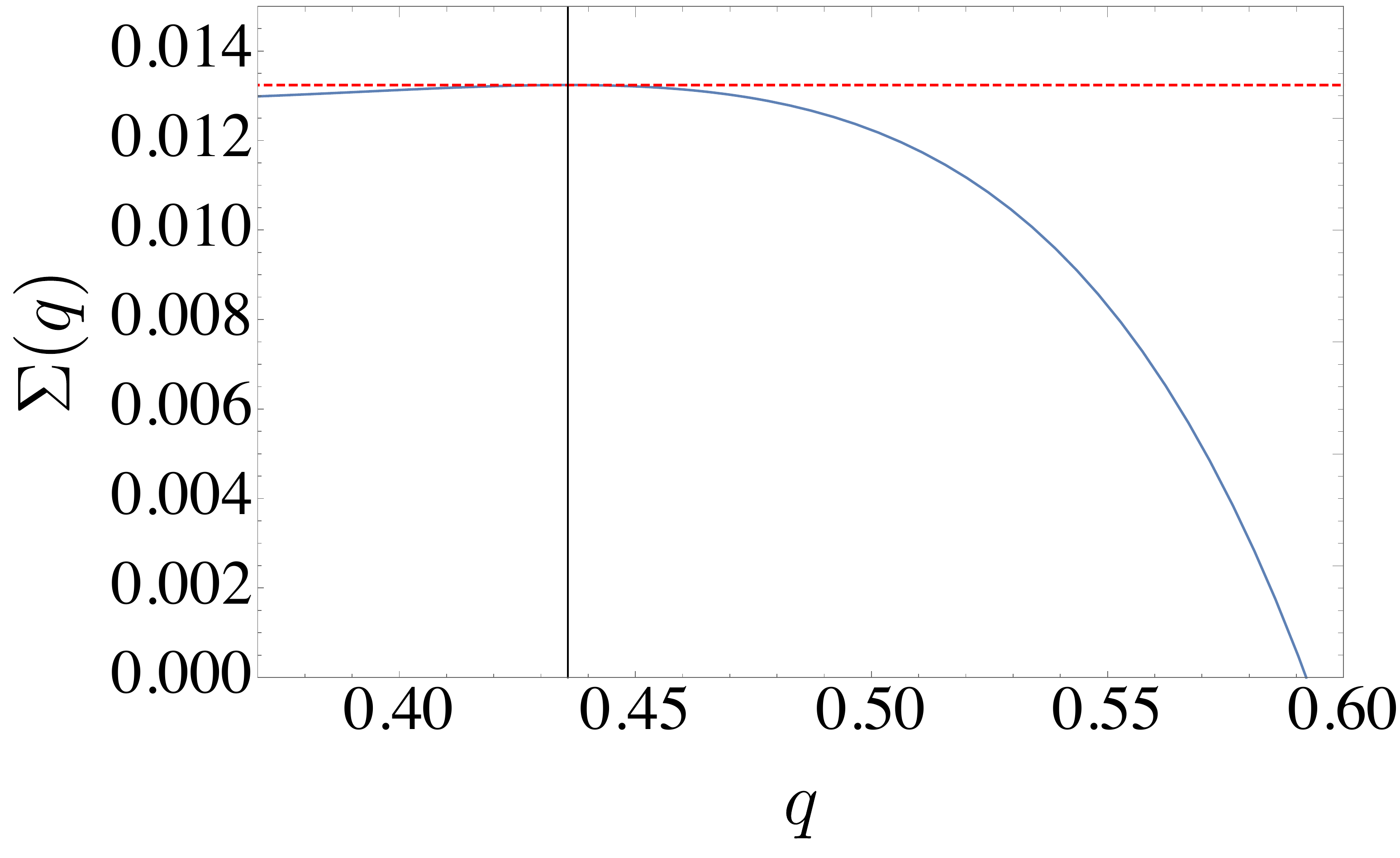} }}%
    \caption{\label{fig:Sq3} $\Sigma(q)$ for the $3$-spin at two temperatures, above the dynamical one. The horizontal line indicates the maximum value attained by the complexity. As long as $\beta>1.5$, in the pure model, the maximum of the complexity does not change. $q_{RS}(T)$, $q_{min}(T'_s,T)$ and $q_{max}(T)$ coincide and are indicated by the vertical line. For $q<q_{RS}$ the solutions are unstable in the replica space and thus unphysical. }%
\end{figure}

\begin{figure}%
    \centering
    \subfloat[\label{Sq34-1} $\beta=1.22$]{{\includegraphics[width=10cm]{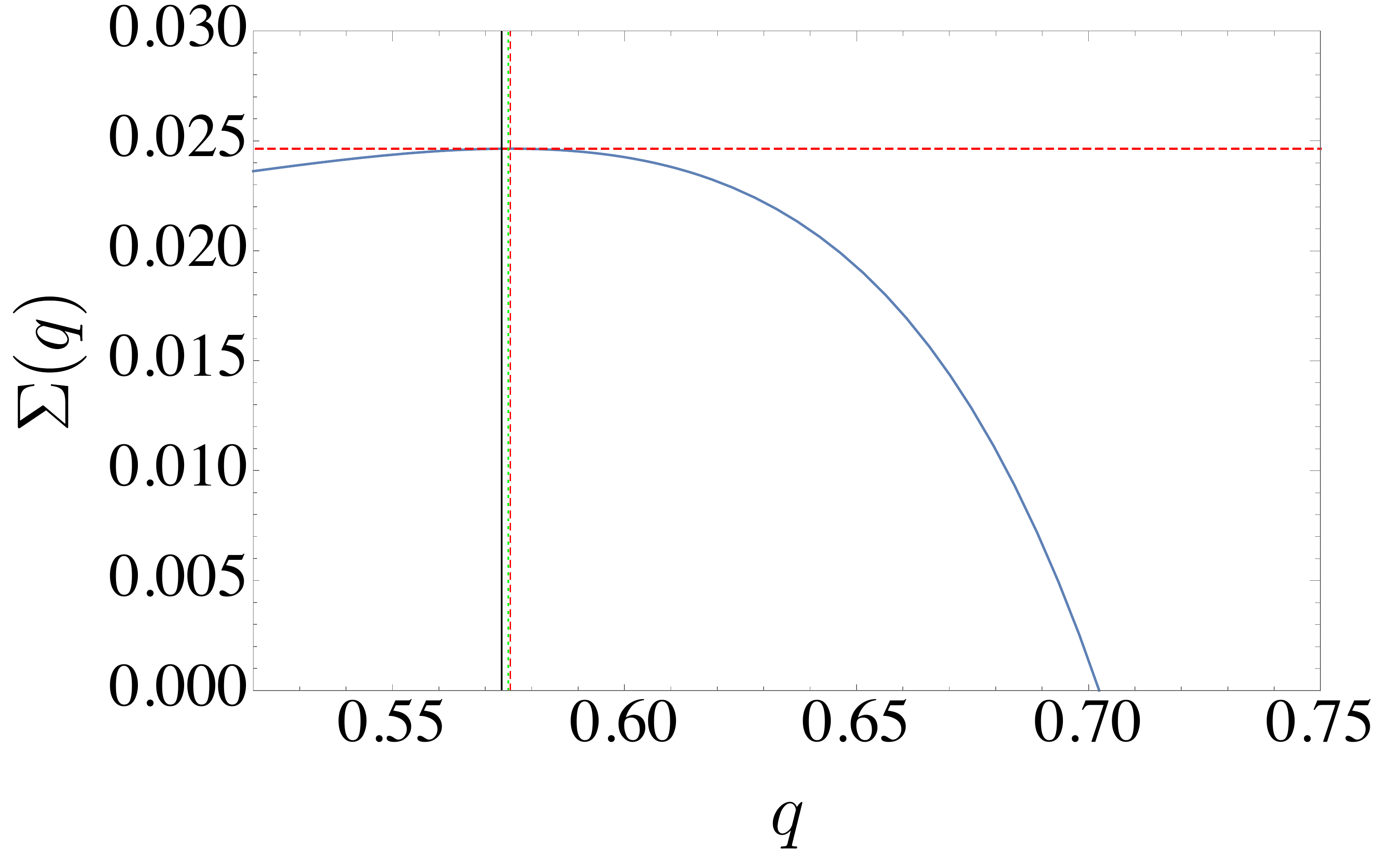} }}%
    \\
    \subfloat[\label{Sq34-2} $\beta=1.16$]{{\includegraphics[width=10cm]{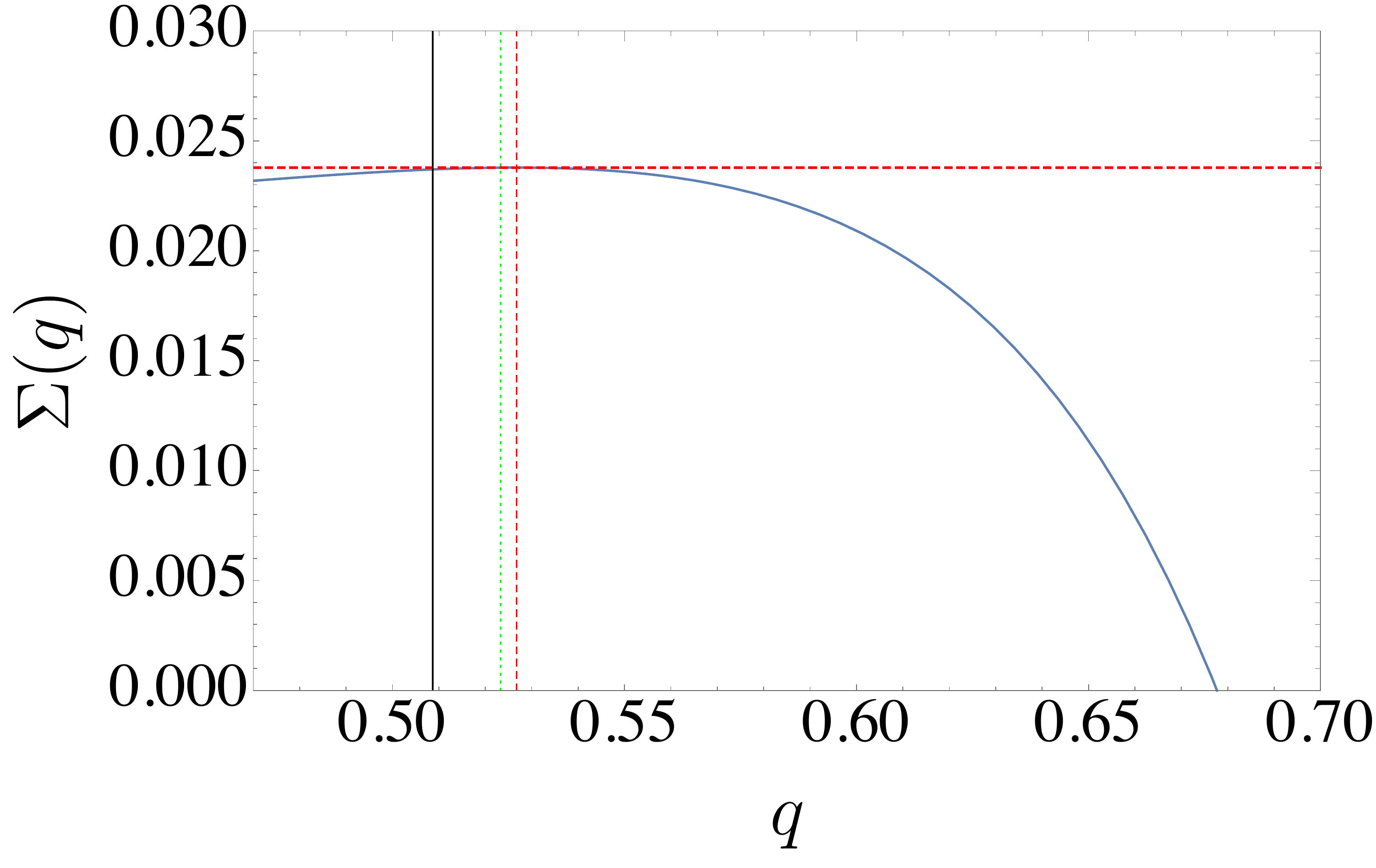} }}%
    \caption{\label{fig:Sq34} $\Sigma(q)$ for the $3+4$-spin at two temperatures, above the dynamical one. In the mixed model, $q_{RS}(T)$, $q_{min}(T'_s,T)$ and $q_{max}(T)$ do not coincide, except at $\beta_d$ ($\beta_d=1.24198$ in the $3+4$-spin) and are indicated by the black, green and red vertical line, respectively. The horizontal line indicates the maximum value attained by the complexity. For $q<q_{RS}$ the solutions are unstable in the replica space and thus unphysical. }%
\end{figure}

For the mixed model, results are more difficult to interpret. This model, contrarily to the pure model, presents the general property of level crossing and the associated temperatures chaos. In other words, following states in temperature does not preserve their order in terms of their free energy \cite{franz1995glassy}.
As described previously, when $T'=T_d$ the secondary stationary point disappears as soon as $T>T_d$. We can again lower $T'$ from $T_d$ in order to make it appear. At this point, we may again look at the value of $p$ where the potential has the secondary stationary point $p_{min}(T'_s,T)$ and we may compare it with $p^{*}(T)$. In this case, there is no match between the two. While $G(p)$ is minimum in $p_{RS}$, as long as it exists for $\beta>\beta_{RS}$ ($\beta_{RS}=1.132$ in the $3+4$-spin model), $p_{min}(T'_s,T)$ is found to be larger: increasing $T$ above $T_d$ in the potential, the secondary minimum disappears before that the states it describes become marginal.
Nevertheless, the large deviation analysis still predicts that, when $\beta_{RS}<\beta<\beta_d$, in order to observe a secondary stationary point in $V(p|\tau)$, the first replica must be chosen in one of the marginal states still existing at $T=1/\beta$, and when $\beta<\beta_{RS}$ in one of the states with overlap $q$ such that $1-\beta^2 f''(q)(1-q)^2$ is minimum.
Looking at the tilted potential, we observe that $\widetilde{V}^*(T)$ coincides with $V^*(T'_s,T)$ as in the pure model, but at difference with the pure model, $V^*(T'_s,T) > V_{min}(T'_s,T)$, since the local minimum is at a value of $p\neq p^*$, as commented previously.
This behavior is analyzed in Fig. \ref{pspm34}-\ref{Vtilt34} in the range $0.9175<\beta<\beta_d$ for the $3+4$-spin. At $\beta=0.9175$ the secondary minimum exists only if the first replica is taken at $\beta'_s=\beta_K$. Decreasing further $\beta$, we should increase the value of $\beta'_s$ of the first replica and our approximation would not be valid anymore.

In order to look for the existence of marginal states both in the pure and in the mixed model, we compute the complexity of states $\Sigma(q)$ as a function of their overlap at temperature $T$, using replicas \cite{monasson1995structural,mezard1999compute}.
We show the results in Figs. \ref{fig:Sq3}-\ref{fig:Sq34}, where we denote by $q$ the self overlap of states $q^1$. 
We denote by $q_{max}(T)$ the value of $q$ where $\Sigma$ is maximum. See the Appendix for more details.
The observation of marginal states above $T_d$ both in the pure and in the mixed model is possible thanks to a non-orthodox prescription in the use of the clone method, used to evaluate the complexity, as discussed in \cite{FRU}. 
In both models, picking the first replica in a marginal state is the key to observe the re-appearance of the secondary stationary minimum for $\beta<\beta_d$. The main difference between the pure and the mixed model is that while, in the former, marginal states can be obtained by following states within the framework of the potential, in the latter they cannot.

\subsection{Zero temperature}

State following for the mixed model when $T<T'$ is a well known open problem. In terms of the potential, when $T'=T_d$, the secondary stationary point disappears not only increasing the second temperature $T$ but also decreasing it \cite{franz1995glassy,barrat1997temperature}. 
One can study the potential with $T=0$, $V_0(p)$, at the RS level and observe that it develops the secondary minimum only for a temperature $T'<T_d$. For comparison, in the pure model, this happen exactly at $T'=T_d$.
The same picture holds at a finite temperature $T$: it is possible to define a $T_{rsb}(T')$ below which states multifurcate and the local minimum of the RS potential disappears \cite{barrat1997temperature}.
The impossibility to follow states by cooling when $T'=T_d$ remains valid also taking into account the 1RSB potential \cite{sun2012following}, with the only difference that the minimum disappears at a larger temperature, still smaller than $T_d$.
Inspired by these anomalies, the long-time limit of the out of equilibrium dynamics of mean-field spherical mixed models has been investigated in \cite{folena2019memories}. 
Interestingly, considering a gradient descent ($T=0$) dynamics starting from a configuration at equilibrium at $T'$, a new description of the dynamical phase transition emerged. Surprisingly, for some temperatures $T'>T_d$, it was shown that the dynamics converges below the energy of the threshold states, where threshold states are the most numerous ones at $T=0$.
This behaviour is radically different from that of the pure model, where starting from $T'>T_d$ the zero-temperature dynamics always converges to the same value $e_{th}$ and memory of the initial condition is lost \cite{cugliandolo1993analytical}. 
In the mixed model, this happens only for $T>T_{onset}>T_{d}$.
Moreover, it was found that it is only for $T<T_{SF}<T_d$ that the dynamics becomes fast and  the long time dynamics converges to the states described by the local minimum of the RS potential with the second temperature equal to zero. 
The existence of a new phase, the \textit{hic sunt leones} phase for $T_{SF}<T<T_{onset}$, is predicted. In this phase, the relation between dynamics and static computation is missing. Differently thermalized configurations lie in basins of attraction of different marginal states. The dynamics does not lose memory of the initial condition and presents the features of aging in metabasins \cite{barrat1997temperature} but the analysis performed in \cite{folena2019memories} seems to exclude this scenario.

\begin{figure}
\centering{}\includegraphics[scale=0.35]{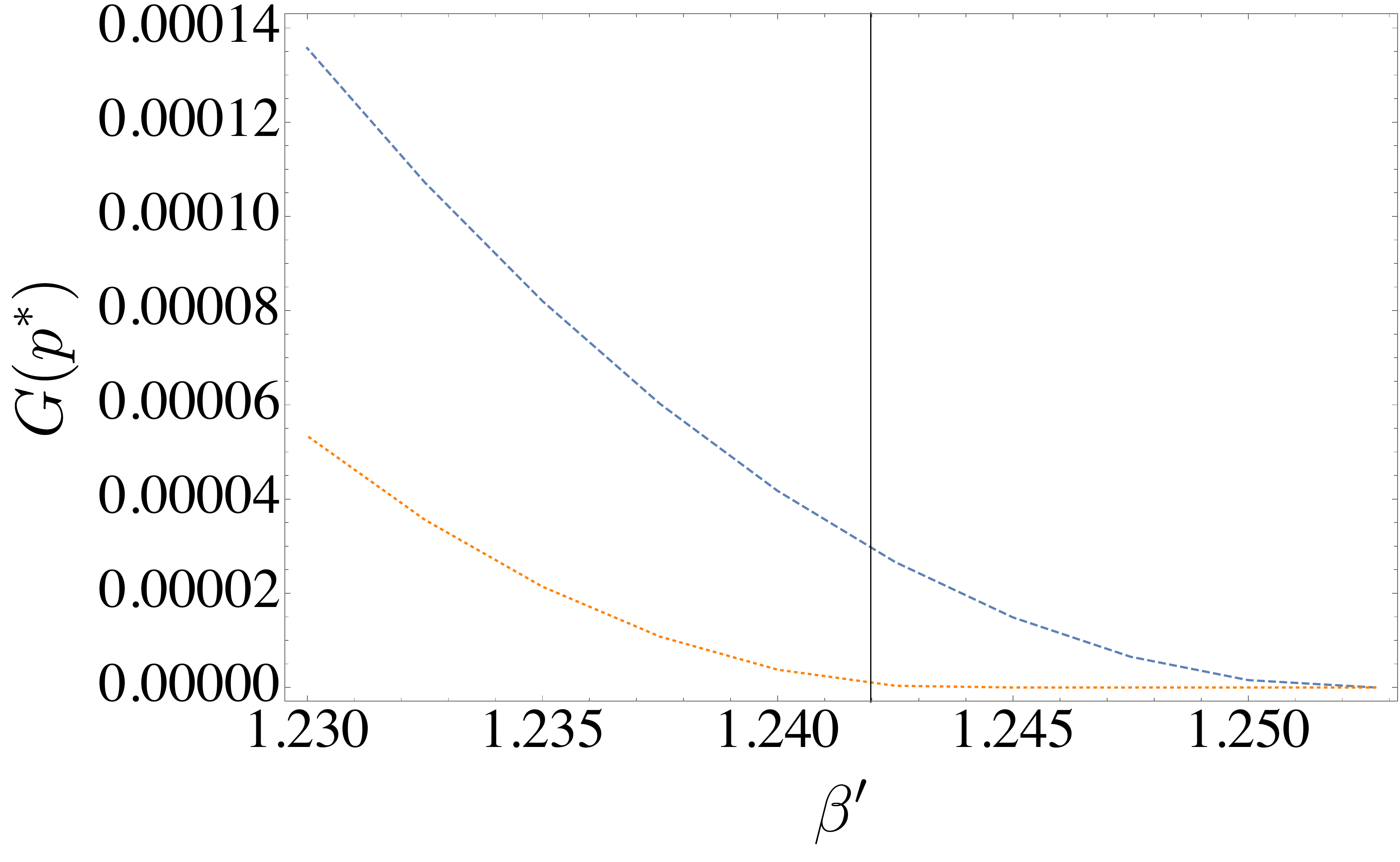}\caption{\label{Gpstar} $G(p^*)$ computed on the RS (blue dashed) and on the 1RSB (orange dotted) potential with $T=0$ per the $3+4$-model. The right most point on the $x-$ axis is $\beta_{SF}$. The RS potential loses its local minimum for $T>T_{SF}$. The 1RSB one lose it at a larger $T=0.8041<T_d$. $\beta_d$ is indicated by the vertical line. As long as $V_0$ has a local minimum, $G_0$ has its local minimum on the same point. $p^*$ computed on the RS expression of $G_0$ does not coincide with the $p^*$ computed with the 1RSB expression of $G_0$. }
\end{figure}

The large deviation analysis can be done at $T=0$ in order to study the rare events of the re-appearance of the local minimum  when typically it does not exists, thus in the \textit{hic sunt leones} phase as well.
The $T=0$ limit $G_0(p)$ of the large deviation function can be done by taking $q^1=1-\chi T$ and $x = y T $ in eq. (\ref{fulldevilargminGp1rsb}), and by taking the $T=0$ limit. The 1RSB expression is valid for $p<p_{RS}$, where RS breaks down. The RS marginality condition $1=\beta^2 f''(q)(1-q)^2$ at $T=0$ is
\begin{equation}
1=\chi^2 f''(1)\:.
\label{eq:marginazerotempdef}
\end{equation}
On the other hand, at $T=0$ the RS saddle point equation relating $q$ and $p$, eq. (\ref{sppqRepsymm}), becomes
\begin{equation}
p=\sqrt{1-\chi^2 f'(1)}\:
\end{equation}
and thus solving for $p$ one finds $p_{RS}=\sqrt{1-f'(1)/f''(1)}$, i.e. $p_{RS}=0.707107$ in the $3$-spin and $p_{RS}=0.781736$ in the $3+4$-spin.
For larger $p$, the RS expression must be used, obtained by taking $q^1=1-\chi T$  in eq. (\ref{rsexpgrandideviazionifinzp}) and by taking the limit $T \rightarrow 0 $,
\begin{eqnarray}
2 G_0(p)= & \Bigg \{ \beta'^2 \chi ^2 f'(p)^2 \left(f'(1)- \left(p^2-1\right) f''(1)\right)+ \nonumber
	\\ & -f'(1) \Bigg[\chi ^2 \left(f''(1)+f'(1)\right) \log \left(\frac{\left(1-p^2\right) \chi }{f'(1)}\right)+ \nonumber
	\\ & \qquad \qquad + \left(\chi ^2 f'(1)-1\right) \left(\chi ^2 f''(1)+1\right)+ \nonumber
	\\ & \qquad \qquad  -3 \chi ^2 \log (\chi ) \left(f''(1)+f'(1)\right) + \nonumber
	\\ & \qquad \qquad - 2 \beta' p \chi  f'(1) f'(p) \left(\chi ^2 f''(1)+1\right) \Bigg] \nonumber \Bigg \} \times 
	\\ & \times \Bigg[ \chi ^2 f'(1) \left(f''(1)+f'(1)\right)\Bigg]^{-1}\:.
\end{eqnarray}
This expression can be obtained from the 1RSB one by setting $q^0=1$.  
While in the pure model, as the local minimum of $V_0$ disappers, $p^*$ coincides with $p_{RS}$, this is not true in the mixed model. 
In the mixed model, $V_0$ has the local minimum in the RS region as long as $T'>T_{SF}$ and in the 1RSB region for larger $T$.
When the 1RSB potential loses its minimum at $T'= 0.8041 <T_d$, the large deviation function $G_0(p)$ has still a local minimum, with $G(p^*)>0$, as can be observed in Fig. \ref{Gpstar}. This implies that, in the \textit{hic sunt leones phase},  results obtained from numerical simulation of the dynamics at $T=0$ in finite systems may be influenced by the rare dynamics where memory of the initial condition is not lost. This new phase is not completely understood yet and a comparison with the results provided in \cite{folena2019memories} in finite systems is still in progress \cite{FRUG}.

\section{Concluding remarks}
\begin{figure}%
    \centering
    \subfloat[$3$-spin.]{{\includegraphics[width=10cm]{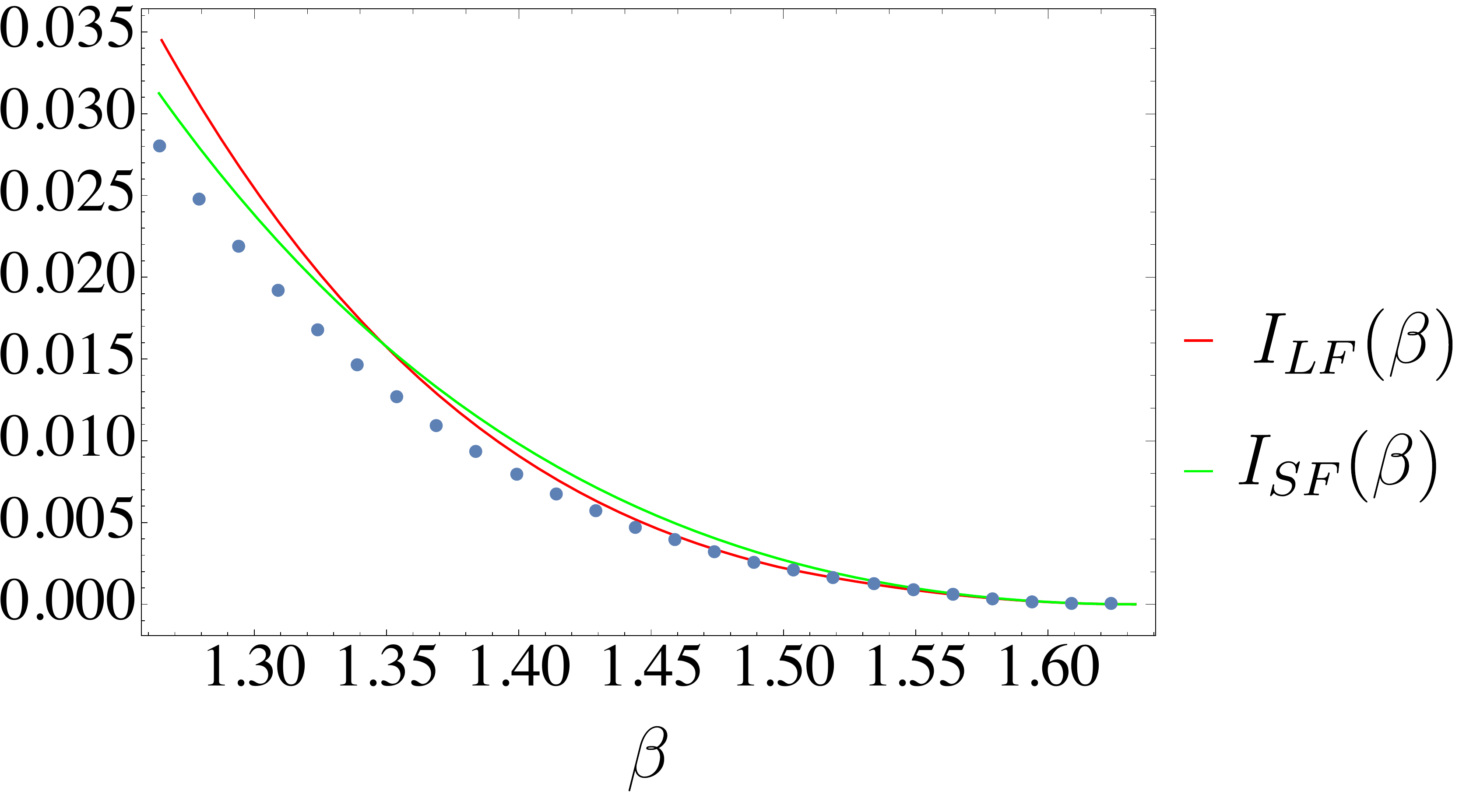} }}%
    \\
    \subfloat[$3+4$-spin.]{{\includegraphics[width=10cm]{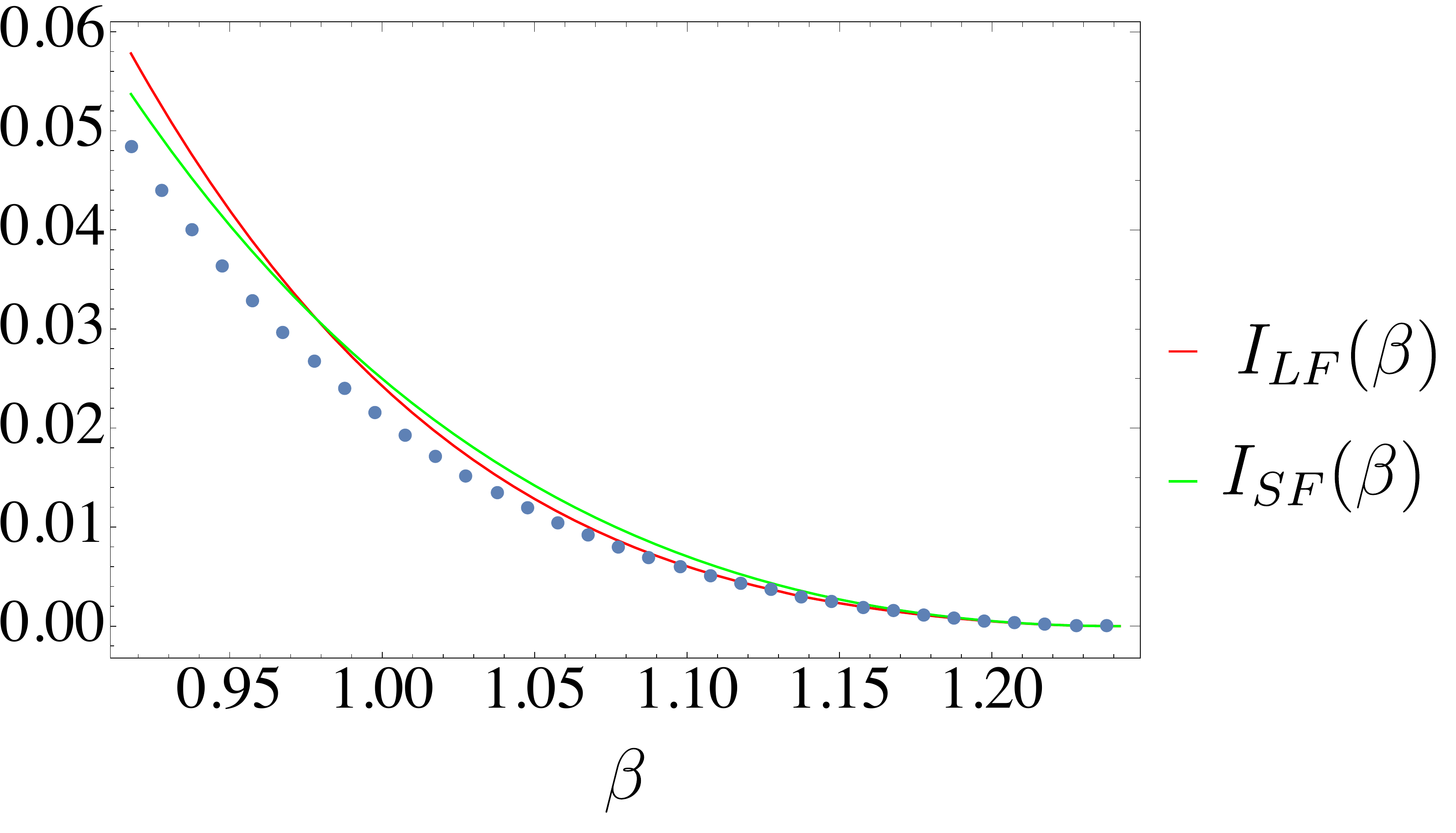} }}%
    \caption{\label{fig:ILFSFfin} Comparison between $I_{SF}$ and $I_{LF}$ for different $\beta<\beta_d$. Dots refers to the second order approximation $I_{LF,2}$ of the large deviation function computed on $p_{RS}$. There are no reasons to compute $I_{LF,2}$ on $p_{RS}$ but, as discussed in the Appendix, this computation gives the same results obtained from the large deviation analysis as long as $\beta>\beta_{RS}$. This property is lost for $\beta<\beta_{RS}$, where $p_{RS}$ is replaced by the value of $p$ associated to the $q$ for which the replicon mode is minimum. The second derivatives of the large deviation functions approaching $\beta_d$, $\lim_{T\rightarrow T_d}I''(T)$, are 0.25 in the $3$-spin and $0.543$ in the $3+4$-spin.}%
\end{figure}

In this paper we provide a first generalization of the theory of fluctuations
in glasses based on the analogy with the spinodal point of the RFIM,
in which the effect of the self-induced disorder in the beta regime
is described by random term in a cubic field theory. This theory provides
a quasi-equilibrium description and it is valid in the vicinity of
the critical point, the mode coupling transition. Moreover, it relies
on the approximation that fluctuations are Gaussian. In our approach,
we keep the first ingredient, which allows to study fluctuations through
the use of constrained equilibrium measures and their associated replica
action. On the other hand, we go beyond the assumption of the vicinity
to the critical point presenting a first principle computation of
a the large deviation function. More precisely we look for the probability
of the existence of the secondary minimum of the potential
in a finite region of the paramagnetic phase above the dynamical temperature.
We consider the case of spherical spin models and we show that in the vicinity of the critical point, we recover the
result implied by the cubic field theory, as observed in Fig. \ref{fig:ILFSFfin} for both the pure and the mixed models.
The large deviation analysis allows to go beyond the vicinity to $T_d$ and to show that the re-appearance of the secondary stationary point 
in a regime where typically it does not exist can be explained in terms of the sampling of the first replica of the potential in 
an out of equilibrium state. As long as marginal states exists, picking the first replica in these states dominates the probability to observe the re-appearance of the local minimum. When they disappear, this probability is dominated by states with an overlap $q$ such that the replicon mode is minimum.
In terms of the dynamics, when $T$ is large and the dynamics is performed at $T=T'$, while typically the system lose memory of its initial condition, our results suggests that on finite $N$ systems it is possible to observe as a rare event that the dynamical correlation does not decay to zero, but to $p^*$.
Finally, extending the analysis at $T=0$, we show that understanding the new phase predicted in \cite{folena2019memories} in numerical simulations may be tricky. In fact, dynamics with memory of the initial condition may be explained in term of atypical events. As with finite temperatures, looking at Fig. \ref{Gpstar}, these events are not so rare: if $N=100$, $e^{-100 G(p^*)} \sim O(1)$ for a wide range of temperatures around $T_d$. 
 
\section{Acknowledgments}
We are very grateful to Giulio Biroli, Valentina Ros, Sungmin Hwang, Giampaolo Folena, Federico Ricci-Tersenghi and Pierfrancesco Urbani for many interesting and stimulating discussions. 
J.R. thanks the Dipartimento di Fisica dell'Universit\`{a} La Sapienza di Roma for the hospitality.
Finally, we acknowledge the support of a grant from the Simons Foundation (No. 454941, Silvio Franz).

\clearpage

\section*{References}

\bibliography{bibsamp}
\bibliographystyle{unsrt}

\section*{Appendix}

\subsection*{Replica computations}

Here we describe the replica computations used to obtain all the results provided in the main text. Using the notation
introduced in the main text, see eq. (\ref{eq:defS2actiondeff}), let us consider 
\begin{equation}
\left<
Z^{n_{1}}[p_1|\tau] Z^{n_2}[p_2|\tau] \ldots  Z^{n_k}[p_k|\tau] \frac{}{}
\right>=e^{\frac{N}{2}S^{(k)}[p_1,p_2,\ldots,p_k]}\:.\label{eq:eqcorrfdefk1}
\end{equation}
Standard manipulations lead to the general result
\begin{equation}
S^{(k)}[Q]=\left(2 \beta \beta' \sum_{k'=1}^k n_{k'} f(p_{k'}) + \beta^{2}\sum_{a,b=1}^{n_1+\ldots+n_k} f(Q_{ab})+\log\det Q \right)\label{eq:actionresultgeneral}
\end{equation}
where the matrix $Q$ is symmetric and $f(q)=q^{p}/2$. The size of $Q$ is $1+n_1 + \ldots + n_k$. For every $k$, the diagonal elements of $Q$ are equal to $1$. For $k=1$, $Q_{1a}=p_{1}$ for $a=1,\ldots, n_1$. For $k=2$, $Q_{1a}=p_{1}$ for $a=1,\ldots, n_1$ and $Q_{1a}=p_{2}$ for $a=n_1+1,\ldots, n_1+n_2$. Generalization to a generic $k$ is straightforward.
The diagonal block matrices are labelled by $Q_{ab}^{1}$, $Q_{ab}^{2}$ and so on. They reflect the replica ansatz of the constrained system. In the RS ansatz they are parametrized by
\begin{equation}
Q_{ab}^{k}=q_{k} + (1-q_k) \delta_{ab}\:,
\end{equation}
and $q_k$ has to be optimized over. For the sake of simplicity, here we describe results obtained with the RS ansatz. Generalization to 1RSB and details on the computation of the determinant are provided later. Both in the RS and in the 1RSB ansatz, the out-of-diagonal block indexed by $lm$ is a rectangular $n_l \times n_m$ matrix with all the elements equal to $q_{lm}$.

The action $S^{(k)}$ can be expanded
in a power series of $n_{1}$, $n_{2}$, ..., $n_{k}$ with the zero
order term being zero. We indicate derivatives with respect to replicas, computed at a number of replicas equal to zero, with a superscript notation. The expression for $k=1$ is
\begin{equation}
S^{(1)}[p_1]=nS^{(1),(1)}[p_1] \label{eq:expansionS1inn}
\end{equation}
neglecting second order terms. Using eq. (\ref{eq:deffcons1}) and (\ref{eq:expansionS1inn}),
the potential is computed from
\begin{equation}
V(p_1)=F(p_1)-F=-\frac{1}{N\beta}\lim_{n\rightarrow0}\frac{\partial}{\partial n}\frac{N}{2}S^{(1)}[p_1]-F\:.
\label{eq:potintermsofS11}
\end{equation}
Setting $p_1=p$ and $q_1=q$ its expression reads 
\begin{equation}
V(p)=\frac{1}{2 \beta}\left( \frac{q-p^2 }{ q-1} - 2 \beta \beta' f(p) + \beta^2 f(q) - \log(1 - q) \right)\:.
\label{potRSexpressionebetabetappp}
\end{equation}
The saddle point value of $q$ at a given $p$ is given by
\begin{eqnarray}
\frac{q -p^2 - (q-1)^2 \beta^2 f'(q)}{ (q-1)^2}& = 0\:. 
\label{eq.qpRSVpot}
\end{eqnarray}
Now we study the term $k=2$. $S^{(2)}[p_1,p_2]$ can be expanded as
\begin{eqnarray}
S^{(2)}[p_1,p_2] = & n_{1}S^{(2),(1,0)}[p_1]+n_{2}S^{(2),(0,1)}[p_2] + \nonumber \\   &  n_{1}^{ 2}S^{(2),(2,0)}[p_1]  +n_{2}^{2}S^{(2),(0,2)}[p_2] + n_{1}n_{2}S^{(2),(1,1)}[p_1,p_2] \label{eq:expansionS2inn}
\end{eqnarray}
neglecting third order terms, and it is has the following properties
\begin{eqnarray}
\partial_{q_{12}}S^{(2),(1,0)}[p_1]& =\partial_{q_{12}}S^{(2),(0,1)}[p_2]&=0\:, \\
\partial_{q_2}S^{(2),(1,0)}[p_1]& =\partial_{q_1}S^{(2),(0,1)}[p_2]&=0\:,
\end{eqnarray}
i.e.: $q_{12}$ appears only in second order terms and the first order terms relative to replica $1(2)$ does not depend on saddle point parameters of replica $2(1)$. 
Moreover, the first order terms $S^{(2),(1,0)}[p]$ and $S^{(2),(0,1)}[p]$ are equal to $S^{(1),(1)}[p]$. $S^{(k)}$ can be expanded in a similar way. 
Using eq. (\ref{Whet0}), fluctuations are computed from
\begin{eqnarray}
W^{(2)}_{p_1,p_2} =  \frac{1}{\beta^2 N^2} \lim_{n_{1}\rightarrow0}\lim_{n_{2}\rightarrow0} & \left( \frac{\partial^{2}\left< Z^{n_{1}}[p_1|\tau]Z^{n_{2}}[p_2|\tau] \right>}{\partial n_{1}\partial n_{2}} + \right. \\ & - \left. \frac{\partial  \left< Z^{n_{1}}[p_1|\tau] \right> }{\partial n_{1} } \frac{\partial \left< Z^{n_{2}}[p_2|\tau] \right>}{\partial n_{2}} \right)\:.
\end{eqnarray}
On the other hand, thanks to eq. (\ref{eq:expansionS2inn}), the first term can be written as
\begin{eqnarray}
& \lim_{n_{1}\rightarrow0}\lim_{n_{2}\rightarrow0}\frac{\partial^{2}}{\partial n_{1}\partial n_{2}}  \left< Z^{n_{1}}[p_1|\tau]Z^{n_{2}}[p_2|\tau] \right> \nonumber \\ 
= &\lim_{n_{1}\rightarrow0}\lim_{n_{2}\rightarrow0}\left[\frac{N}{2}\frac{\partial^{2}S^{(2)}[p_1,p_2]}{\partial n_{1}\partial n_{2}}+\left(\frac{N}{2}\right)^{2}\frac{\partial S^{(2)}[p_1,p_2]}{\partial n_{1}}\frac{\partial S^{(2)}[p_1,p_2]}{\partial n_{2}}\right]\nonumber \\ 
= &\lim_{n_{1}\rightarrow0}\lim_{n_{2}\rightarrow0} \left[\frac{N}{2} S^{(2),(1,1)}[p_1,p_2] + \left(\frac{N}{2}\right)^{2} S^{(2),(1,0)}[p_1] S^{(2),(0,1)}[p_2] \right]
\label{eq:2pfrepexpconn}
\end{eqnarray}
and we observe that the last term is equal to the disconnected part. Thus, as stated in eq. (\ref{Whet11}), fluctuations are computed from
\begin{equation}
W^{(2)}_{p_1,p_2} =  \frac{1}{2 \beta^2 N}  \lim_{n_{1}\rightarrow0}\lim_{n_{2}\rightarrow0} \frac{\partial^{2}}{\partial n_{1}\partial n_{2}}  S^{(2)}[p_1,p_2]
\end{equation}
and their expression reads
\begin{equation}
W^{(2)}_{p_1,p_2} = \frac{1}{2 \beta^2 N} \left( -\frac{( q_{12}-p_1 p_2)^2}{( q_1-1) ( q_2-1)} + 2 \beta^2 f(q_{12}) \right)\:.
\end{equation}
We use $z$ to denote the set of all the order parameters over which we have to optimize $S^{(k)}$. For $k=2$, this set contains $q_1$, $q_2$ and the mixed term $q_{12}$. The optimization over $z$ leads to
\begin{eqnarray}
\frac{q_1 -p_1^2 - (q_1-1)^2 \beta^2 f'(q_1)}{ (q_1-1)^2}& = 0 \:,\\
\frac{q_2 -p_2^2 - (q_2-1)^2 \beta^2 f'(q_2)}{ (q_2-1)^2}& = 0 \:, \\ 
 \frac{ p_1 p_2 - q_{12}}{(q_1-1) ( q_2-1)} +   \beta^2 f'(q_{12}) & = 0 \:.
 \label{eq:speqrsmixedtermexpression}
\end{eqnarray}
We observe that $q_1$ and $q_2$ are found optimizing over $S^{(2),(1,0)}[p_1]$ and $S^{(2),(0,1)}[p_2]$, respectively, and that they coincide with eq. (\ref{eq.qpRSVpot}). The mixed term, on the other hand, is given by the optimization over $S^{(2),(1,1)}[p_1,p_2]$. 

\subsection*{Fluctuations}

As observed above, while the $k=1$ term is relevant for the potential, the $k=2$ term is relevant for the computation of the fluctuations and of the rate function $I(T)$.
While the large deviation case, eq.  (\ref{eq:computeLDFfull}), has already been discussed in the main text, here we focus on the small fluctuations, providing details on the computation of eq.  (\ref{eq:computemuIsd}) and (\ref{eq:computeILDSDl1}).
We focus on the computation of $G_2(p)$, defined in eq. (\ref{eq:deffirstg2tobegeneral}), from which both of them can be retrieved, as shown below. From now on this quantity is called $G_2(p)$ and contains the first derivative of the potential and the connected correlation function of the derivative of the potential, that can be written as
\begin{equation}
\left<\left(\frac{}{}V'(p|\tau) \right)^{2}\right>_{c} = \frac{1}{\beta^2 N^2}\lim_{p_1\rightarrow p} \lim_{p_2\rightarrow p} \frac{d}{d p_1} \frac{d}{d p_2} \left<\log Z[p_1|\tau] \log Z[p_2|\tau]\right>_c
\end{equation}
Using replicas and repeating the manipulations of eq. (\ref{eq:2pfrepexpconn}), we obtain
\begin{equation}
\left<\left(\frac{}{}V'(p|\tau) \right)^{2}\right>_{c} = \frac{1}{2 \beta^2 N}\lim_{p_1\rightarrow p} \lim_{p_2\rightarrow p} \frac{d}{d p_1} \frac{d}{d p_2} S^{(2),(1,1)}[p_1,p_2] \:.
\label{eq:var2Vpexpf}
\end{equation}
At this point, by looking at eq. (\ref{eq:computemuIsd}), we observe the similarity between $\left<\left(\frac{}{}V'(p|\tau) \right)^{2}\right>_{c}$ and $\sigma$. In fact, the two rate functions in eq. (\ref{eq:smallflucIT}) and (\ref{eq:computeILDSDl1}) differ only for the point $p$ in which we compute fluctuations. In the first case, this point is the $p$ where the potential has the secondary minimum at $T_d$ ($\forall T \neq T_d$). In the second case, this point is chosen by looking at the minimum of $G_2(p)$ over $p$ ($\forall T>T_d$).
In both cases, the derivatives with respect to $p_1$ and $p_2$ require some care. In fact, they are total derivatives and saddle point parameters ($z$) are sensitive to shits in  $p_1$ and $p_2$.
Moreover, as observed in the main text (see discussion below eq. (\ref{eqduetouno})), when $p_1$ and $p_2$ are chosen close enough to $p$, we expect the saddle point values of $e^{ \frac{N}{2} S^{(2)}[p_1,p_2]} $ to be close to those of  $e^{\frac{N}{2}S^{(1)}[p]}$. Their small perturbations can be computed taking the derivative of saddle point equations with respect to $p_k$, $k=1,2$,
\begin{equation}
\left. \partial_{p_k} \nabla_z S^{(2)}[p_1,p_2]+ H_z S^{(2)}[p_1,p_2] \partial_{p_k} z \right|_* = 0\:,
\end{equation}
where $\nabla_z$ denotes the gradient with respect to $z=\{q_1,q_2,q_{12}\}$, $H_z$ denotes the Hessian and $*$ denotes the saddle point values at $p_1=p_2=p$. We obtain 
\begin{equation}
\partial_{p_k} z = - \left. [H_z S^{(2)}]^{-1} \partial_{p_k} \nabla_z S^{(2)} \right|_*.
\end{equation} 
The explicit form of these derivatives is provided below. When $p_k$ is shifted, $z$ changes in the following way 
\begin{eqnarray}
\left. \frac{\partial q_k}{\partial p_k} \right|_{p_1=p_2=p}& = -\frac{2 \sqrt{q - ( q-1)^2 \beta^2 f'(q)}}{  2 ( q-1) \beta^2 f'(q) + ( q-1)^2 \beta^2 f''(q)-1} \:, \label{shiftsfinalexpressions1} \\
\left. \frac{\partial q_k}{\partial p_{k'}}  \right|_{p_1=p_2=p}& = 0\:,\qquad k' \neq k,\: \\
\left. \frac{\partial q_{12}}{\partial p_k}  \right|_{p_1=p_2=p} & = -\frac{ \sqrt{q - ( q-1)^2 \beta^2 f'(q)}}{  2 ( q-1) \beta^2 f'(q) + ( q-1)^2 \beta^2 f''(q)-1}  \:. \label{shiftsfinalexpressions3}
\end{eqnarray}
We observe that order parameters relative to replica $1$ ($2$) are insensitive to small changes in $p_2$ (resp. $p_1$).
Given the expansion in eq. (\ref{eq:expansionS2inn}), changes induced in $q_k$ by $p_k$ are determined by first order terms and thus by $S^{(1)}$.
We also notice that the shifts in the mixed term due to a change in $p_1$ and $p_2$ are the same. This information is very important in the large deviation computation, when we set $p_1=p-\delta p/2$ and $p_2=p+\delta p/2$, and thus we can set the shift in the mixed term to be $O(\delta p ^2)$.
Once obtained the shifts in the order parameters due to shifts in $p_1$ and $p_2$ around $p$, we may
define the two following auxiliary functions and compute the variance in eq. (\ref{eq:var2Vpexpf}) as
\begin{eqnarray}
D_1[p_1,p_2] & = \partial_{p_2} S^{(2),(1,1)}[p_1,p_2] + \nabla_{z} S^{(2),(1,1)} [p_1,p_2] \partial_{p_2} z \nonumber \\
D_2[p_1,p_2] & = \partial_{p_1} D_1 [p_1,p_2] + \nabla_{z} D_1[p_1,p_2] \partial_{p_1} z \nonumber \\
\left<\left(\frac{}{}V'(p|\tau) \right)^{2}\right>_{c} & = \left. \frac{1}{2 \beta^2 N} \lim_{p_1\rightarrow p} \lim_{p_2\rightarrow p}      D_2[p_1,p_2]\right|_{*}\:.
\end{eqnarray}
The final expression of  $G_2(p)$ reads
\begin{eqnarray}
G_2(p) & =  -\frac{1}{2}(1-q)^2 \left[ ( q-1) \beta^2 [2 f'(q) + (q-1) f''(q)] -1 \frac{}{} \right]^2 \times \nonumber \\
 & \quad  \times  \left[p + (q-1) \beta^2 f'(p) \frac{}{} \right]^2 \left \{ \left [ \frac{}{} 4 (2 p^2 - q) (q-1)^2 \beta^2 f'(q)^2  + \right. \right. \nonumber \\
 & \quad + f'(q) \left[ \frac{}{} ( 8 p^2 - 3 q-1) (-1 +  q)^3 \beta^2 f''(q)  + p^2 (8 - 12 q) + \right. \nonumber \\
 & \qquad \qquad \quad \left.  + 6 p^4  + q ( 5 q-6) \frac{}{} \right] + \nonumber \\
 & \quad + f''(q) \left[ \frac{}{}  p^2 (2 - 6 q)  + (2 p^2 - q) ( q-1)^3 \beta^2 f''(q) + \right.   \nonumber \\
 & \left. \qquad \qquad \quad \left. + 4 p^4 + ( q-1) q \frac{}{} \right] \right ] \times \nonumber \\
 & \left. \quad \times (q-1)^2 \beta^2  -2 (p^2 - q)^2 (q-2) \frac{}{} \right \}^{-1}
\label{finaleq2smexprRS}
\end{eqnarray}
where $q$ is given by eq. (\ref{eq.qpRSVpot}) at any value of $p$ and $T>T_d$. As discussed above, this quantity gives both eq. (\ref{eq:smallflucIT}) and eq. (\ref{eq:computeILDSDl1}) at the RS level, by taking the appropriate $p$, at any $T>T_d$.

\begin{figure}
\centering{}\includegraphics[scale=0.5]{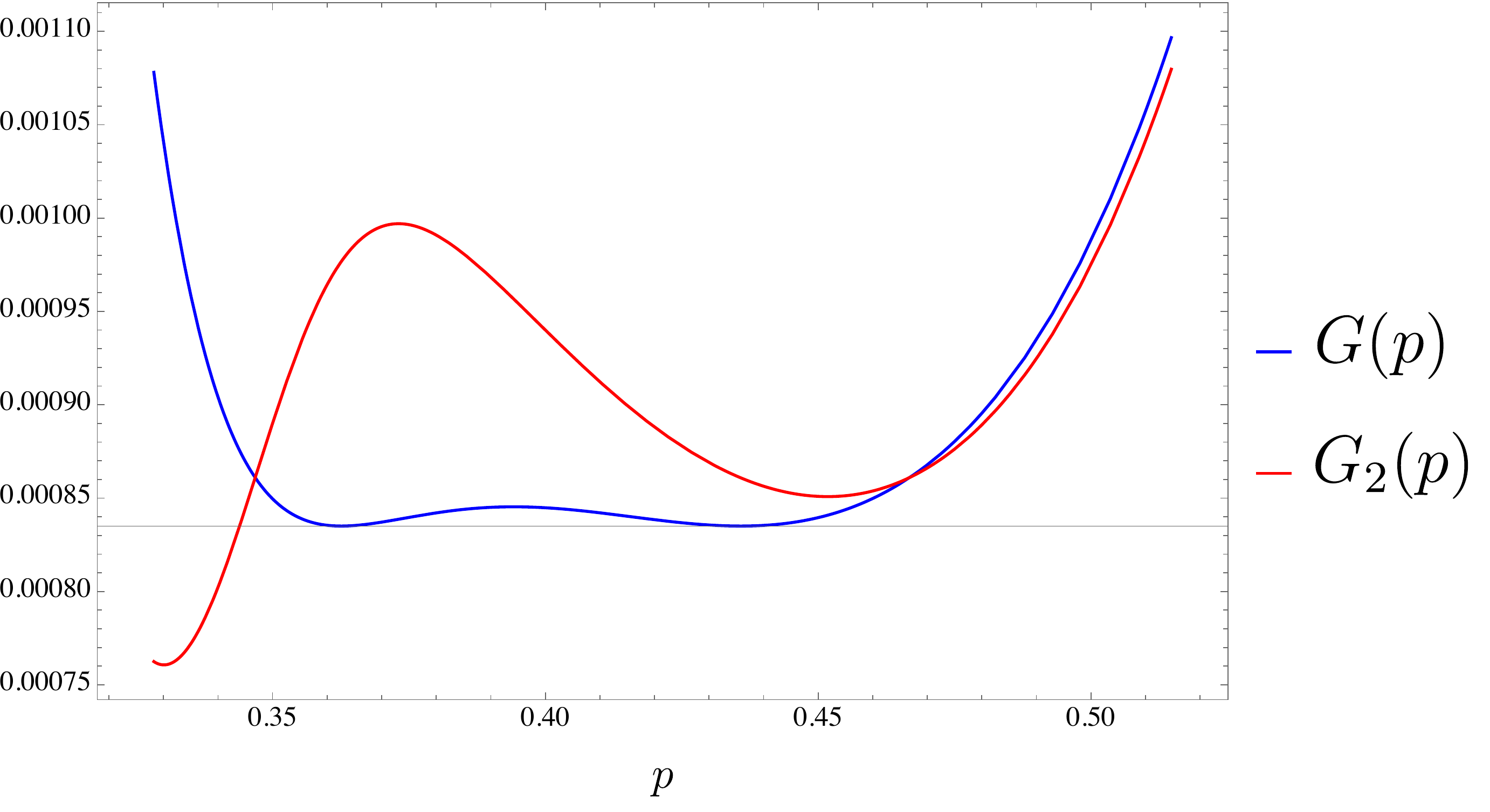}\caption{\label{GG2}$G(p)$ and $G_2(p)$ at $\beta=\beta'=1.55$ at the RS level for the $3$-spin. The two lines cross at $p_{RS}$.}
\end{figure}

This result can be recovered by taking a perturbative expansion of the large deviation function computed up to the second order, see eq. (\ref{eq:gamma2expsecord}). Using $u=m/\delta p$ as in the main text
\begin{equation}
\Gamma_2\left(m,p-\frac{\delta p}{2},p+\frac{\delta p}{2}\right)=  u \left< \frac{}{} V'(p|\tau) \right> + \frac{N u^{2}}{2}\left<\frac{}{}V'(p|\tau)^{2}\right>_{c} \nonumber
\end{equation}
and so, using eq. (\ref{eq:computeLDFfull}), we obtain
\begin{equation}
\left< \frac{}{} V'(p|\tau)^2 \right> =\left.  \frac{1}{N} \lim_{u \rightarrow 0} \lim_{\delta p \rightarrow 0}    \frac{1}{2} \frac{d^2}{d u^2} S^{(2)}\left[p-\frac{\delta p}{2},p+\frac{\delta p}{2}\right]\right|_{n_1=+\frac{m}{\beta } \atop n_2=-\frac{m}{\beta}}\:.
\label{otromodoconn2punt}
\end{equation}
Due to shifts in $p_1$ and $p_2$ in $S^{(2)}$, $\delta q_1$, $\delta q_2$ and $\delta q_{12}$ are shifted as in eq. (\ref{eq.rescspparS2}). For simplicity, similarly to eqs. (\ref{eq:computeLDFfull}) and (\ref{eq:computeLDFfull2}), we set
\begin{eqnarray}
\Gamma_{2} (m,p_1,p_2) &= \left.  \frac{S_2^{(2)}[p_1, p_2]}{2} \right|_{n_1=+\frac{m}{\beta } \atop n_2=-\frac{m}{\beta}} \:, \\
S^{(2)}_2(u,p) &=  \lim_{\delta p \rightarrow 0} \left. S^{(2)}\left[p-\frac{\delta p}{2},p+\frac{\delta p}{2}\right]\right|_{n_1=+\frac{u}{\beta \delta p} \atop n_2=-\frac{u}{\beta \delta p}}\:.
\end{eqnarray}
This expansion leads to
\begin{eqnarray}
S^{(2)}_2(u,p) = & u \left( \frac{2p}{\beta(1-q)} -2 \beta' f'(p) \right) + \frac{u^2 }{4(q-1)^2\beta^2} \times \nonumber \\
  & \times \Bigg(  4 p \delta q -4 q - \delta q^2 - 
 2 ( q-1)^2 \beta^4 \delta q^2 f'(q)^2 +  \nonumber
 \\ & + 4 ( q-1) \beta^2 (-2 + 2 q - 2 p \delta q + \delta q^2) f'(q) + \nonumber
 \\ & + ( q-1)^2 \beta^2 \delta q^2 f''(q) \Bigg)
\end{eqnarray}
which is the Taylor expansion of eq. (\ref{eq:exprescompletefullladevfuff}) till the second order in $u$. The $O(u)$ term is $V'(p)$, see eq. (\ref{potRSexpressionebetabetappp}), while the $O(u^2)$ gives the connected correlation function as expressed in eq. (\ref{otromodoconn2punt}). 
The agreement between $-S^{(2)}_2(u^*,p)/2$ and $G_2(p)$, see eq. (\ref{finaleq2smexprRS}), is obtained by replacing $\delta q$ with eq. (\ref{shiftsfinalexpressions1}), and optimizing over $u$. 
A fundamental difference with the large deviation computation is that, in the last case, the optimization over $\delta q$, $\delta q_{12}$ are not done in the limit $n_1 \rightarrow 0$, $n_2 \rightarrow 0$: saddle point equations need to be satisfied while adapting also the value of $u$, see eq. (\ref{eq:saddlepGLFD11}).
Here, on the other hand, the saddle point value for $\delta q$ ($\delta q_{12}$ does not appear in the second order expansion of $\Gamma$) is computed in the limit of the number of replicas going to zero and, thus, it is determined by the first term (the potential) of the expansion of $\Gamma$, see eqs. (\ref{eq:expansionS1inn})-(\ref{eq:expansionS2inn}).

\subsection*{Higher order expansion}

\begin{figure}
\centering{}\includegraphics[scale=0.5]{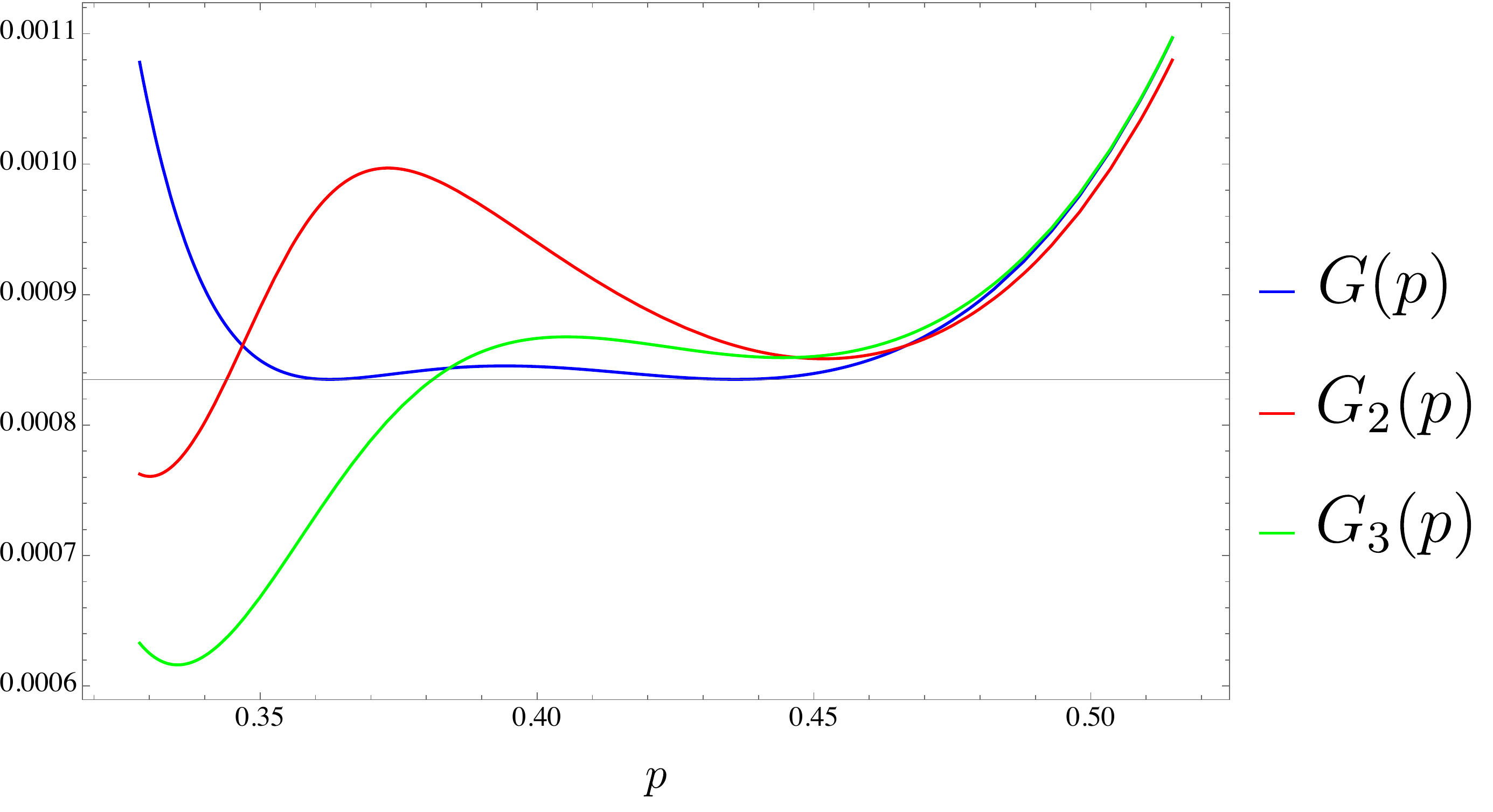}\caption{\label{GG2G3}$G(p)$ and $G_2(p)$ and $G_3(p)$ for the $3-$spin at $\beta=\beta'=1.55$ at the RS level for the $3$-spin. While $G(p)$ and $G_2(p)$ cross at $p_{RS}$, $G(p)$ and $G_3(p)$ do not.}
\end{figure}

Eq. (\ref{eq:computeILDSDl1}) is an expression for the large deviation function obtained expanding $\Gamma$ at the second order in $m$ and $\delta p$, see eq. (\ref{eq:gamma2expsecord}).
Here we discuss how to generalize this computation to the third order,
\begin{eqnarray}
 \Gamma_{3}\left(m,p+\frac{\delta p}{2},p-\frac{\delta p}{2}\right)= & m\left<\frac{}{}V'(p|\tau)\delta p+\frac{V'''(p|\tau)}{6}\delta p^{3}\right> + \nonumber  \\
& + \frac{N m^{2}}{2}\left<\left(\frac{}{}V'(p_{1}|\tau)\delta p \right)^{2}\right>_{c} + \nonumber \\ & + \frac{N^2 m^{3}}{6}\left<\left(\frac{}{}V'(p_{1}|\tau)\delta p\right)^{3}\right>_{c}+\ldots\:.
\end{eqnarray}
Similarly to what has been done previously in eq. (\ref{eq:deffirstg2tobegeneral}), eq. (\ref{eq:lagrangeTV2V1p2p1}) defines $G_3(V_2-V_1,p_1,p_2)$.
The derivative over $m$ leads to a second order expression, i.e. the optimization over $m$ produces two solutions. We take the solution that matches eq. (\ref{eq:gamma2solmstar}) in the limit when the
third term is zero.
Inserting this solution in $\Gamma_3$ and setting
$V_{2}-V_{1}=0$ in $G_3$ we obtain another expression for the large
deviation function, to be compared with eqs. (\ref{eq:computeILDSDl1})-(\ref{eq:computeILDSDl2}),
\begin{eqnarray}
G_3(p) = & \Bigg \{ 
\Bigg[    \left<\frac{}{}V'(p|\tau)^{2}\right>_{c}^{2} -2\left<\frac{}{} V'(p|\tau)\right>\:\left<\frac{}{} V'(p|\tau)^{3}\right>_{c} \Bigg]^{\frac{3}{2}} +
 \nonumber  \\
& + 3\left<\frac{}{}V'(p|\tau)\right>\:\left<\frac{}{}V'(p|\tau)^{2}\right>_{c}\:\left<\frac{}{}V'(p|\tau)^{3}\right>_{c}+ \nonumber \\
&  -\left<\frac{}{}V'(p|\tau)^{2}\right>_{c}^{3} \Bigg \} \left(3\left<\frac{}{}V'(p|\tau)^{3}\right>_{c}^{2}\right)^{-1}\:, \label{exp3mexplargedev} \\
N I_{3}^{LF}(T) = &  \min_{p} G_3(p).
\end{eqnarray}
Connected correlation functions are defined from derivatives of the
generating function $\Gamma$. From eq. (\ref{eq:gammaJexpansionmpp})
we have 
\begin{equation}
N^{k-1}\left<\frac{}{}V'(p|\tau)^{k}\right>_{c}=\lim_{\delta p \rightarrow 0}\frac{\partial^{k}\overline{\Gamma_{J}(0,p-\delta p /2,p+\delta p /2)}}{\partial m^{k}}\:.
\end{equation}
Below we write down the expression for the two and three point connected
correlation functions,
\begin{eqnarray}
\left<\frac{}{}V'(p|\tau)^{2}\right>_{c}= & \left< \frac{}{}V'(p|\tau)^{2}\right> - \left< \frac{}{} V'(p|\tau)\right>^{2}\:,\\
\left<\frac{}{}V'(p|\tau)^{3}\right>_{c}= & \left<\frac{}{}V'(p|\tau)^3\right> +2\left<\frac{}{}V'(p|\tau)\right>^{3} +
\nonumber \\ & -3  \left<\frac{}{}V'(p|\tau)^{2}\right>\left<\frac{}{}V'(p|\tau)\right>\:, 
\end{eqnarray}
and while the first one is $O(1/N)$, the second one is $O(1/N^{2})$.
They are defined by
\begin{equation}
\left<\frac{}{}V'(p|\tau)^{k}\right>_{c}=\lim_{\delta p\rightarrow0}\frac{1}{\delta p^k}\left<\left[\frac{}{}V\left(p+\frac{\delta p}{ 2}\Bigg|\tau\right)-V\left(p-\frac{\delta p}{ 2}\Bigg|\tau\right)\right]^{k}\right>_{c}\:
\label{eq:expkcorrfuncex}
\end{equation}
and, as illustrated below, they can be computed knowing 
\begin{eqnarray}
W^{(k)}_{p_1\ldots p_k} & =\left(-\frac{T}{N}\right)^k\left.\frac{\partial^{k} \log\left<\frac{}{}Z^{n_{1}}[p_{1}|\tau]\ldots Z^{n_{2}}[p_{k}|\tau]\right>_c}{\partial n_{1}\ldots \partial n_{k}}\right|_{n_k=0 \atop n_1=0}\:.
\end{eqnarray}
In fact, using eq. (\ref{eq:eqcorrfdefk1}), the last equation reads
\begin{equation}
W^{(k)}_{p_1\ldots p_k}=\left(-\frac{T}{N}\right)^k \frac{N}{2}\lim_{n_{1}\rightarrow0}\ldots \lim_{n_{k}\rightarrow0}\frac{\partial^{k}S^{(k)}[p_{1},\dots,p_{k}]}{\partial n_{1}\ldots \partial n_{k}}\:,\label{eq:eqW2pqres}
\end{equation}
and, expanding  eq. (\ref{eq:expkcorrfuncex}), leads to 
\begin{equation}
\left<\frac{}{}V'(p|\tau)^{2}\right>_{c}=\lim_{\delta p \rightarrow0} \frac{1}{\delta p^{2}}  \left(W_{p+\frac{\delta p}{2},p+\frac{\delta p}{2}}^{(2)}+W_{p-\frac{\delta p}{2},p-\frac{\delta p}{2}}^{(2)}-2W_{p+\frac{\delta p}{2},p-\frac{\delta p}{2}}^{(2)}\right)
\label{eq:Vp2finalresulttocompare}
\end{equation}
\begin{eqnarray}
\left<\frac{}{}V'(p|\tau)^{3}\right>_{c}=  \lim_{\delta p \rightarrow0} \frac{1}{\delta p^{3}}  \left(  T^{(3)}_{p+\frac{\delta p}{2},p-\frac{\delta p}{2}} - T^{(3)}_{p-\frac{\delta p}{2},p+\frac{\delta p}{2} } \right) \:,
\label{eq:Vp3finalresult}
\end{eqnarray}
where $T^{(3)}_{qt} = W_{q,q,q}^{(3)} -  3W_{q,q,t}^{(3)}$. The $k=3$ term can be computed generalizing eq. (\ref{eq:defS1firstappea}) and eq. (\ref{eq:defS2actiondeff}), 
\begin{eqnarray}
e^{\frac{N}{2}S^{(3)}[p_1, p_2, p_3]}= & \int\mathcal{D}Q_{ab}e^{\frac{N}{2}S[Q,n_{1}+n_{2}+n_{3}]}\prod_{a=1}^{n_{1}}\delta(p_{1}-q(s^{a},s^{0})) \nonumber \\ &
\prod_{a=n_{1}+1}^{n_{1+}n_{2}}\delta(p_{2}-q(s^{a},s^{0}))\prod_{a=n_{2}+1}^{n_{1+}n_{2}+n_{3}}\delta(p_{3}-q(s^{a},s^{0}))\:.
\end{eqnarray}
Using eq. (\ref{eq:eqW2pqres}), it is possible to see that eq. (\ref{eq:Vp2finalresulttocompare}) is equal to eq. (\ref{eq:var2Vpexpf}), while eq. (\ref{eq:Vp3finalresult}) provides the connected correlation function at three points. 
In both of them, when $p_k$ is shifted from $p$ by an $O(\delta p)$ term, the order parameters change according to eq. 
(\ref{shiftsfinalexpressions1})-(\ref{shiftsfinalexpressions3}). These equations have been derived for $S^{(2)}$ but they are valid for any $S^{(k)}$ because the expansion done in eq. (\ref{eq:expansionS2inn}) can be generalized for any $k$. 
While the first order terms of this expansion coincide with those of eq. (\ref{eq:expansionS2inn}),
\begin{eqnarray}
S^{(3),(1,0,0)}[p_1] = & S^{(1),(1)}[p_1]\:, \\
S^{(3),(0,1,0)}[p_2] = & S^{(1),(1)}[p_2]\:, \\
S^{(3),(0,0,1)}[p_3] = & S^{(1),(1)}[p_3]\:,
\end{eqnarray}
the second order terms coincide with the mixed term of $S^{(2)}$, 
\begin{eqnarray}
S^{(3),(1,1,0)}[p_1,p_2] = & S^{(2),(1,1)}[p_1,p_2]\:, \\
S^{(3),(1,0,1)}[p_1,p_3] = & S^{(2),(1,1)}[p_1,p_3]\:, \\
S^{(3),(0,1,1)}[p_2,p_3] = & S^{(2),(1,1)}[p_2,p_3]\:.
\end{eqnarray}
Thus, when evaluating the saddle point equations of $S^{(k)}$, $q_{lm}$ is only determined by $p_l$ and $p_m$ for any $k$, not only for $k=2$, according to eq. (\ref{eq:speqrsmixedtermexpression}). Similarly, the shift in $ q_{lm}$, $\delta q_{lm}$, due a shift in $p_l$ or $p_m$ has the same form derived in eq. (\ref{shiftsfinalexpressions3}), while it is zero when shifts are made on $p_q$ with $q \neq l,m$.
Using these expressions for the connected correlation functions in eq. (\ref{exp3mexplargedev}) we obtain $G_3(p)$ and we may compare it with $G_2(p)$ and $G(p)$. This is done in Fig. \ref{GG2}.
We notice that $G_2(p)$ and $G(p)$ cross in $p_{RS}$ $\forall T$. This may suggest the peculiarity of this point, where a second order Taylor expansion equals the non-perturbative result. Nevertheless this does not happen for $G_3(p)$ and $G(p)$. 
 
\section*{Further details on the replica computation}

In this section we give more details on the computation of the determinant in eq. (\ref{eq:actionresultgeneral}) and on the 1RSB ansatz used to evaluate both the potential and the large deviation function.
In the 1RSB ansatz, the diagonal block matrix $Q^k_{ab}$ described previously, reads
\begin{equation}
Q_{ab}^{k}=q_{k}^{0}+(q_{k}^{1}-q_{k}^{0})\epsilon_{ab}^{k}+(1-q_{k}^{1})\delta_{ab}\:,
\end{equation}
where the matrix $\epsilon_{ab}^{k}$ has ones
on the diagonal block of size $x_{k}$ and zeros elsewhere, with
$n/x_{k}\in\mathbb{N}$. To ease the notation we use $y_k$ to denote the saddle point parameters relative to replica $k$. For each $k$, in the RS ansatz $y$ is just the overlap $q$, while in the 1RSB ansatz it is the collection $\{q^1, q^0, x\}$.
On the other hand, the elements of the out-of-diagonal blocks are taken to be the same.
The saddle point values of $q^0$, $q^1$ and $x$ at a given $p$ are given by
\begin{eqnarray}
\frac{x [p^2 - q^0 + (1 + q^1 (x-1) - q^0 x]^2 \beta^2      f'(q^0)}{ [1 + q^1 ( x-1) - q^0 x]^2}& = 0\:, \label{eq.qpRSVpot1}
\\
\left[\frac{(p^2-q^1) (q^1-1) + (q^0-q^1)^2 x}{( q^1-1) [1 + q^1 ( x-1) - q^0 x]^2
}  + \beta^2 f'(q^1)\right](x-1) & = 0 \:,  \label{eq.qpRSVpot2} \\ 
 \frac{(q^0-p^2)(q^0-q^1)}{[1+q^1(x-1)-q^0x]^2}+\frac{q^1-q^0}{x[1+q^1(x-1)-x q^0]} + \nonumber \\
+\beta^2[f(q^1)-f(q^0)] - \frac{1}{x^2} \log\left[1+\frac{(q^1 - q^0)x}{1-q^1} \right]& = 0 \:.
\label{eq.qpRSVpot3}
\end{eqnarray}
When to use the RS or the 1RSB scheme will be discussed later. 

\begin{figure}
\centering{}\includegraphics[scale=0.8]{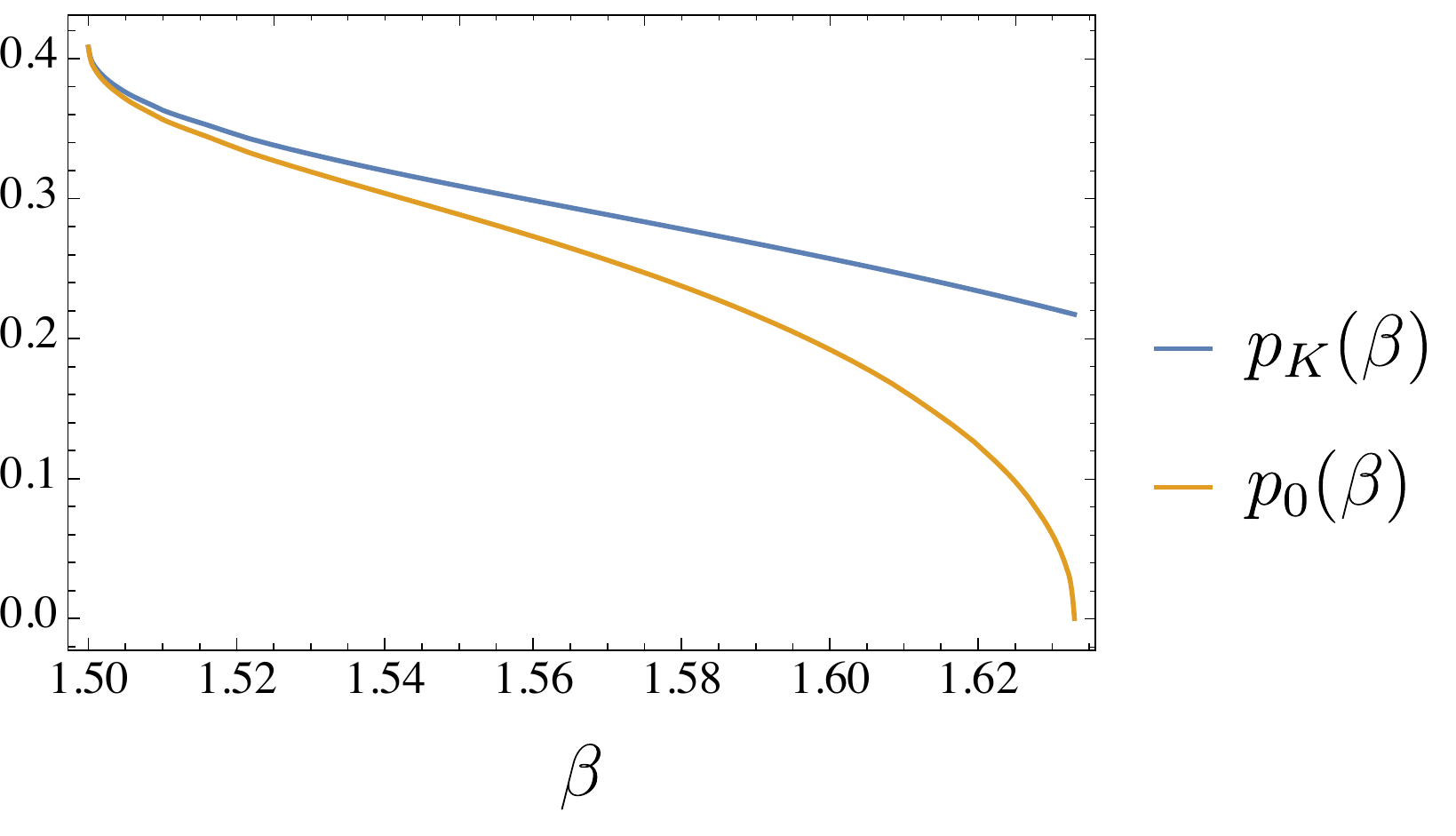}\caption{\label{p0pK3}Values of $p_0$ and $p_K$ for the $3$-spin for $\beta<\beta_d$. The complexity $\Sigma(p)$ is positive for $p_0<p<p_K$, being zero at $p_K$.}
\end{figure}

\begin{figure}
\centering{}\includegraphics[scale=0.8]{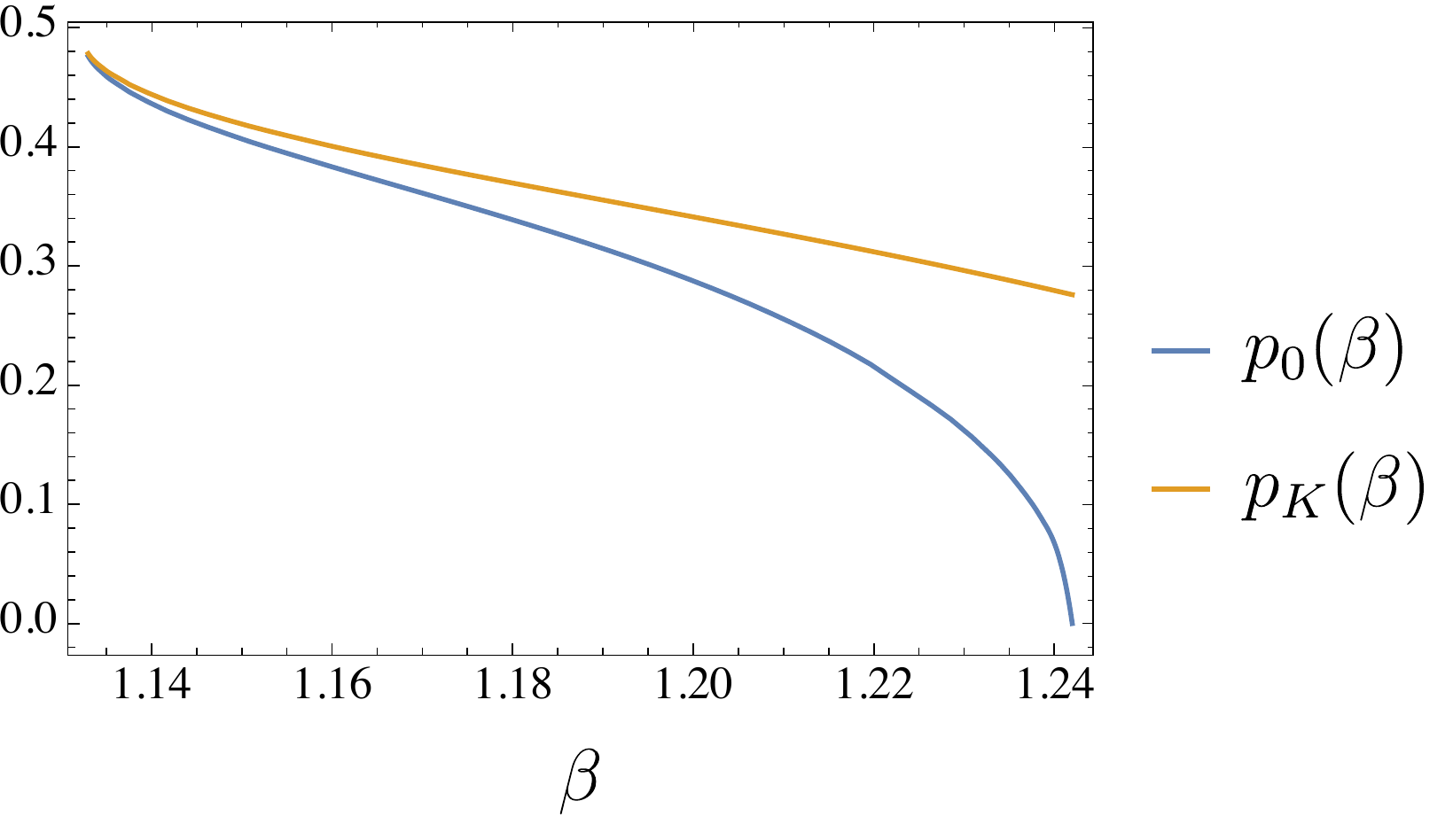}\caption{\label{p0pK34}Values of $p_0$ and $p_K$ for the $3+4$-spin for $\beta<\beta_d$. The complexity $\Sigma(p)$ is positive for $p_0<p<p_K$, being zero at $p_K$.}
\end{figure}

In order to obtain the expression of $S^{(k)}[p_1,p_2,\ldots,p_k]$ we need to compute
the determinant of the $Q$ matrix, thus its spectrum. Let us first
discuss the simple case of $k=1$. In this case, the spectrum of $Q$ is
found by looking for eigenvectors that have a $1+n$ structure. We call these components $u$, $v_{a},\;a=1,\ldots,n$, obtaining the following eigenvalue equations
\begin{equation}
\left\{ \begin{array}{ll}
u+p\sum_{b}v_{b}=\lambda u\\
p u+\sum_{b}Q_{ab}v_{b}=\lambda v_{a}
\end{array}\right.\:.
\end{equation}
Let us set $u=0$, $\sum_{a}v_{a}=0$, leading to $n-1$ eigenvectors
of $Q_{ab}$. Given the 1RSB structure of $Q_{ab}$, the
equation $\sum_{b}Q_{ab} v_{b}=\lambda v_{a}$ is equal to 
\begin{equation}
q_{}^{0}\sum_{b}v_{b}+(q_{}^{1}-q_{}^{0})\sum_{b}\epsilon_{ab}^{}v_{b}+(1-q_{}^{1})v_{a}=\lambda v_{a}\:.
\end{equation}
We look for two different kinds of solutions:
\begin{enumerate}
\item we may require that $\sum_{b}\epsilon_{ab}^{2}v_{b}=0\:\forall a$,
i.e. the partial sum $\sum_{b\in block_{a}}v_{b}=0$, where $block_{a}$
is the block to which replica $a$ belongs to. We obtain eigenvectors
with eigenvalue $\lambda=1-q_{}^{1}$. Since we have $n/x_{}$
blocks and each one has size $x_{}$, the multiplicity of this eigenvalue
is $n/x_{}(x_{}-1)$;
\item we may consider the case when in each of $n/x_{}$ blocks we have
a distinct value of $v_{b}$ and their sum is zero. This condition
leads to eigenvectors with eigenvalue equal to $\lambda=(q_{}^{1}-q_{}^{0})x_{}+(1-q_{}^{1})$,
whose multiplicity is equal to $n/x_{}-1$. 
\end{enumerate}
Counting the degeneracies of the eigenvalues obtained up to now, we
see that we have got $n-1$ eigenvectors of $Q_{ab}^{}$ . Thus,
we see that we still miss one eigenvector of $Q_{ab}^{}$. Clearly
the constant vector $\overline{v}=(v_{a=1,\ldots,n})$ is the last
eigenvector of $Q_{ab}^{}$, with eigenvalue equal to $\sum_{b}Q_{ab}^{}$,
ruled out by the condition $\sum_{a}v_{a}=0$. Neverthless this vectors
is not an eigenvector of the matrix $Q$, and we observe that in order
to find the 2 missing ones we need to solve the reduced eigenvalue
problem

\begin{equation}
\left\{ \begin{array}{ll}
u+pnv=\lambda u\\
p_{}u+\sum_{b}Q_{ab}^{}v=\lambda v
\end{array}\right.\:.
\end{equation}
$F$ can be easily computed from the condion $V(0)=0$ and realizing
that in this case the overlap matrix $Q$ reduces to a single number,
1. We find $F=-\beta f(1)/2=-\beta/4$. While $p_{}$ is a control
parameter of the problem, $q_{}^{1}$, $q_{}^{0}$ and $x_{}$
are parameters we need to optimize over. This task cannot be accomplished
in a straightforward way \cite{barrat1997temperature}. In fact, for $\beta$
in the dynamical phase, physical intuition suggests the existance
of three regions: 
\begin{itemize}
\item one at very small values of $p_{}$, where the second replica is
weekely constrained to the first one, having the possibility to explore
an exponential number of states and leading thus to a dynamical 1RSB
phase, where the $x_{}=1$ and $q_{}^{1}\neq q_{}^{0}$;
\item one at intermediate values of $p_{}$, where the second replica
can explore $O(N)$ metastable states, leading to a static 1RSB phase
with $x_{}<1$ and $q_{}^{1}\neq q_{}^{0}$;
\item one at very large values of $p_{}$ where the second replica is
forced to be very close to the first one, exploring only the state
where the first one is and leading thus to an RS phase where $x_{}=1$
and $q_{}^{1}=q_{}^{0}$.
\end{itemize}
The vanishing of the complexity at $p_{K}$ identifies the transition
between the dynamical 1RSB region and the static one, while the instability
of the replicon at $p_{RS}$ identifies the transition between the
static 1RSB region and the RS one. 
The RS instability is detected by looking at the largest solution of the replicon equation
$\beta^2(1-q)^2 f''(q)=1$. This solution survives as long as $\beta>\beta_{RS}$.
$\beta_{RS}$ is equal to $1.5$ in the $3$-spin and equal to $1.13234$ in the $3+4$-spin. This point coincides also with the point where $p_0$ and $p_K$ cross, see Fig. \ref{p0pK3}-\ref{p0pK34}. 
Thus, for $\beta<\beta_{RS}$ the potential is always RS.


\begin{figure}
\centering{}\includegraphics[scale=0.5]{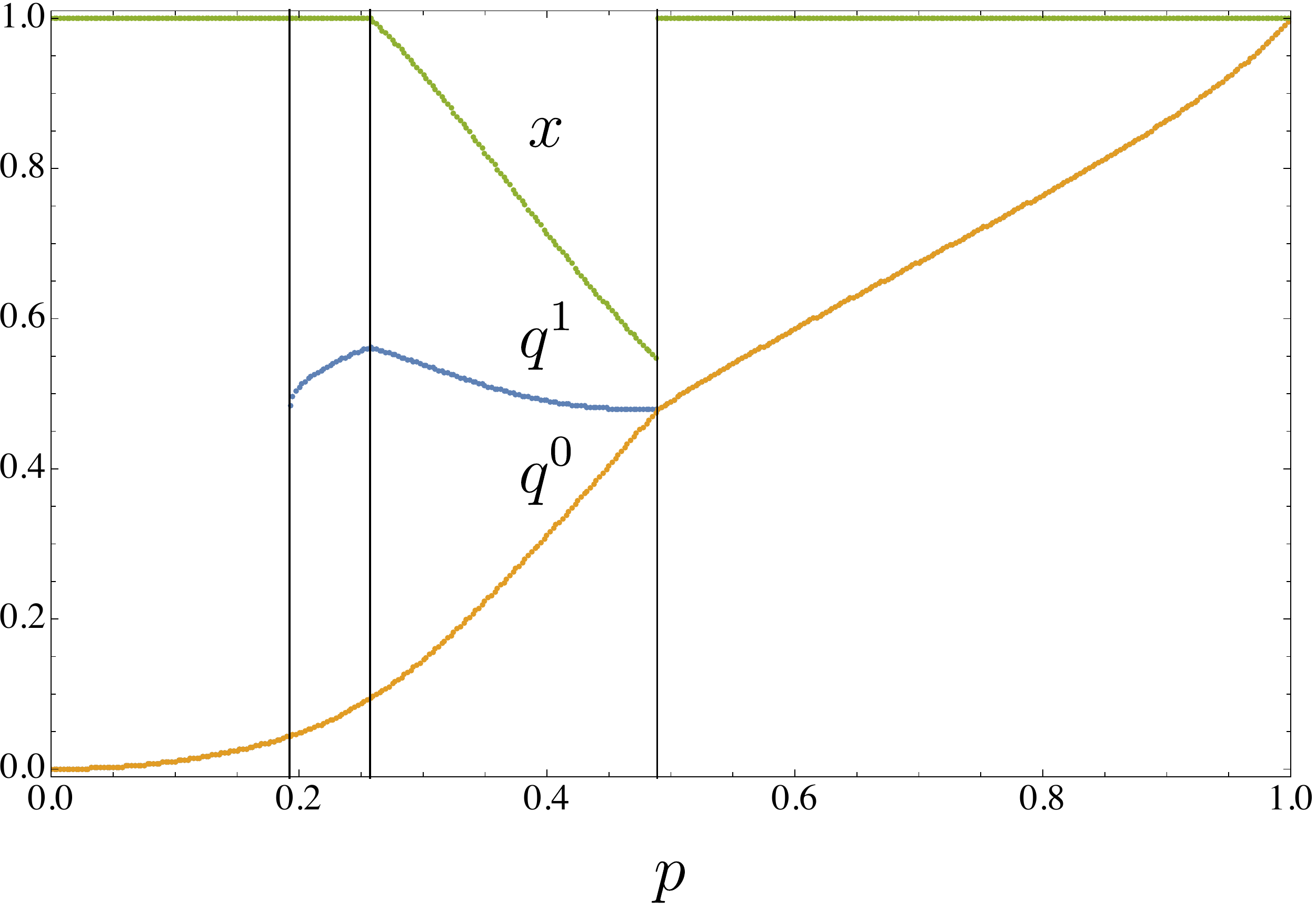}\caption{\label{gifspvluesT}Saddle point values for the $3$-spin for $q_{}^{1}$ (blue), $q_{}^{0}$ (orange),
and $x_{}$ (green) for different values of $p_{}$ at $\beta=1.6$,
smaller than $\beta_{d}$. We observe the four phases discussed in
the text: from left to right the preliminar RS phase, the dynamic
1RSB phase, the static 1RSB phase and the final RS phase. The three vertical lines from left to right are $p_0$, $p_K$ and $p_{RS}$.}
\end{figure}

When $\beta<\beta_{d}$,
already in the limiting case $p_{}=0$, the second replica cannot
explore an exponential number of states. This implies the existence
of a preliminar RS phase at low values of $p_{}<p_{0}$. It turns
out that the RS ansatz is locally stable even for $p_{}>p_{0}$
and thus, in order to detect $p_{0}$ we study the stability of the
1RSB solution in the dynamical 1RSB region. Thus, in general we
have four regions.
While in the first and in the last regions we have
to fix $x_{}=1$, $q_{}^{1}=q_{}^{0}$ and solve the RS saddle
point equations, in the other ones we need to solve the 1RSB saddle
point equations. Anyway, in the dynamical 1RSB region, we need to
fix $x_{}=1$. It is worth noticing that the dynamical 1RSB region
is characterized by a finite complexity $\Sigma(p_{})$ and that
the point at which the complexity goes to zero separates this region
from the static 1RSB region. The computation of the complexity in
the dynamical 1RSB region will be addressed later. In Fig. \ref{gifspvluesT} we show the value of the saddle point values at a particular temperature in the paramagnetic phase. 

When dealing with $k>1$ the same reasoning can be
repeated, looking respectively for eigenvectors with a $1+n_{1}+n_{2}$ 
and $1+n_{1}+n_{2}+n_{3}$ structure for $k=2$ and $k=3$, and setting $u=0,\sum_{a}v_{a}^{1}=0,\sum_{a}v_{a}^{2}=0,\sum_{a}v_{a}^{3}=0$
. This lead to $n_{1}+n_{2}-2$ and $n_{1}+n_{2}+n_{3}-3$ eigenvalues,
respectively, with the same multiplicity discussed above. In the first
case we find, besides those mentioned above, an eigenvalue equal to
$1-q_{2}^{1}$ with multiplicity $n_{2}/x_{2}(x_{2}-1)$
and another one equal to $(q_{2}^{1}-q_{2}^{0})x_{2}+(1-q_{2}^{1})$
with multiplicity $n_{2}/x_{2}-1$. These are $n_{2}-1$ eigenvalues
of the matrix $Q_{ab}^{2}$. Similarly, in the second case, we
also have an eigenvalue equal to $1-q_{3}^{1}$ with multiplicity
$n_{3}/x_{3}(x_{3}-1)$ and another one equal to $(q_{3}^{1}-q_{3}^{0})x_{3}+(1-q_{3}^{1})$
with multiplicity $n_{3}/x_{3}-1$, which are $n_{3}-1$ eigenvalues
of the matrix $Q_{ab}^{3}$. Finally, for $k=2$ we need to solve the system

\begin{eqnarray}
\left\{ \begin{array}{cc}
u+p_{1}n_{1}v^{1}+p_{2}n_{2}v^{2}&=\lambda u\\
p_{1}u+\sum_{b}Q_{ab}^{1}v^{1}+n_{2}q_{12}v^{2}&=\lambda v^{1}\\
p_{2}u+n_{1}q_{12}v^{1}+\sum_{b}Q_{ab}^{2}v^{2}&=\lambda v^{2}
\end{array}\right.\:,
\end{eqnarray}
and
while for $k=3$ we need to solve the system
\begin{eqnarray}
\left\{ \begin{array}{cc}
u+p_{1}n_{1}v^{1}+p_{2}n_{2}v^{2}+p_{3}n_{3}v^{3}&=\lambda u\\
p_{1}u+\sum_{b}Q_{ab}^{1}v^{1}+n_{2}q_{12}v^{2}+n_{3}q_{13}v^{3}&=\lambda v^{1}\\
p_{2}u+n_{1}q_{12}v^{2}+\sum_{b}Q_{ab}^{2}v^{2}+n_{3}q_{23}v^{3}&=\lambda v^{2}\\
p_{3}u+n_{1}q_{13}v^{1}+n_{2}q_{23}v^{2}+\sum_{b}Q_{ab}^{3}v^{3}&=\lambda v^{3}
\end{array}\right.\:.
\end{eqnarray}

\subsection*{Complexity}

We describe the analysis of the number of equilibrium states accessible to a system
constrained to be at a given distance from a reference configuration, using the formalism developed in  \cite{monasson1995structural,barrat1997temperature,mezard1999compute}. It is convenient
to start from the situation when $p=0$, i.e. when the second replica of the potential, is not constrained by the first one. For $\beta>\beta_{d}$, the system can explore an exponential number
of TAP states, but increasing $p$ this number decreases. The
complexity $\Sigma(p_{})$ is the logarithm of this number and can
be computed introducing the free energy of $m$ coupled replicas,
constrained to be at a fixed overlap $p_{}$ from a reference configuration

\begin{equation}
Z_{m}^{c}=\sum_{\alpha}e^{-\beta mNf_{\alpha}^{c}}=e^{-N\beta \Phi_{c}(m,T)}\:.
\end{equation}
In fact, 

\begin{equation}
Z_{m}^{c}=\int df_{c}\sum_{\alpha}\delta(f_{c}-f_{\alpha}^{c})e^{-\beta mNf^{c}}=e^{N\left(\Sigma(p_{},f_{c}^{*}(m,T),m)-\beta mf_{c}^{*}\right)}\:,
\end{equation}
where $f_{c}^{*}$ denotes the saddle point value of the constrained
free energy, and thus we see that $T/m$ plays the role of an effective
temperature in the relation between $m^{-1}\Phi_{c}(m,T)$ and
the complexity:

\begin{equation}
m^{-1}\Phi_{c}(m,T)=f_{c}^{*}-\frac{T}{m}\Sigma(p_{},f_{c}^{*}(m,T),m)\:.
\end{equation}
As the entropy can be computed from the free energy as
$S=-dF/dT$, the complexity reads

\begin{eqnarray}
\Sigma(p_{})&\equiv\Sigma(p_{},f_{c}^{*}(m,T),m)= \nonumber \\& =-\frac{\partial(m^{-1}\Phi_{c}(m,T))}{\partial(T/m)}=m^{2}\frac{\partial(\beta m^{-1}\Phi_{c}(m,T))}{\partial m}
\label{defSigmadausarepfin}
\end{eqnarray}
and so all we need to do is compute $\Phi_{c}(m,T)$. By definition,
\begin{equation}
-\beta N\Phi_{c}(m,T)=\overline{\mathbf{\mathbb{E}_{\tau}}\log\left[\int D\sigma e^{-\beta H(\sigma)}\delta\left(Np_{}-\sum_{i}\sigma_{i}\tau_{i}\right)\right]^{m}}
\end{equation}
and thus, introducing $n$ replicas to deal with the logarithm, and
using that $\lim_{n\rightarrow0}\overline{\log x}=\lim_{n\rightarrow0}\partial_{n}\overline{x^{n}}$
we obtain

\begin{equation}
-\beta N\Phi_{c}(m,T)=\lim_{n\rightarrow0}\partial_{n}\overline{\mathbf{\mathbb{E}_{\tau}}\left[\int D\sigma e^{-\beta H(\sigma)}\delta\left(Np_{}-\sum_{i}\sigma_{i}\tau_{i}\right)\right]^{nm}}
\end{equation}
Using eq. (\ref{eq:eqcorrfdefk1}) with $k=1$,
we may write the r.h.s. of this equation as 

\begin{equation}
-\beta N\Phi_{c}(m,T)=\lim_{n\rightarrow0}\partial_{n}e^{\frac{N}{2}\left[\beta^{2}\sum_{ab}f(Q_{ab})+\log\det Q\right]_{n\rightarrow mn,\:x_{}\rightarrow m}}
\end{equation}
where $Q_{ab}$ has size $1+m n$. In fact, when $k=1$, the matrix $Q$ in eq. (\ref{eq:actionresultgeneral}) has size $1+n$ and in the 1RSB ansatz, the $n \times n$ sub-matrix has a block structure with block-size equal to $x$. Here the role of $x$ is played by $m$. A comparison with eq. (\ref{eq:expansionS1inn}) leads to
\begin{eqnarray}
-\beta N\Phi_{c}(m,T)&=\left. \lim_{n\rightarrow0}\partial_{n}\frac{N}{2}S^{(1)}[p] \right|_{\;\: n \rightarrow nm \atop x_{}\rightarrow m}=\\ &=  \left. \frac{N}{2}mS^{(1),(1)}[p]\right|_{ x_{}\rightarrow m}\:,
\end{eqnarray}
Thus we get
\begin{equation}
m^{-1}\Phi_{c}(m,T)=-\left.\frac{T}{2} S^{(1),(1)}[p]\right|_{ x_{}\rightarrow m}\:.
\end{equation}
Free energy $F$ apart, the expression for $\Phi_{c}(m,T)$ is equivalent to that of the potential replicated $m$ times where $x \rightarrow m$,  see eq. (\ref{eq:potintermsofS11}). 
Since $\Phi_{c}(m,T)$ is the free energy of $m$ copies of the system constrained to be at distance $p$ from a reference configuration, this could be guessed from its definition.
Finally, using eq. (\ref{defSigmadausarepfin}), the constrained complexity reads
\begin{equation}
\Sigma(p_{})=-\frac{1}{2}\frac{\partial \left. S^{(1),(1)}[p]\right|_{ x_{}\rightarrow m} }{\partial m}\:,
\end{equation}
i.e.
\begin{eqnarray}
2 \Sigma(p)=&\frac{q^0 - q^1}{m [1 + q^1 (m-1) - q^0 m]}  +\frac{(p^2 - q^0) (q^0 - q^1)}{[q^1 -1 +  m (q^0- q^1) ]^2} + \nonumber \\
& +\beta^2 f(q^0) - \beta^2 f(q^1) + \frac{1}{m^2}\log \left( 1 + \frac{(q^0 - q^1) m)}{ q^1-1} \right)\:.
\label{eq:fullsigmapgeneral}
\end{eqnarray}
For $\beta>\beta_d$, the complexity is larger than zero for small values of $p$, i.e. when the system is not very constrained. 
For all the values of $p_{}<p_{RS}$ $\Sigma(p_{})$ as well as
$V(p_{})$ has to be computed on the solutions $q_{}^{1}$, $q_{}^{0}$
of the 1RSB saddle point equations where $x_{}=1$ ($m=1$ in $\Sigma(p)$).
For $p_{}>p_{RS}$,
in order to get $\Sigma(p_{})=0$, $x_{}$ cannot be taken equal to $1$ anymore
and the system enters in a static 1RSB phase. As explained previously, this region lasts until
when the 1RSB saddle point equations give $q_{}^{1}\neq q_{}^{0}$
and corresponds to intermediate values of $p_{}$ after which a
RS region appears, because the constraint is so strong that the second
replica can only explore the state of the first one. At $p_{}=0$
this expression gives $\Sigma(0)=TV(q_{EA})$, the unconstrained complexity.
This equality comes from the observation that in the dynamical phase one has $F=F_{RS}=f-\Sigma(0)T$, where $f=F(q_{EA})$ is the TAP free energy of the system in one equilibrium TAP state, whose number is $e^{N\Sigma(0)}$, and $F_{RS}$ denotes the free energy computed at the RS level. On the other hand, by definition $F(q_{EA})=V(q_{EA})+F$. Thus $\Sigma(0)=TV(q_{EA})$.
Finally we notice that the expression of the saddle point eq. (\ref{eq.qpRSVpot3}) obtained optimizing over $x$ is equal to the expression of the complexity, see  eq. (\ref{eq:fullsigmapgeneral}). This means that when in the static 1RSB regime we impose eq. (\ref{eq:fullsigmapgeneral}), we are forcing the complexity to be zero, as explained above. 

When $p=0$, $q^0$ is zero as can be observed in Fig. \ref{gifspvluesT}. The complexity is thus a function of $q^1$ and $m=x$ only.
Having set $p=0$, there no more constraints and thus we remove the subscript $c$ from $\Phi_c$
The optimization of $\Phi$ over $q^1$ produces 
\begin{equation}
\frac{q^1}{( q^1-1) (1 + (m-1) q^1)} + \beta^2 f'(   q^1) = 0
\end{equation} 
and it is possible to check that this equation corresponds to eq. (\ref{eq.qpRSVpot2}) in the limit $q^0=0$.
Solving this equation for $m$ and using this solution in eq. (\ref{eq:fullsigmapgeneral}), setting $p=q_0=0$, we obtain
\begin{eqnarray}
2 \Sigma(q) =& [\chi(q) q^2 f'(q) ]^{-1} \Bigg[ -f(q) [q - \chi(q)]^2  +q \beta^2 f'(q)^2 \times \nonumber \\
& \times  \Bigg( ( q-1 )^2 q \left( \log\frac{q}{\chi(q)}-1 \right) + ( q-1 )^2 \chi(q) \Bigg) \Bigg]
\end{eqnarray}
where we set $q^1=q$ and $\chi(q)=\beta^2 (q-1)^2 f'(q)$. This expression gives the complexity of states as a function of their overlap $q$ at any given $T$.  

\end{document}